\documentclass[journal,10pt,twoside,twocolumn]{./IEEEtran}

\usepackage{epsfig,endnotes}
\usepackage{color}
\usepackage{subfig}
\usepackage{amsmath}
\usepackage{bm}
\usepackage{url}

\usepackage{amsmath}

\usepackage{bm}
\usepackage{amsfonts}

\usepackage{acronym}

\usepackage{verbatim}

\usepackage{hyperref}

\usepackage[svgnames]{xcolor}
\usepackage{todonotes}

\usepackage{color}
\usepackage{colortbl}
\definecolor{light-gray}{gray}{0.45}

\usepackage{ragged2e}

\usepackage{amssymb}%
\usepackage{mathtools}

\DeclareSymbolFont{cmsymbols}{OMS}{cmsy}{m}{n}
\SetSymbolFont{cmsymbols}{bold}{OMS}{cmsy}{b}{n}
\DeclareSymbolFontAlphabet{\mathcal}{cmsymbols}

\DeclareMathOperator{\Tr}{Tr}

\newcommand{\tnsr}[1]{\mathcal{\uppercase{#1}}}					% a discrete tensor command
\newcommand{\mtrx}[1]{\textbf{\uppercase{#1}}}					% a discrete matrix command with capital boldface
\newcommand{\vect}[1]{\textbf{#1}}                              % a discrete vector command with boldface instead of arrow
\newcommand{\expectation}{\boldsymbol{\mathsf{E}}\!}            % expectation command
\newcommand{\revdots}{\mathinner{\mkern1mu\raise1pt\vbox{\kern7pt\hbox{.}}\mkern2mu\raise4pt\hbox{.}\mkern2mu \raise7pt\hbox{.}\mkern1mu}}

\colorlet{olivegreen}{green!50!black}
\colorlet{highlight_clr}{black}

\usepackage{xspace}

\usepackage{caption}
\usepackage{subfig}

\usepackage{pgfplots}
\pgfplotsset{compat=newest}
\pgfplotsset{width=0.75\textwidth}
\usepgflibrary{patterns}

\usepackage{tikz}
\usetikzlibrary{patterns}
\usetikzlibrary{calc}
\usetikzlibrary{arrows,shapes}
\usetikzlibrary{shapes.callouts,decorations.pathmorphing}
\usetikzlibrary{shapes.arrows}

\usepackage{varwidth}

\usepackage{ctable}

\usepackage{siunitx}

\usepackage{chngpage}

\usepackage{nth}

\usepackage{fancyvrb}

\usepackage{balance}

\begin{document}

\bstctlcite{ieee_bst_cntrl}

%don't want date printed
\date{}

%make title bold and 14 pt font (Latex default is non-bold, 16 pt)
\title{On Tracking the Physicality of Wi-Fi: \\
A Subspace Approach}

\author{Mohammed~Alloulah, %~\IEEEmembership{Member,~IEEE,}
        Anton~Isopoussu,
        Chulhong~Min,
        and~Fahim~Kawsar% <-this % stops a space
\thanks{M. Alloulah, A. Isopoussu, C. Min, and F. Kawsar are with Nokia Bell Labs, Cambridge,
UK (e-mail: first.second@nokia-bell-labs.com).}% <-this % stops a space
\thanks{The authors would like to thank Howard Huang for the helpful comments on this manuscript.}
}

\maketitle

% Use the following at camera-ready time to suppress page numbers.
% Comment it out when you first submit the paper for review.
%\thispagestyle{empty}

\begin{abstract}
Wi-Fi channel state information (CSI) has emerged as a plausible modality for sensing different human activities as a function of modulations in the wireless signal that travels between wireless devices.  
Until now, most research has taken a statistical approach and/or purpose-built inference pipeline. 
Although interesting, these approaches struggle to sustain sensing performances beyond experimental conditions. 
As such,  the full potential of CSI as a general-purpose sensing modality is yet to be realised. 
We argue a universal approach with well-grounded formalisation is necessary to characterise the relationship between wireless channel modulations (spatial and temporal) and human movement. 
To this end, we present a formalism for quantifying the changing part of the wireless signal modulated by human motion. 
Grounded in this formalisation, we then present a new subspace tracking technique to describe the channel statistics in an interpretable way, which succinctly contains the human modulated part of the channel. 
We characterise the signal and noise subspaces for the case of uncontrolled human movement, \textcolor{highlight_clr}{and show that these subspaces are dynamic}.
Our results demonstrate that proposed channel statistics alone can robustly reproduce state-of-the-art application-specific feature engineering baseline, however, across multiple usage scenarios. 
We expect, our universal channel statistics will yield an effective general-purpose featurisation of wireless channel measurements and will uncover opportunities for applying CSI for a variety of human sensing applications in a robust way.
\end{abstract}


%%%%%%%%%%%%%%%%%%%%%%%%%%%%%%%%%%%%%%%%%%%%%%%%%%%%%%%%%%%
\section{Introduction}
%%%%%%%%%%%%%%%%%%%%%%%%%%%%%%%%%%%%%%%%%%%%%%%%%%%%%%%%%%%

Due to the ubiquity and penetration of Wi-Fi in our homes, workplaces and cities, Wi-Fi traffic can be repurposed as a sensing modality for many potential applications beyond the original intended data-carrier functionality. Indeed, recent compelling research has reimagined a commodity Wi-Fi device as a multi-purpose sensor capable of turning \emph{Wi-Fi traffic}---that is, packets transmitted over a wireless communication channel for either data transfer and/or the judicious probing of the channel---into a rich source of computational information explaining space dynamics, assessing the social environment and even tracking people's posture, and gestures~\cite{Yousefi17_SurveyOnWifiBehaviourRecognition, Xi14_ElectronicFrogEyeCrowdCounting, Depatla15_OccupancyEstimationUsingWifi,Wang18_LowEffortDfl,Wang15_UnderstandingAndModelingWiFiHumanActivity}.  

However, human-perturbed Wi-Fi channels remain ill-understood. Despite prior art showcasing compelling use cases, ad hoc inference pipeline and careful parameter tuning are commonplace for arriving at sensing recipes that yield good performance. Essentially, conventional approaches seek to associate patterns in Wi-Fi channel state information (CSI)  with human activity through training classifiers on top of often bespoke featurisation e.g. statistical distributions in~\cite{Depatla15_OccupancyEstimationUsingWifi} and Doppler variations in~\cite{Wang15_UnderstandingAndModelingWiFiHumanActivity}. Although these sensing approaches demonstrated the potential of CSI sensing in a brand new class of applications, often, they are sensitive to environmental conditions and thus require controlled setup and development of pre-processing and inference pipelines which do not generalise across tasks (i.e. applications) and deployment environments. As such, CSI as a general-purpose sensing modality has not been adopted widely. 

We argue for unleashing the true potential of CSI as a general-purpose human sensing modality; we need to turn our attention to developing sound theories explaining the relationship between spatiotemporal wireless channel modulations and human movement. 
Such characterisation will assist in designing the future Wi-Fi network with stack layers augmented with annotations derived from the wireless propagation medium.
\textcolor{highlight_clr}{These annotations would describe the physicality induced by the dynamic human movement which accompanied the delivery of data, thereby providing added \emph{context}.}

To this end, in this paper we present a first formalisation for quantification of the changing part of the wireless signal modulated by human motion. 
Based on established channel models we devise new channel statistics that succinctly characterise the signal modulated by dynamic human movement. 
We then demonstrate that these channel statistics carry enough information to describe spatiotemporal human movement when observed continuously. 
This leads us to develop a novel subspace tracking algorithm that continuously analyses signal subspace as a function of dynamic human movement. 
The application of such metric enables us to precisely describe a set of human movement primitives including presence, motion activities, etc. 
As a step towards realising CSI as a general-purpose sensing modality, we showcase how features extracted from the evolution of these subspaces can robustly reproduce \textcolor{highlight_clr}{state-of-the-art} application-specific feature engineering baseline, however, across multiple usage scenarios and environmental conditions. 
Our research contributions are three-fold:

\begin{enumerate}
    \item Statistical analysis of the signals to formally devise new statistics characterising human-perturbed Wi-Fi channels.
    \item Formalisation of CSI sensing as a subspace tracking problem, demonstrating that the analysis of the dynamics of a signal subspace is the equivalent of sensing human movements.
    \item Quantification of the benefits of using features derived from the proposed statistics and corresponding tracking technique concerning bleeding-edge CSI sensing applications.
\end{enumerate}

We start by reviewing the required mathematical background of channel modelling in Section~\ref{sec:mimo_model}. How the channel model can be used for sensing is explained in Section~\ref{sec:observation_model}.
We use subspace based statistics to analyse human modulation of wireless channels in Section~\ref{sec:subspace_characterisation}. 
We show that the analysis of the dynamics of a signal subspace is equivalent to sensing human movements.
We show by way of example how features extracted from subspace evolution can be used to solve sensing tasks in Section \ref{sec:subspace_tracking}. 
\textcolor{highlight_clr}{We evaluate our subspace tracking featurisation for two applications in Section~\ref{sec:evaluation}, provide a discussion in Section~\ref{sec:discussion}, and conclude with Section~\ref{sec:conclusion}.}

%%%%%%%%%%%%%%%%%%%%%%%%%%%%%%%%%%%%%%%%%%%%%%%%%%%%%%%%%%%
\section{Measurement Model} \label{sec:measurement_model}
%%%%%%%%%%%%%%%%%%%%%%%%%%%%%%%%%%%%%%%%%%%%%%%%%%%%%%%%%%%

\vspace{-0.75cm}
%++++++++++++++++++++++++++++++++++++++++++++++++++++++++++
\textcolor{highlight_clr}{\subsection{Notation}}\label{desc:notation}
%++++++++++++++++++++++++++++++++++++++++++++++++++++++++++

\textcolor{highlight_clr}{
Vectors and matrices are denoted in bold lowercase $\vect{a}$ and bold uppercase $\mtrx{A}$, respectively.
We use $||\vect{a}||$ to denote the Euclidean norm of a vector and $\angle (\vect{a}, \vect{b})$ to denote the angle between two vectors.
The operator $\expectation\{\cdot\}$ represents the expectation. The operator $\Tr\{\cdot\}$ represents the trace. The superscript $^H$ denotes the Hermitian transpose. The $i$th row and $j$th column entry of a matrix $\mtrx{A}$ is $a_{ij}$. The all-ones matrix is denoted by $\mtrx{1}$.
Higher-order tensors are denoted by uppercase calligraphic letters $\tnsr{A}$. The symbol $\sim$ means statistically distributed as. The complex normal distribution is referred to as $\mathcal{CN}$.
}

%++++++++++++++++++++++++++++++++++++++++++++++++++++++++++
\subsection{Problem Statement}\label{desc:assumptions}
%++++++++++++++++++++++++++++++++++++++++++++++++++++++++++

Our goal is to take steps towards a systematic study of the human-modulated subspace of CSI measurements.
To this end, suppose we have a collection of CSI measurements $\tnsr{H}$. 
We postulate the existence of a universal decomposition
\begin{equation}\label{eq:decomp}
  \tnsr{H} = \tnsr{S} + \tnsr{N},
\end{equation}
where $\tnsr{S}$ contains all information of human modulation. 
Moreover, we assume that at each $k=1,2,3,\ldots$, the signal subspace $\tnsr{S}[k]\subset\tnsr{H}[k]$ at time step $k$ can at least in principle be computed from measurements $\tnsr{D}[l]$, with $l$ ``not too far'' from $k$.
What this means in practice, is that it is possible to \emph{filter} out the noise subspace $\tnsr{N}$ and to \emph{track} the human modulated subspace $\tnsr{S}[k]$ as time, represented by $k$, evolves.
We make two further hypotheses about the decomposition~\eqref{eq:decomp}:
\begin{description}
  \item[Sufficiency of covariance statistics.] \hfill \\ \emph{It is sufficient to consider the covariance statistics of $\tnsr{H}$ along different measurement dimensions independently}.
  \item[Dominance of the signal subspace.] \hfill \\ \emph{The human modulation is characterised by magnitudes of variation of the covariance statistics at the appropriate time scales}.
\end{description}
Considering the measurement axes independently leads to an interpretable and effective dimensionality reduction on $\tnsr{H}$.
Our approach is to use the eigendecomposition of the covariance matrices derived from the tensor $\tnsr{H}$.

We introduce the \emph{structured channel model} and the \emph{observation model} used in the rest of the paper in Sections \ref{sec:mimo_model} and \ref{sec:observation_model}, respectively.
The channel model provides a mathematical description of the measurement data.
The observation model will be used to derive pre-processing techniques that have a sound physical justification for sensing tasks in Sections \ref{sec:subspace_characterisation} and \ref{sec:subspace_tracking}.

%++++++++++++++++++++++++++++++++++++++++++++++++++++++++++
\subsection{Wideband MIMO Channel Model}\label{sec:mimo_model}
%++++++++++++++++++++++++++++++++++++++++++++++++++++++++++

The Structured channel model we use belongs to a class of correlative wideband MIMO channel models \cite{Costa08_NovelWidebandMimoChannelModel}.
Our starting point is the eigendecompositions of the channel model. 
This approach was first developed by Weichselberger~\cite{Weichselberger06_StochasticMimoChannelModelWithJointEndCorrelations}, although it had also been developed independently by other works e.g.,~\cite{Tulino06_Capacity-achievingMimoCovariance}.

In the general case, we assume that CSI data forms a four dimensional dataset, with the four axes being the choice of receive antenna, transmit antenna, delayspread tap during one transmission step in time. 
\textcolor{highlight_clr}{We denote these \emph{measurement dimensions} by the subscripts $\text{Rx}$, $\text{Tx}$, and $\text{Dy}$, respectively.}
We arrange CSI measurements into a tensor $\tnsr{H}$, and for each time step $k$, we write $\tnsr{H}[k]$ for the three dimensional tensor of measurements at that timestep.
The $m$th unfolding $\tnsr{H}_{(m)}$ is defined to be a matrix, whose columns are formed of the $m$th index of $\tnsr{H}$, and rows are formed by ordering the rest of the dimensions lexicographically.
The three tensor unfoldings $\tnsr{H}_{(1)}[k]$, $\tnsr{H}_{(2)}[k]$, and $\tnsr{H}_{(3)}[k]$ describe the channel behaviour in single-input multiple-ouput (SIMO), multiple-input single-ouput (MISO), and delayspread, respectively.
Basically, the dimensions in the chosen axis are considered to be variables, and the other dimensions are \textcolor{highlight_clr}{collapsed in a lexicographical order.}
%simply considered

We treat the measurement of the tensor $\tnsr{H}[k]$ as an unknown, random, and \emph{non-stationary} complex Gaussian process. Concretely, in the absence of prior formalism on human-modulated wireless channels, we assume a generally evolving random process at two different time instances of the form~\cite{Lopez-Martinez15_EigenvalueDynamicsOfCentralWishartForMimo}
\begin{align}
  \tnsr{H}_{(m)}(t+\tau) = \mtrx{W} \tnsr{H}_{(m)}(t) + \bar{\mtrx{W}} \boldsymbol{\Xi}
\end{align}
where $m \in [1,2,3]$, $\mtrx{W}$ (and its complement $\bar{\mtrx{W}} = \sqrt{\mtrx{1} - \mtrx{W}}$) is a correlation matrix whose elements $w_{i,j}$ are the correlation coefficients between the $i$th and $j$th RV-modelled entries of $\tnsr{H}_{(m)}$, and the auxiliary matrix $\boldsymbol{\Xi}$ is independent of $\tnsr{H}_{(m)}$ with i.i.d. entries $\sim \mathcal{CN}(0, \sigma^2_\xi)$.

We now suppress the time step $k$ for simplicity, and assume that all calculations are done at a fixed time.
The one-sided correlation matrices at the receive-, transmit-, and delayspread-side are then computed as 
\begin{align}\label{eq:corrs}
  \mtrx{R}_{\text{Rx}} = \expectation \left\{ \tnsr{H}_{(1)} \tnsr{H}^H_{(1)} \right\}, \notag \\
  \mtrx{R}_{\text{Tx}} = \expectation \left\{ \tnsr{H}_{(2)} \tnsr{H}^H_{(2)} \right\}, \notag \\
  \mtrx{R}_{\text{Dy}} = \expectation \left\{ \tnsr{H}_{(3)} \tnsr{H}^H_{(3)} \right\},
\end{align}
where the expectation is in practice computed by averaging samples across a short time interval, during which human modulation is assumed to be static i.e. wide-sense stationary (WSS).

Eigendecomposition is then applied to Equation \eqref{eq:corrs} in order to extract the channel eigenbases in space (receive and transmit dimensions) and in delay spread according to
\begin{align}
  \mtrx{R}_{\text{Rx}} &= \mtrx{U}_{\text{Rx}} \boldsymbol{\Delta}_{\text{Rx}}\mtrx{U}^H_{\text{Rx}}, \notag \\
  \mtrx{R}_{\text{Tx}} &= \mtrx{U}_{\text{Tx}} \boldsymbol{\Delta}_{\text{Tx}}\mtrx{U}^H_{\text{Tx}}, \notag \\
  \mtrx{R}_{\text{Dy}} &= \mtrx{U}_{\text{Dy}} \boldsymbol{\Delta}_{\text{Dy}}\mtrx{U}^H_{\text{Dy}}. 
  \label{eq:widebandMimoEigendecomposition}
\end{align}

%++++++++++++++++++++++++++++++++++++++++++++++++++++++++++
\subsection{Observation Model}\label{sec:observation_model}
%++++++++++++++++++++++++++++++++++++++++++++++++++++++++++

We observe a sequence of channel tensors $\tnsr{H}[k] \in \mathbb{C}^N, k=1,2,3,\ldots,T$. We define a decomposition
\begin{align}
    \tnsr{H}[k] &= \tnsr{S}[k] + \tnsr{N}[k]
    \label{eq:observation model}
\end{align}
where $\tnsr{S}[k]$ and  $\tnsr{N}[k]$ are the \emph{latent} human-modulated channel component, and additive noise tensor uncorrelated with human activity, respectively. 
The noise tensor need not be additive Gaussian and could account for many effects ranging from wireless SNR variations\footnote{A ``watery'' human body in motion gives rise to complex and unconventional wireless propagation properties.}, suboptimal channel estimation, and/or quantisation.

We write the observation model in terms of unfolding matrices as
\begin{align}
    \tnsr{H}_{(i)}[k] = \tnsr{S}_{(i)}[k] + \tnsr{N}_{(i)}[k]
\end{align}
for $i=1,2,3$.

In what follows, we look at the \textcolor{highlight_clr}{third} unfolding, which in our setup corresponds to the ${\text{Dy}}$ dimension.
Similar treatment applies to the ${\text{Rx}}$ and $\text{Tx}$ dimensions.
Since the human-induced modulation and noise are uncorrelated, we can rewrite the one-sided correlations of equation~\eqref{eq:corrs} as
\begin{align}
    \mtrx{R}_{\text{Dy}}[k] &:= \expectation \left\{ \tnsr{H}_{(3)}[k] \tnsr{H}^H_{(3)}[k] \right\} \\ 
    &= \mtrx{C}_{\text{Dy}}[k] + \sigma^2_{\text{Dy}}[k] \mtrx{I}_{M_h}
    \label{eq:channel-covariance}
\end{align}
where $\mtrx{C}_{\text{Dy}}$ is the rank deficient covariance arising from the human-modulating effect, $\sigma^2_{\text{Dy}}$ is AWGN noise power, and $\mtrx{I}_{M_h}$ is an identity matrix~\cite{Delmas10_SubspaceTrackingForSigProc}. 

Dropping $k$ for brevity, the eigendecomposition in equation~\eqref{eq:widebandMimoEigendecomposition} can now be rewritten for the \emph{observed covariance} $\mtrx{R}_{\text{Dy}}[k]$ as
\begin{align}
  \mtrx{R}_{\text{Dy}} &= \mtrx{U}_{\text{Dy}} \boldsymbol{\Delta}_{\text{Dy}} \mtrx{U}^H_{\text{Dy}} \notag \\
    \mtrx{R}_{\text{Dy}} &= 
    \begin{bmatrix}
        \mtrx{U}^s_{\text{Dy}} \ \mtrx{U}^n_{\text{Dy}}
    \end{bmatrix}
    \begin{bmatrix}
        \hat{\boldsymbol{\Delta}}^s_{\text{Dy}}&    \mtrx{0} \\
        \mtrx{0}&    \boldsymbol{\Delta}^n_{\text{Dy}}
    \end{bmatrix}
    \begin{bmatrix}
        {\mtrx{U}^s_{\text{Dy}}}^H \\ {\mtrx{U}^n_{\text{Dy}}}^H
    \end{bmatrix}
    \label{eq:sig-noise-subspaces}
\end{align}
where $\mtrx{U}^s_{\text{Dy}} \in \mathbb{C}^{M_h \times M_s}$ is an orthonormal \emph{signal subspace} basis, $\mtrx{U}^n_{\text{Dy}} \in \mathbb{C}^{M_h \times M_n}$ is an orthonormal \emph{noise subspace} basis, $M_s$ and $M_n$ are respectively the signal and noise subdimensions of the $M_h$-dimensional channel (i.e. $M_n = M_h-M_s$), $\hat{\boldsymbol{\Delta}}^s_{\text{Dy}} = \boldsymbol{\Delta}^s_{\text{Dy}} + \sigma^2_{s,\text{Dy}} \mtrx{I}_{M_s} \in \mathbb{R}^{M_s \times M_s}$ is a noisy estimate diagonal eigenvalue matrix for $\mtrx{C}_{\text{Dy}}$ the covariance matrix arising from the \emph{true} human-modulating effect, and $\boldsymbol{\Delta}^n_{\text{Dy}} = \sigma^2_{n,\text{Dy}} \mtrx{I}_{M_n}$ is a diagonal noise eigenvalue matrix.

Each of the eigendecompositions in Equation \eqref{eq:widebandMimoEigendecomposition} define a natural \emph{filtration}, that is, a succession of growing subspaces $V_0\subset V_1\subset \cdots\subset V = \mathbb{C}^N$ spanned by the first $i$ eigenvectors $\vect{u}_{{\text{Dy}},j}[k]$, where $j\leq i$ and $i=1,\ldots,N$.
Here $N$ is the dimensionality of the chosen measurement dimension, i.e. the number of delayspread taps (or by duality, frequency bins). 
By our assumptions in~\ref{desc:assumptions}, we may use the subspace $V_i$ as a sufficient statistic for the signal subspace of $\tnsr{H}$ for some $i<N$.
For each measurement dimension, we call the subspace defined here the \emph{Tx/Rx/Dy-projected instantaneous signal subspace}, and we denote it by $V_{\text{Tx}}$, $V_{\text{Rx}}$, and $V_{\text{Dy}}$, respectively.

%%%%%%%%%%%%%%%%%%%%%%%%%%%%%%%%%%%%%%%%%%%%%%%%%%%%%%%%%%%
\section{Subspace characterisation} \label{sec:subspace_characterisation}
%%%%%%%%%%%%%%%%%%%%%%%%%%%%%%%%%%%%%%%%%%%%%%%%%%%%%%%%%%%
In this section we hope to justify the claim that the projected signal subspaces introduced in the previous section are useful statistics which preserve human channel-modulating effects, while simultaneously being minimally diluted by noise. 
This claim is clearly non-trivial: human movements in the signal locale exert unconventional effects on the wireless channel which have not seen similar formal treatment in literature compared to more established channel models adopted widely by industry, say typical urban cellular fading channels~\cite{3GPP_TU6}.
The closest kin to human-modulated Wi-Fi channels in prior literature are perhaps body area network (BAN) channel models; consult~\cite{Smith15_ChannelModelingForWBAN,Smith13_PropagationModelsForBAN,Fort07_IndoorBanModelForNarrowbandComms} and literature therein for further detail.
Specific characteristics of the wireless standard 802.11g/n/ac such as bandwidth, carrier frequencies, and air interface, impart modulating effects well beyond those studied for BANs. 

\subsection{CSI Sensing Model}%
\begin{figure}[h]
  \vspace{-0.10in}
  \centering
    \includegraphics[width=0.24\textwidth]{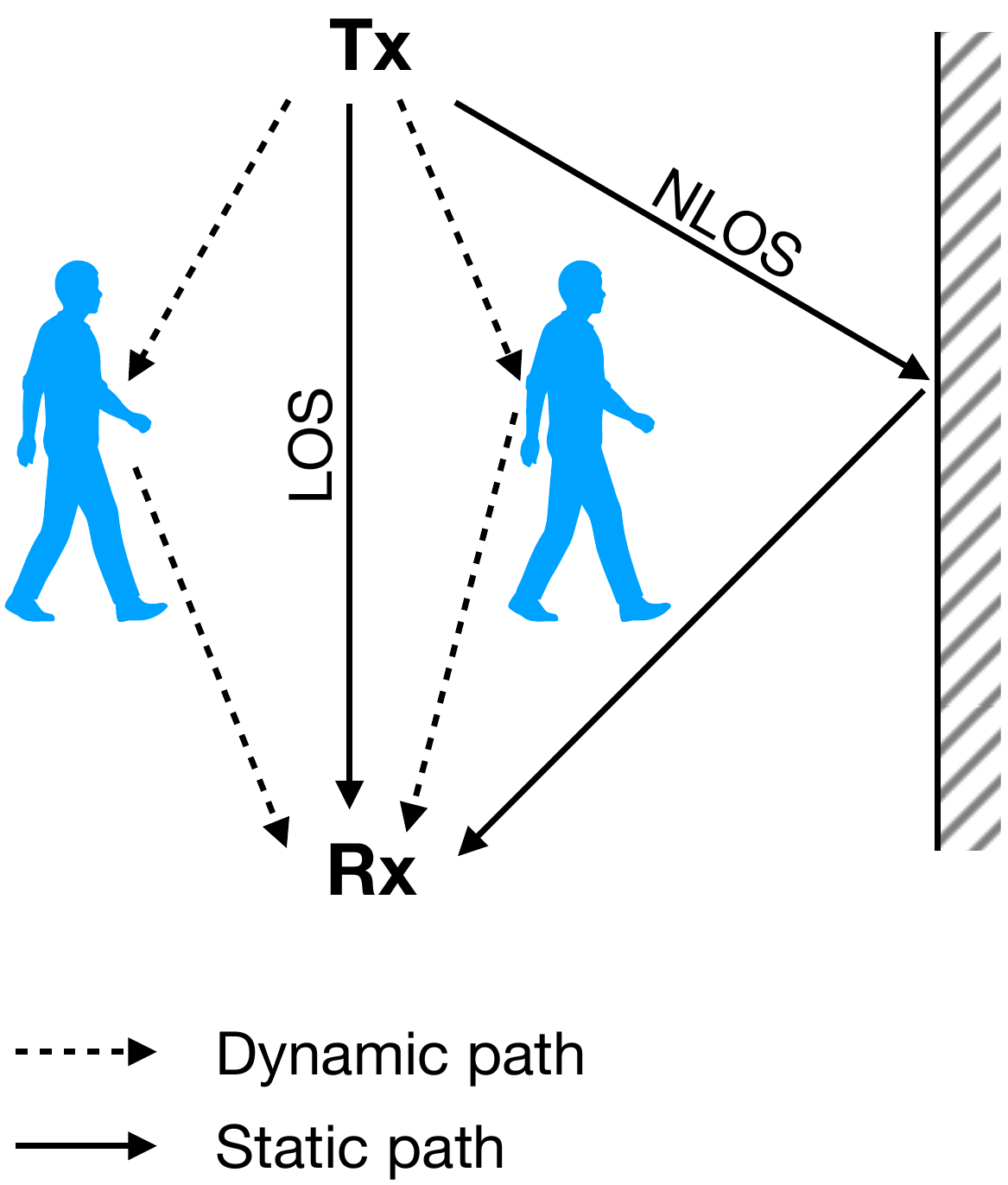}
  \vspace{-0.10in}
  \caption{Human channel modulation}
  \label{fig:human_channel_modulation}
\end{figure}

As illustrated in figure~\ref{fig:human_channel_modulation}, the Wi-Fi-based sensing model consists of placing a pair of transmitter and receiver devices in the environment. 
There are many paths by which electromagnetic energy travels between the transmitter and receiver. When people move, they disturb the multipath profile in the environment. 
The multipath profile is the linear superposition of a number of paths. For instance, figure~\ref{fig:human_channel_modulation} shows two static paths: a direct \textcolor{highlight_clr}{line-of-sight} (LOS) and a reflected non-line-of-sight (NLOS) paths. 
When a human subject walks from left to right in the figure, a dynamic path is \emph{modulated} by this movement. 
By analysing the temporal pattern of these dynamic paths at the receiver, we are able to build sensing applications.

For each transmitter-receiver pair, the superposition of multipaths in the time domain is described by a $N_{sc}$-dimensional frequency-domain CSI $H$ corresponding to a sampling of OFDM subcarriers across the bandwidth.\textcolor{highlight_clr}{\footnote{e.g. $N_{sc} = 30$ for the widely used Intel 5300 chip \url{https://dhalperi.github.io/linux-80211n-csitool/}.}}

As such, the transmitted signal $X$ can be related to the received signal $Y$ through this input-output channel response relationship according to $Y = H X$. 
A MIMO system generalises this input-output relationship for $N_{tx}$ transmitters and $N_{rx}$ receivers. 
For instance, if we have 3 transmitters and 3 receivers, the channel is described as a $3 \times 3 \times 30$ tensor.

We ask some basic questions:
\begin{itemize}
  \item How can we characterise the human modulated subspace of the channel?
  \item How do the dimensionality and direction of the subspace vary in time as a result of human movement?
\end{itemize}
We take a first step towards providing a formal treatment of these key questions, and present a semi-analytical analysis of the projected signal subspace.

We discuss the theoretical underpinnings of our approach in Section \ref{sub:background}, particularly with a view towards contrasting to seminal prior work in Wi-Fi sensing.
We then study data on uncontrolled human movement in Section \ref{sub:empirical}

\subsection{Background on subspace tracking for wireless signals}%
\label{sub:background}
\textcolor{highlight_clr}{In classic signal processing}, estimating the relevant subspace of variation in data is a basic building block of a data processing pipeline ~\cite{Scharf94_MatchedSubspaceDetectors,Kraut01_AdaptiveSubspaceDetectors}.
In the context of an indoor wireless channel, the human modulated portion of the correlation data (cf. Equation \eqref{eq:corrs}) is unknown with complex temporal dynamics.

Wang et al. \cite{Wang15_UnderstandingAndModelingWiFiHumanActivity} obtain good sensing results using an ad hoc pipeline starting with the full wideband covariance matrix (cf. \cite{Costa08_NovelWidebandMimoChannelModel}).
We believe that this choice necessitates the use of excessive time-averaging of the CSI data.
Furthermore, the resulting signal subspace is not easy to interpret. In contrast, the Rx, Tx and Dy correlations defined in Equation~\eqref{eq:corrs} are interpretable low dimensional representations.
Despite the pioneering sensing approach, two drawbacks come to mind:
\begin{itemize}
  \item the spatial and temporal behaviour of the channel are not easily exposed, \textcolor{highlight_clr}{and}
  \item the temporally highly averaged subspaces are less reactive to human activities.
\end{itemize}

The good sensing results aside, the approach of \cite{Wang15_UnderstandingAndModelingWiFiHumanActivity} does not conform to wireless theory, according to which human modulation should be quantifiable using subspace tracking.
Correlative MIMO subspace-based channel models have been shown to estimate capacity~\cite{Weichselberger06_StochasticMimoChannelModelWithJointEndCorrelations,Costa08_NovelWidebandMimoChannelModel, Tulino06_Capacity-achievingMimoCovariance}, and therefore the physicality of the medium.
Intuitively, a model able to conform with a universal information-theoretic measure such as capacity is bound to convey fundamental information about the state of the channel irrespective of what modulates the channel. 
Further, recent theoretical results suggest that the rate of change of a MIMO OFDM channel can be inferred from the statistical analysis \textcolor{highlight_clr}{of its} first and last eigenvectors~\cite{Lopez-Martinez15_EigenvalueDynamicsOfCentralWishartForMimo}, which can be viewed as canonical representatives of the signal and noise subspaces, respectively.
\begin{figure}
  \centering
    \subfloat[strong reflection\label{fig:destructiveStrongMultipath}]{\includegraphics[width=0.225\textwidth]{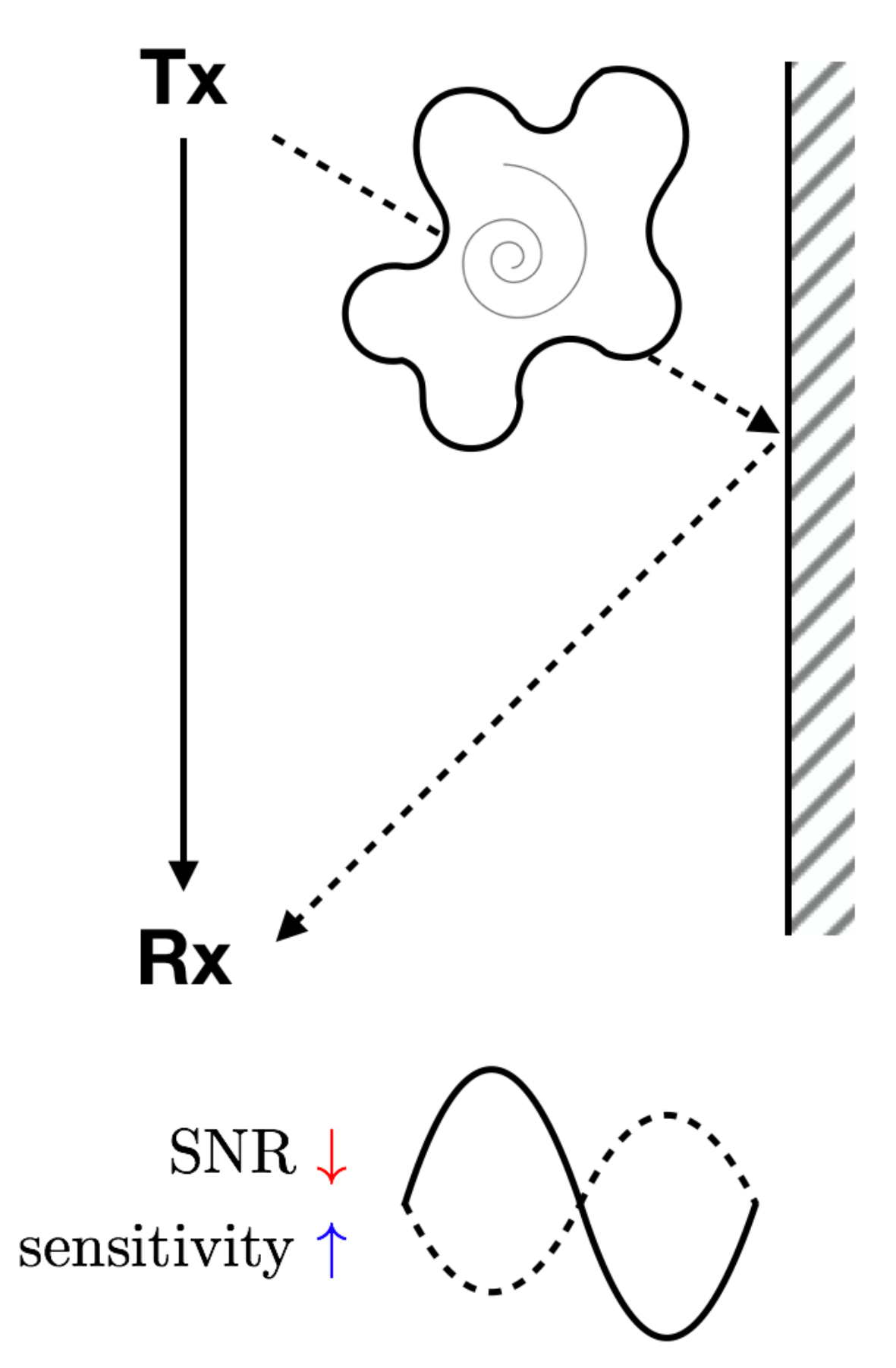}} \ \
    \subfloat[weak reflection\label{fig:destructiveWeakMultipath}]{\includegraphics[width=0.225\textwidth]{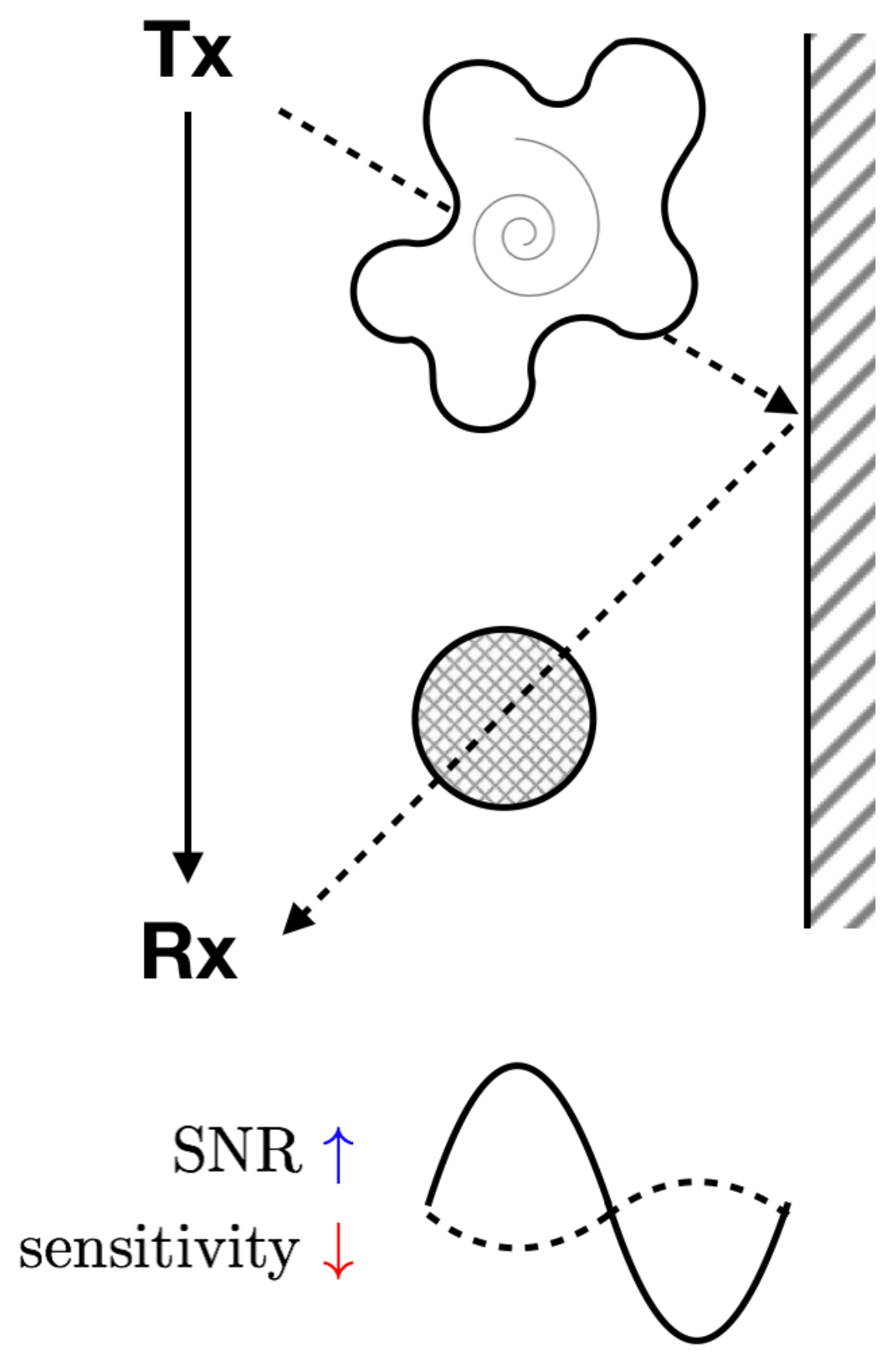}}
  \vspace{-0.10in}
  \caption{Good wireless SNR does not necessarily translate into good sensing sensitivity. For sensing, there is more to designating signal and noise subspaces than meets the eye.}
  \label{fig:multipath_thought_experiment}
\end{figure}

To elaborate on the dynamic nature of the signal subspace, consider a multipath component whose phase adds destructively to a main cluster of multipath, as depicted in Figure~\ref{fig:destructiveStrongMultipath}. 
If the single multipath were to be shadowed as a result of a transient movement as in Figure~\ref{fig:destructiveWeakMultipath}, it is clear that SNR would increase momentarily commensurate with the gain in total multipath arrivals energy.
However, the sensing scene could have further nuances that are not captured by this simple SNR enhancement.
As a further thought experiment, let the single multipath component be probing of a spatial sector in the environment in which a physical activity is unfolding---denoted by a spiral in Figure~\ref{fig:multipath_thought_experiment}. 
That is, the single multipath component disproportionately delivers added movement sensitivity over that delivered by the main cluster of multipath. 
Despite the transient shadowing effect resulting in a \emph{boost} in SNR, the instantaneous combined channel response is rendered \emph{less sensitive} to activities occurring in the aforementioned spatial sector. 
The reduced motion modulation is manifested in reduced correlation structure in the regions of covariance matrix. 
Consequently---and perhaps counter-intuitively given the SNR gain---the signal subspace would necessarily ``shrink'' and noise subspace would ``expand'' momentarily. 
Therefore, robust sensing requires that the signal and noise subspaces be tracked explicitly in order to account for nuanced instantaneous channel effects. 

The above contrived discussion suggests that a sensing system is required to adapt to dynamic channel effects in order to sustain optimal performance. 
Until provision for such adaptation is made in CSI-based sensing systems, we argue that models will fall short at being generalisable with guaranteed performance bounds irrespective of the nuances encountered in real-world deployment environments.

\noindent \textbf{Stationarity period.} \label{para:stationarity-period}
The evolution of the signal subspace can be monitored at different granularities depending on the end-user application. 
An example of this scenario may be seen in activity recognition applications. 
Activity recognition requires deriving channel signatures of sufficient discriminatory power as to allow for the unambiguous separation of activities potentially similar in their broad nature e.g. walking versus running. 
The stationarity period is affected by, besides the application, the sensor configurations such as sampling rate.
 
For example, while $25$ms may be necessary for responsive activity recognition applications, a $100$ms or more may suffice for the much coarser presence detection.
Note that sensing models may also be possible to realise even with ``aliased'' channel statistics akin to \emph{compressive} sensing. However, we will not discuss this further here.

\subsection{Sensing complexity}%
\label{sub:sensing_complexity}

The trade-off between sensing sensitivity and generality is a key question when it comes to designing any data processing pipeline.
Generality implies flexibility for applying techniques from one sensing application to another.
Sensitivity refers to optimality for a fixed sensing task.
These are affected mainly by
\begin{itemize}
  \item sensing pipeline configuration alongside its parameters, \textcolor{highlight_clr}{and} 
  \item the dimensionality of the signal subspace of the data, as it travels through the pipeline.
\end{itemize}
The latter is of particular importance because the size of the signal subspace allows for a controlled grading of sensing sophistication from the simplest (i.e. a one-dimensional subspace) to the most general (i.e. the entire signal subspace). 
The simplest extreme is particularly useful when out-of-the-box flexibility and ease of realisation are desirable.
When optimal performance and sensitivity are required, more elaborate and intricate sensing models can be used on a larger portion of the signal subspace.

We next shed light on the complexity of the human-modulated Wi-Fi signal subspace by way of an empirical study. The aim is to establish that there is more to designating signal and noise subspaces than meets the eye. Future research ought to take this complexity into consideration if Wi-Fi sensing were to be transitioned from controlled setups and into the wild.

\begin{figure*}[t]
    \centering
    \mbox{
        \subfloat[Fractional energy $E_s$ evaluated \textcolor{highlight_clr}{across the signal dimensionality}.\label{fig:fractEnergy}]{\includegraphics[width=0.245\textwidth]{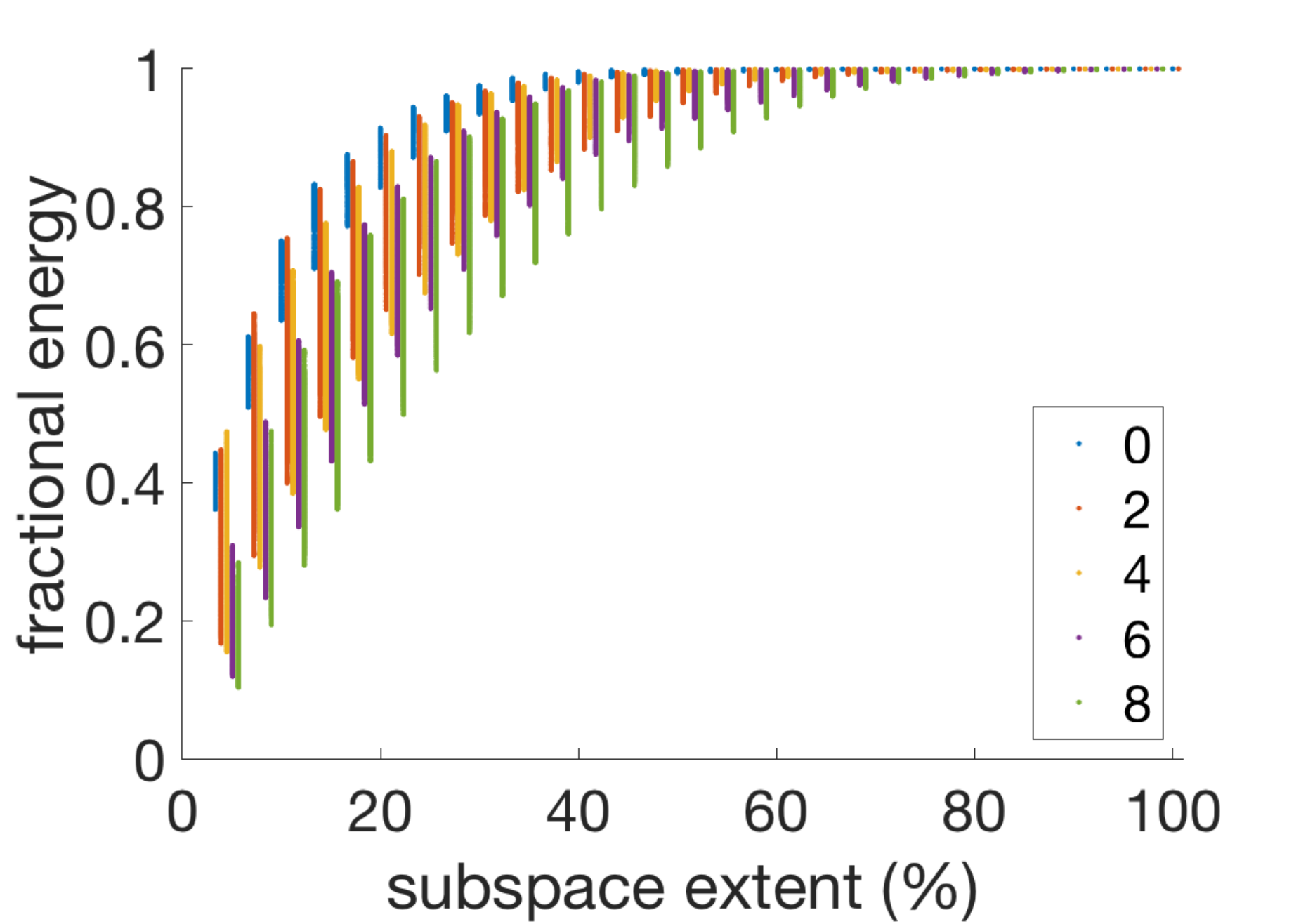}} \ \
        \subfloat[MSE-based subspace boundary at -6 dB reconstruction objective.\label{fig:mseSubspacesBoundary_m6dB}]{\includegraphics[width=0.245\textwidth]{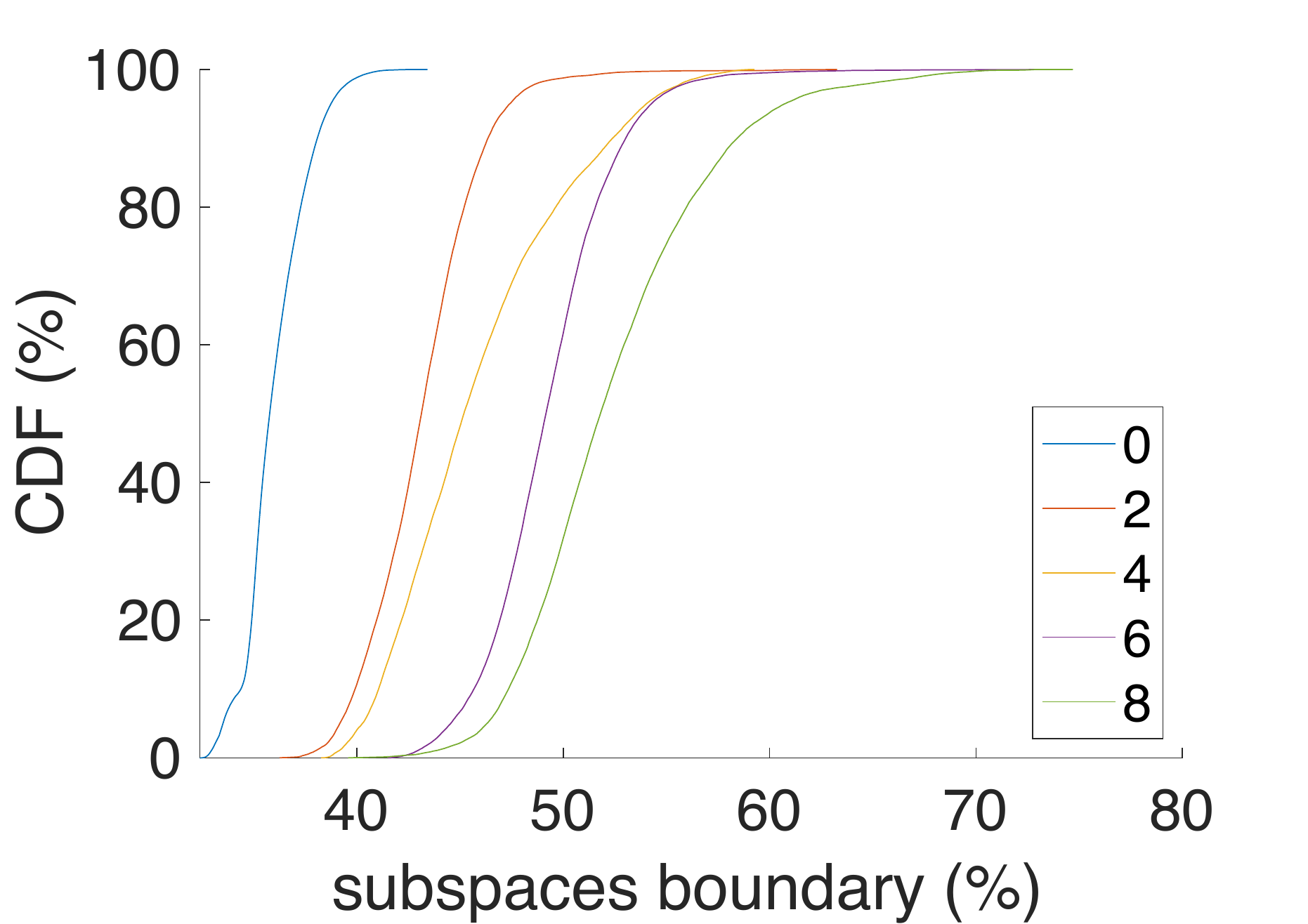}} \ \
        \subfloat[MSE-based subspace boundary at -12 dB reconstruction objective.\label{fig:mseSubspacesBoundary_m12dB}]{\includegraphics[width=0.245\textwidth]{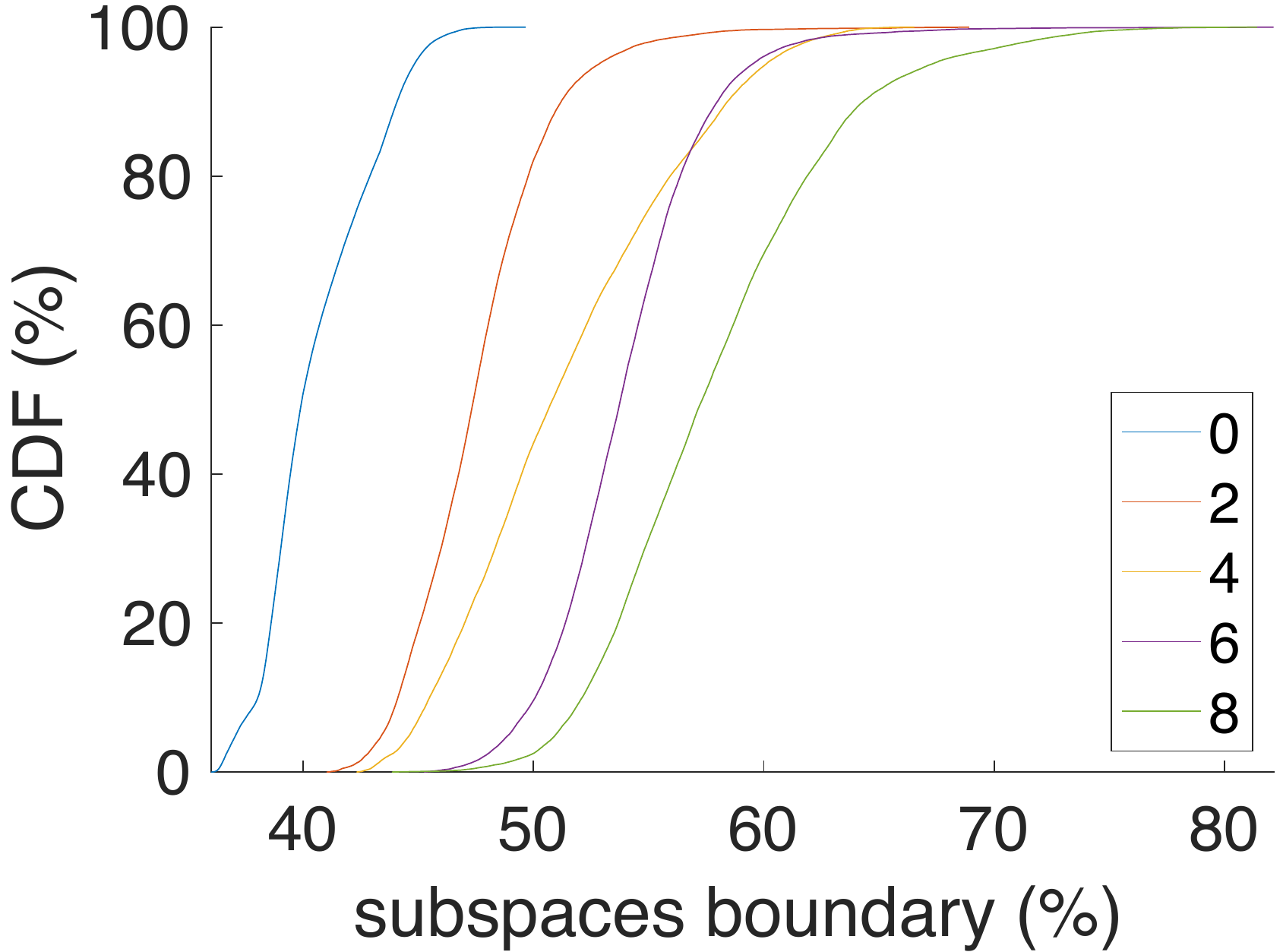}} \ \
        \subfloat[fractional energy at subspace boundary\label{fig:fractEnergySubBoundary}]{\includegraphics[width=0.245\textwidth]{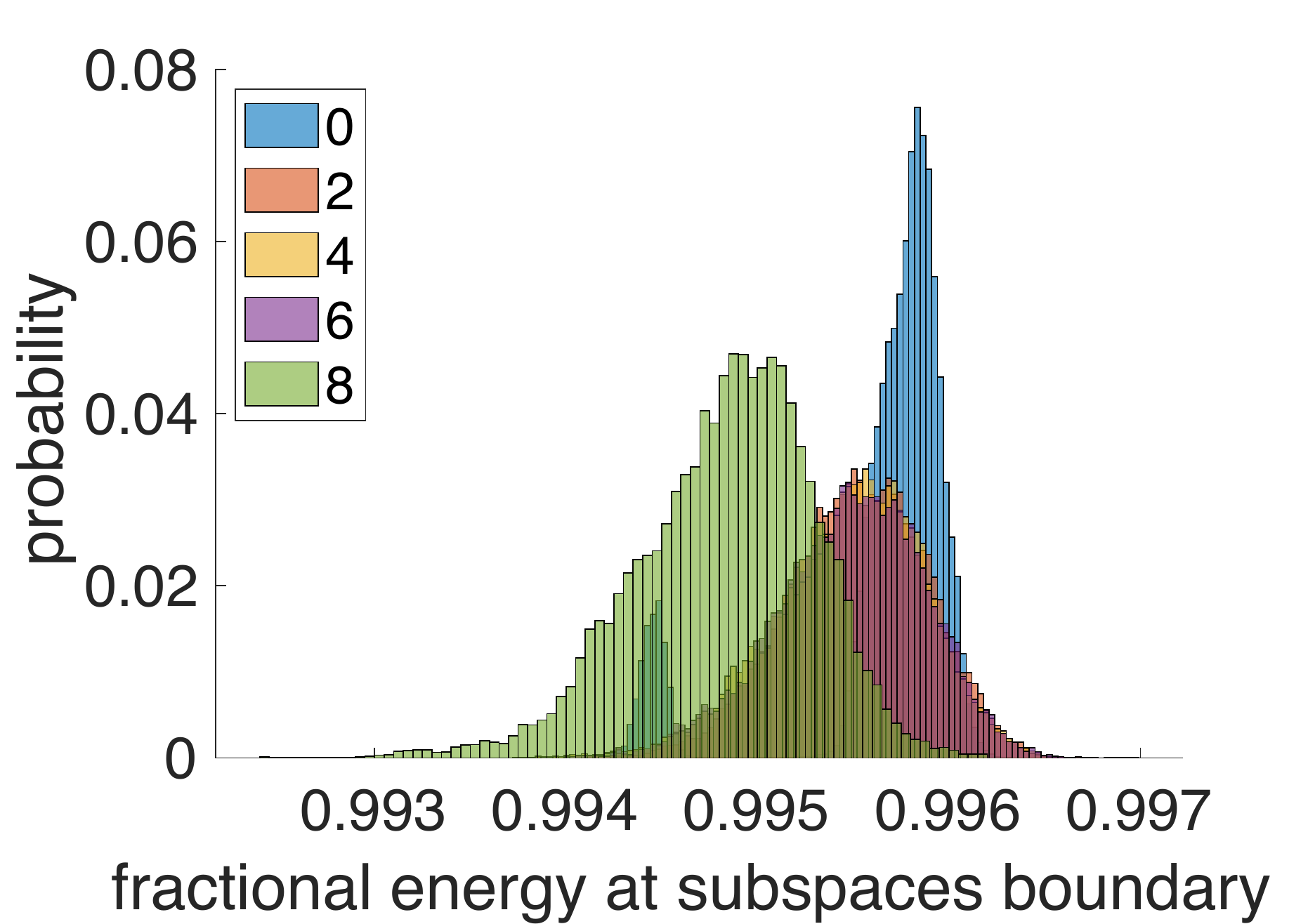}}
    }
    \vspace{-0.10cm}
    \caption{Characterising signal and noise subspaces through an MSE search procedure for 5 uncontrolled human movement scenarios. The numbers in legend 0, 2, 4, 6, and 8 denote how many moving people are present.}
    \label{fig:varyingPeoplePresence}
\end{figure*}

\subsection{Empirical study of uncontrolled human movement}
\label{sub:empirical}

We proceed to study the statistical effects of human activities on the channel covariances.
Specifically, we study the projected signal subspaces, our putative proxies for the signal subspace for human modulation.
To this end, we first quantify the information about the physical environment contained in the covariance matrix.
This information is dynamic in nature and needs to be quantified instantaneously. 
One approach to gauging the information content in a series of covariances is to monitor the distortion contributed by the constituent eigenvectors. 
That is, by successively nulling the respective eigenvectors and measuring the fidelity of the covariance matrix reconstruction, we can quantify in the mean squared error-sense (MSE) the signal and noise boundaries at a given target distortion level (e.g. -12dB).

Concretely, let $\mtrx{R} = \mtrx{U} \boldsymbol{\Delta} \mtrx{U}^H$ be the eigendecomposition of one of the channel correlation matrices in equation~\eqref{eq:widebandMimoEigendecomposition}. Define $\mtrx{R}_i^\prime = \mtrx{U} \boldsymbol{\Delta}_i^\prime \mtrx{U}^H$ as a reconstructed channel correlation matrix whose modified diagonal eigenvalue matrix $\boldsymbol{\Delta}_i^\prime$ nullifies all diagonal entries beyond index $i$ i.e.
\begin{align}
  \boldsymbol{\Delta}^{\prime}_{i} = 
  \begin{bmatrix}
  \ddots         &0              &0           &\ldots   &\ldots   &0        \\
  0              &\delta_{i-1}  &0           &0        &\ldots   &0        \\
  0              &0              &\delta_{i} &\ddots   &         &\vdots   \\
  \vdots         &               &\ddots      &0        &\ddots   &\vdots   \\
  \vdots         &               &            &\ddots   &\ddots   &\vdots   \\
  0              &0              &\ldots      &\ldots   &\ldots   &0
  \end{bmatrix}
\end{align}
The reconstruction error matrix is $\mtrx{R}^\epsilon = \mtrx{R} - \mtrx{R}_i^\prime = \textcolor{highlight_clr}{[r_{kl}^\epsilon]}$. The reconstruction MSE error can then be described as
\begin{align}
  \mathrm{MSE} = \frac{1}{N^2} \textcolor{highlight_clr}{\sum_{k,l} (r^{\epsilon}_{kl})^2}
\end{align}

The above MSE search allows us to build a time-series picture of the dynamic partitioning of the covariance into signal and noise subspaces. 
This evolution of signal and noise subspaces is indicative of the evolution in the corresponding propagation conditions and also necessarily human movement. 
Intuitively, the harsher the dynamics of wireless propagation conditions, the more fluctuating the boundary between signal and noise subspaces is.

Having arrived at a statistical picture of subspaces boundary, we can utilise this knowledge to examine how the fractional signal subspace energy changes throughout human movement. 
We define the \emph{fractional signal subspace energy} as the ratio between energy in the signal subspace to total energy contained in the channel. 
Thus, the fractional energy can be written as $E_s = \Tr(\boldsymbol{\Delta}^s_{\textrm{x}})/\Tr(\boldsymbol{\Delta}_{\textrm{x}})$, where $\Tr$ is the trace operator, and $\boldsymbol{\Delta}$ is the unitary eigenvalue matrix (cf. Equation \eqref{eq:sig-noise-subspaces}), and $\textrm{x} \in [\textrm{Rx}, \textrm{Tx}, \textrm{Dy}]$.
As such, $E_s$ conveys information about optimum \emph{sensing} SNR dynamics. 
A \emph{parsimonious} suboptimal sensing system that utilises instantaneously less of the available $E_s[k]$ at the $k$th time is effectively throwing away information.

The following discourse considers uncontrolled indoor human movement. 
This is perhaps the most generic form of activities likely to occur indoors. 
Naturally, uncoordinated motion components superimpose to modulate the signal subspace in random ways. 
Stronger motion components could also mask much weaker ones.

We begin by examining what effect increased human movements has on the signal and noise subspaces. 
We conduct an experiment in which participants were asked to walk randomly in a room. 
The number of moving people present was varied from 0 (i.e. empty) to 8. 
The duration of movement per session was 5 to 10 minutes. An 802.11n 3$\times$3 MIMO transmitter node was placed outside the room and a receiver node was placed inside. 
The CSI was sampled at a nominal sampling rate of 500Hz using a 5GHz carrier and 40MHz channel bandwidth. 
The reported CSI is 30 dimensional for each transmitter-receiver pair sampling the available 40MHz bandwidth coarsely but equidistantly. That is, 1-in-4 OFDM subcarriers are reported, resulting in a measured MIMO CSI $3 \times 3 \times 30$ tensor. 

We investigate the effect of increased human movement on signal and noise subspaces by way of searching for the subspaces boundary yielding an objective MSE distortion as outlined earlier.
Eigenvectors contributing less to the fidelity of covariance reconstruction will fall within the noise subspace. 
Conversely, eigenvectors impacting the fidelity of reconstruction more pronouncedly belongs to the signal subspace. 
The MSE-guided search finds the subspaces boundary that satisfies a desired distortion level in the MSE sense. 
Owing to the finite subspace resolution of a practical system, we interpolate between two MSE distortion levels produced at adjacent eigenvectors in order to simulate the effect of a smoothly varying MSE distortion and its respective ``fractional'' subspace index.

Figure~\ref{fig:fractEnergy} illustrates the variability in the fractional subspace energy $E_s$ within the signal subspace extent and across movement scenarios---as denoted by the vertical scatter points. 
It is evident that the variability increases towards the lower-end of the subspace extent, reflecting the poor SNR contributed. 
Further, the variability increases markedly with the number of moving people i.e. fractional energy is more diffused in higher occupancy classes. Figure~\ref{fig:mseSubspacesBoundary_m12dB} shows the result of the MSE search procedure on the demarcation of the boundary between the signal and noise subspaces. 
Note, however, the statistical variability corroborating the earlier hypothesis; namely, that dynamic stresses on the wireless channel would result in equivalent shrinkage or expansion of the signal subspace as needed to satisfy the target reconstruction distortion level.
Similar subspace dynamic behaviour can be seen when doubling the objective MSE distortion in figure~\ref{fig:mseSubspacesBoundary_m6dB}; the subspaces boundary demarcation is insensitive to the chosen MSE level. 
It is further interesting to observe the accompanied effects in figure~\ref{fig:fractEnergySubBoundary} on the fractional energy at the very same instantaneous demarcations of the signal and noise boundary established by the MSE search. 
The fractional energy at the \emph{true}\footnote{we will return to address this claim in due course} instantaneous subspaces boundary is unable to provide a faithful statistical account on the expansion/shrinkage of the signal subspace at least for scenarios 2, 4, and 6 as evident by their density overlap. 
That is, the fractional energy cannot be called upon to optimally partition the covariance matrix.

\begin{figure}[h]
    \centering
        \subfloat[Instantaneous subspace boundary\label{fig:instantaneousSubspace}]{\includegraphics[width=0.33\textwidth]{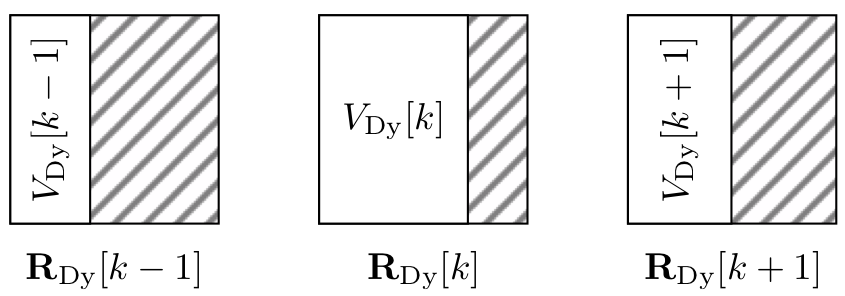}} \\
        \vspace{-0.25cm}
        \subfloat[Normalised MI versus MSE reconstruction objective.\label{fig:miVsMse}]{\includegraphics[width=0.24\textwidth]{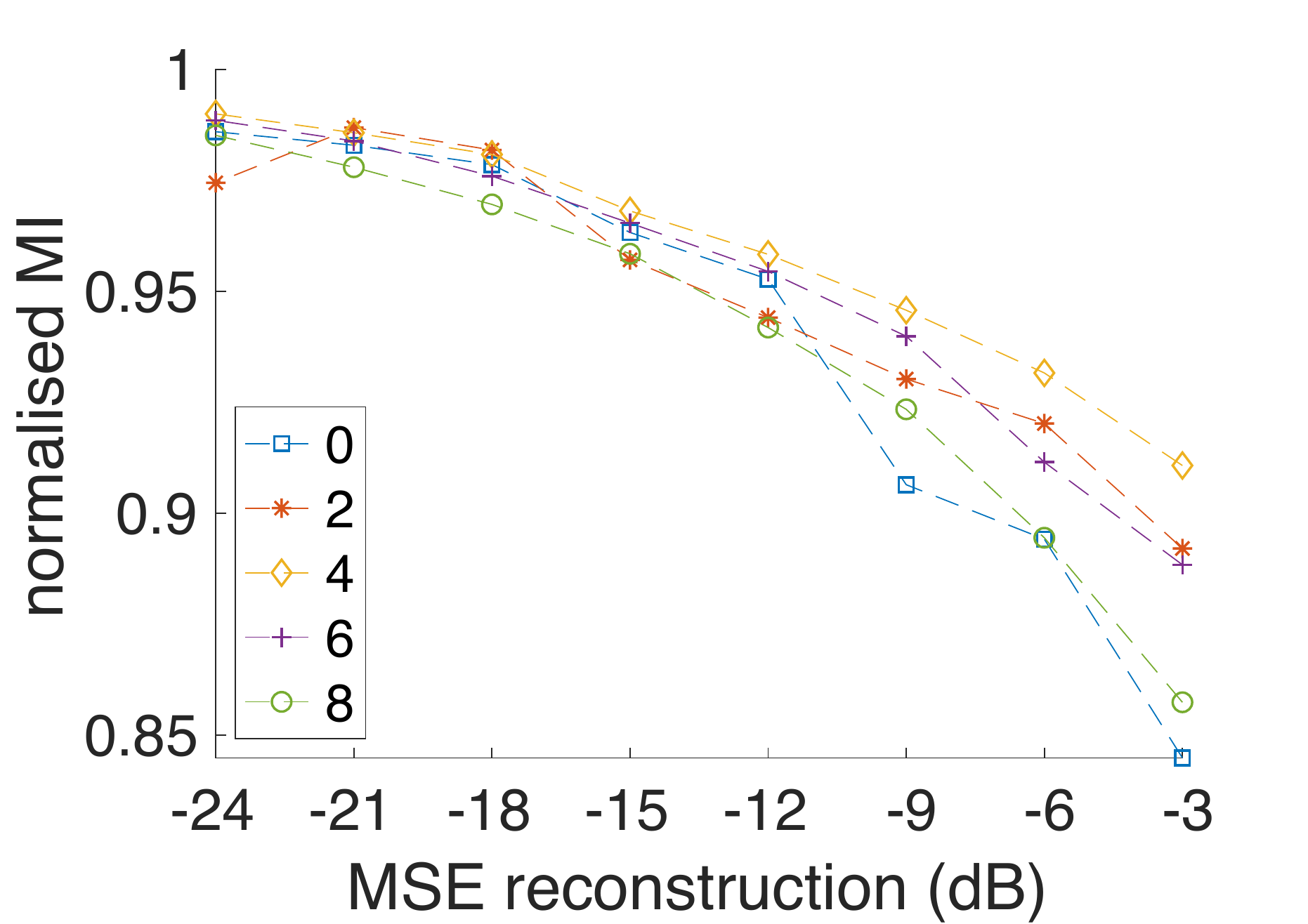}} \ 
        \subfloat[Normalised MI versus subspace extent.\label{fig:miVsSubspaceExtent}]{\includegraphics[width=0.24\textwidth]{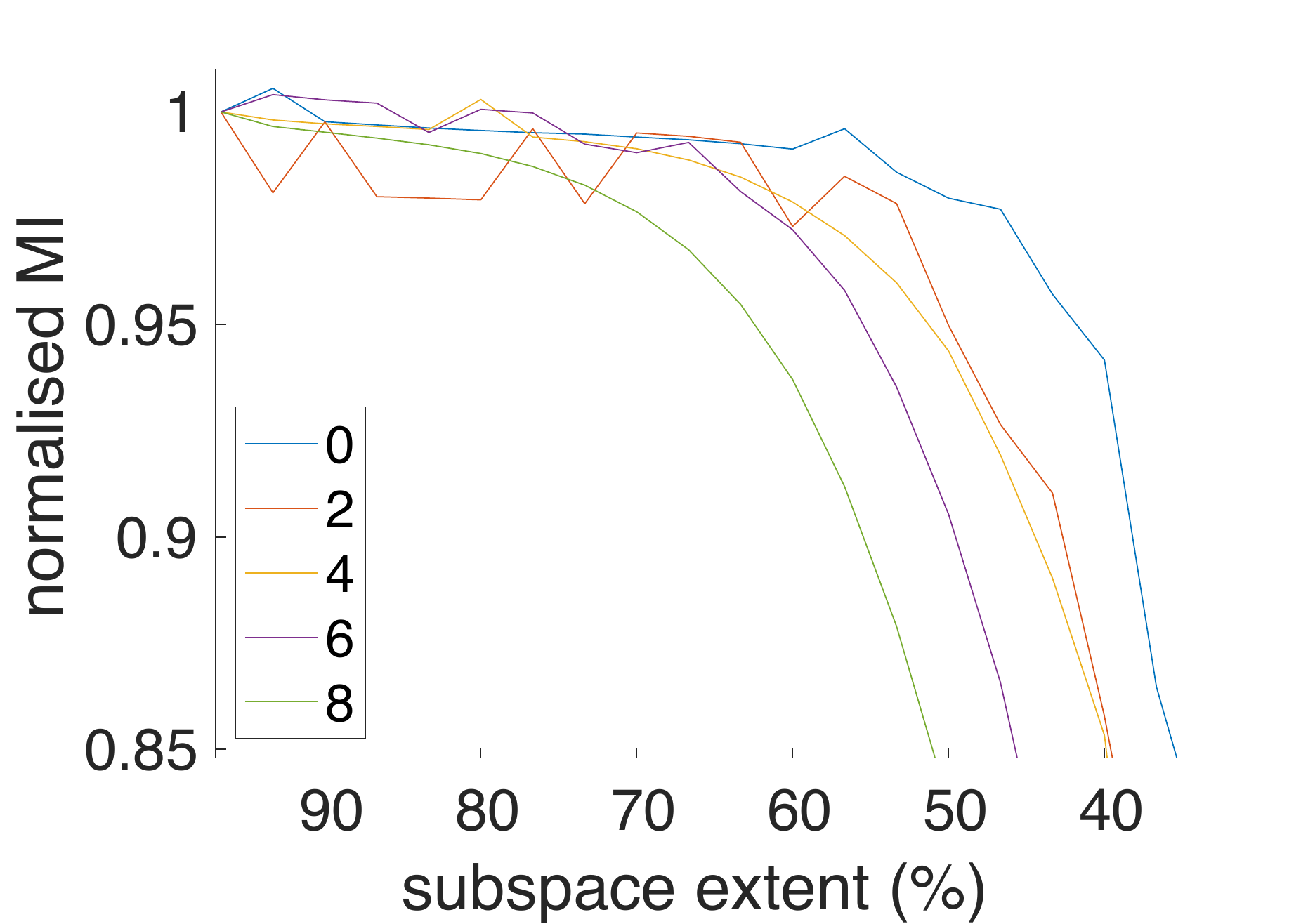}}
        \vspace{-0.1cm}
        \caption{Normalised mutual information between covariances and their imperfect reconstructions over time and across 5 uncontrolled human presence scenarios, highlighting that the signal subspace is dynamic in nature. Legend denotes how many moving people are present.}
        \label{fig:normalisedMiVersusMse}
\end{figure}

We conclude this section by qualifying our MSE-search methodology using mutual information (MI). 
The \emph{instantaneous} subspaces boundary is used to agglomerate series of reconstructions of the covariance matrix as to compare against the groundtruth covariance distribution. 
We sweep the objective MSE distortion between $-24$ dB and $-3$ dB in $3$ dB increments. 
We then measure the normalised mutual information between $V_{\text{Dy}}[k]$ and $R_{\text{Dy}}[k]$ for different occupancy cases as illustrated in figure~\ref{fig:instantaneousSubspace}.
Figure~\ref{fig:miVsMse} shows that, for all human presence scenarios, the normalised MI at the instantaneous subspaces boundary steadily approaches unity as MSE reconstruction fidelity increases towards $-24$ dB. 
We observe that in terms of mutual information, our MSE-reconstruction based methodology is consistent under different channel conditions. 
In order to corroborate this observation, we compute the same normalised mutual information metric for the static (i.e. truncated) subspace extent across occupancy cases. 
Figure~\ref{fig:miVsSubspaceExtent} depicts such MI between $R_{\text{Dy}}[k]$ on the one hand, and $V_0 \subset V_1 \subset \cdots \subset V = \mathbb{C}^N$ on the other \textcolor{highlight_clr}{hand.\footnote{see Section~\ref{sec:observation_model} for a reminder on the definition of the succession of growing subspaces.}} 
A ``waterfall'' effect can be seen whereby more truncated static subspace is needed at higher occupancy classes in order for MI to approach unity. 
Such MI waterfall effect is equivalent to the MSE-based subspace boundary shown earlier in figures~\ref{fig:mseSubspacesBoundary_m6dB}~\&~\ref{fig:mseSubspacesBoundary_m12dB}, reaffirming the notion of instantaneous subspace expansion and shrinkage as a function of the intensity of human movements.

%%%%%%%%%%%%%%%%%%%%%%%%%%%%%%%%%%%%%%%%%%%%%%%%%%%%%%%%%%%
\section{Subspace Tracking} \label{sec:subspace_tracking}
%%%%%%%%%%%%%%%%%%%%%%%%%%%%%%%%%%%%%%%%%%%%%%%%%%%%%%%%%%%

In Section~\ref{sec:subspace_characterisation}, we established and characterised the notions of signal and noise subspaces within the context of human-induced channel perturbations. 
We now turn to examples of how to derive features for sensing tasks. 
Our approach is to track the evolution of the projected signal subspaces (cf. Section \ref{sec:observation_model}).

The subspace-based human sensing we advocate for is in line with foundational work in wireless channels~\cite{Weichselberger06_StochasticMimoChannelModelWithJointEndCorrelations,Costa08_NovelWidebandMimoChannelModel, Tulino06_Capacity-achievingMimoCovariance, Lopez-Martinez15_EigenvalueDynamicsOfCentralWishartForMimo}, which is in contrast to prior work on wireless Wi-Fi sensing (see e.g. \cite{Wang15_UnderstandingAndModelingWiFiHumanActivity}).
We show that with a good enough instantaneous estimate of the covariances described in Section \ref{sec:measurement_model} this tracking can be used to capture the effects of human modulation.
We present our analysis of the $\text{Dy}$-projected signal subspace, but the same can be easily repeated for the $\text{Rx}$ dimension.

%++++++++++++++++++++++++++++++++++++++++++++++++++++++++++
\subsection{A geometric view of subspace evolution}
%++++++++++++++++++++++++++++++++++++++++++++++++++++++++++ 
As an example of subspace tracking, we present the trajectories of the eigenvectors of the covariance matrices (cf. Equation \eqref{eq:corrs}).

\begin{figure}[htb]
    \centering
    \mbox{
        \subfloat[Subspace component 0\label{fig:subspace0Trajectory}]{\includegraphics[width=0.285\textwidth]{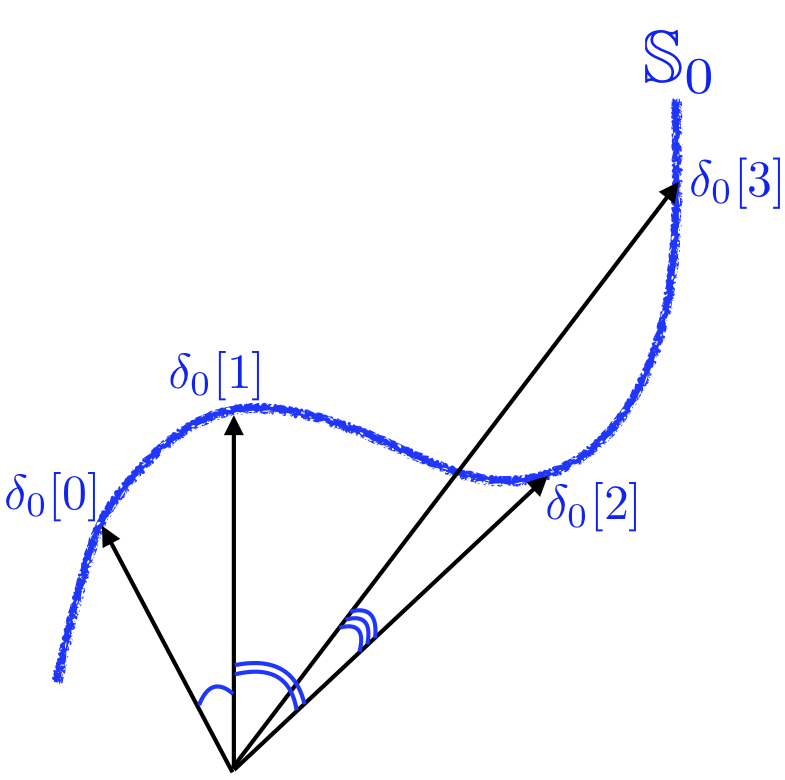}} 
        \subfloat[Subspace component 1\label{fig:subspace1Trajectory}]{\includegraphics[width=0.205\textwidth]{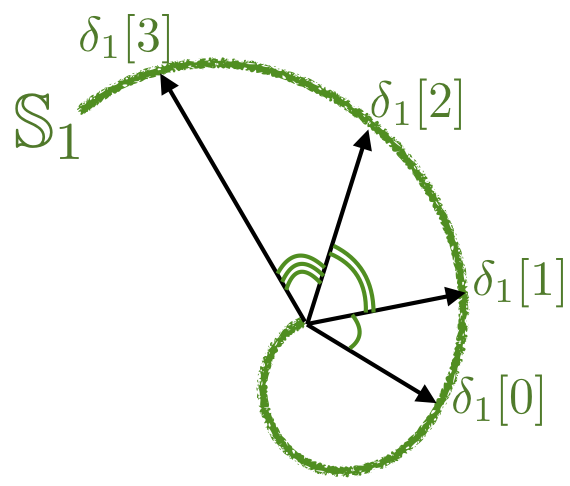}}
    }
    \vspace{-0.0in}
    \caption{\small Geometric interpretation of subspace evolution.}
    \label{fig:geometricSubspaceInterpretation}
    \vspace{-0.0in}
\end{figure}

Consider the time evolution of subspaces spanned by the first two (unnormalised) eigenvectors $\delta_0[k]$ and $\delta_1[k]$ of $\mtrx{R}_{\text{Dy}}[k]$.
Let $\mathbb{S}_0$ and $\mathbb{S}_1$ be subspaces spanned by $\delta_0[k]$ and $\delta_1[k]$ for $k=1,2,3,\ldots$---depicted in local coordinate systems---respectively.
See figure~\ref{fig:geometricSubspaceInterpretation} for a geometric interpretation.

That is, recalling equation~\eqref{eq:sig-noise-subspaces}, each of these subspace components at the $k$th discrete time would correspond to (1) an $i$th eigenvector $\vect{u}_{{\text{Dy}},i}[k] \in \mtrx{U}^s_{\text{Dy}}[k]$ (i.e. belonging to the signal subspace), and (2) a scaling eigenvalue $\delta_i[k] \in \hat{\boldsymbol{\Delta}}^s_{\text{Dy}}[k]$. 
The empirical signal and noise characterisation study reported in Section~\ref{sec:subspace_characterisation} has concluded that the fractional energy $E_s$ evaluated at the subspaces boundary is less able to reveal increased multi-user channel variations. 
That is, when considering the movement of the signal subspace as a result of human-induced channel stresses, less stock should be put in the eigenvalues $\delta$'s.
This is also intuitive to communications practitioners because phase-modulation, when combined with amplitude-modulation, is what really allows for packing more information efficiently within a finite stretch of bandwidth. 
The equivalence to the unreliability of power (i.e. eigenvalues) has also been echoed in prior art; namely, that ``wireless internal state transitions result in high amplitude impulse and burst noises in CSI streams''~\cite{Wang15_UnderstandingAndModelingWiFiHumanActivity}.
As an example of this noisy state transition, note the bimodal nature of the 0 occupancy density of the fractional energy $E_s$ in figure~\ref{fig:fractEnergySubBoundary}---as indicated by the transparent underlaying behind the 8 occupancy case density. \textcolor{highlight_clr}{Another example is recent work on \emph{channel charting} in the context of urban CSI measurements from basestations wherein Studer et al. propose CSI scaling part of their feature mapping procedure~\cite{Studer18_ChannelCharting}.} 

Therefore, referring to figure~\ref{fig:geometricSubspaceInterpretation} again, a critical insight emerges: \emph{human effects on the wireless channel can be ``demodulated'' by observing the corresponding angular movements of the signal subspace}.
%++++++++++++++++++++++++++++++++++++++++++++++++++++++++++
\subsection{Differential subspace evolution}
\label{sec:trackers-innerworkings}
%++++++++++++++++++++++++++++++++++++++++++++++++++++++++++

The time dependency of the angular movements of the subspace is visualised in figure~\ref{fig:geometricSubspaceInterpretation}. 
The (complex) angles, which can be computed as the real part of Hermitian inner product, 
$\psi[1] = \angle (\vect{u}_{\text{Dy},i}[0], \vect{u}_{{\text{Dy}},i}[1])$, $\dots$, and $\psi[3] = \angle (\vect{u}_{{\text{Dy}},i}[2], \vect{u}_{{\text{Dy}},i}[3])$ are depicted for both subspace component 0 and 1. 

These angles signify the \emph{differential} movement of a certain signal subspace component between the $k-1$ and $k$ discrete times. 
Incidentally, these angles have also another interpretation. 
Note that the diagonalisation of the the covariance matrix of equation~\eqref{eq:sig-noise-subspaces} will produce eigenvectors which are by construction unitary i.e. $\vect{u}_{\text{Dy},i}^{H} \ \vect{u}_{\text{Dy},i} = 1 = \cos(0)$. However, a human movement will cause the channel's signal subspace to \emph{evolve} out of its ``rest'' condition. The resultant deviation in the subspace will be manifested in equivalent deviation in the \emph{unitarity} of its constituent, evolved eigenvectors w.r.t. their original ``rest'' conditions. 
Thus, the successive change in \emph{unitarity} for the $i$th subspace component between time $k-1$ and $k$ is \emph{quantified} by $\vect{u}^{H}_{\text{Dy},i}[k] \ \vect{u}_{\text{Dy},i}[k-1] = \cos(\psi[k])$ which coincides with the angular movement of the subspace. 
Hence we term this angular metric the \emph{differential unitarity}.

In general, our proposed differential unitarity feature for tracking human-modulated signal subspaces is applicable to any channel eigendecomposition formulation commonly encountered in literature.
Denote by $\vect{u}_{\text{Dy},i}[k]$ the $i$th delayspread eigenvector at time $k$. Then the differential unitarity $\hat{u}_{\text{Dy},i}[k] = \cos(\psi_{\text{Dy},i}[k])$ between time $k$ and $k-1$ is formulated as
\begin{align}
    \hat{u}_{\text{Dy},i}[k] &= \vect{u}_{\text{Dy},i}^{H}[k] \ \vect{u}_{\text{Dy},i}[k-1]
    \label{eq:dirrentialUnitarity_rx}
\end{align}

Similarly, for the receive-side eigenbasis
\begin{align}
    \hat{u}_{\text{Rx},i}[k] &= \vect{u}_{\text{Rx},i}^{H}[k] \ \vect{u}_{\text{Rx},i}[k-1]
    \label{eq:dirrentialUnitarity_dy}
\end{align}

\begin{figure*}[t]
    \centering
        \subfloat[$||\hat{u}_{\textrm{Rx},1}||$ (dB) \label{fig:rxDiffUnit}]{\includegraphics[width=0.245\textwidth]{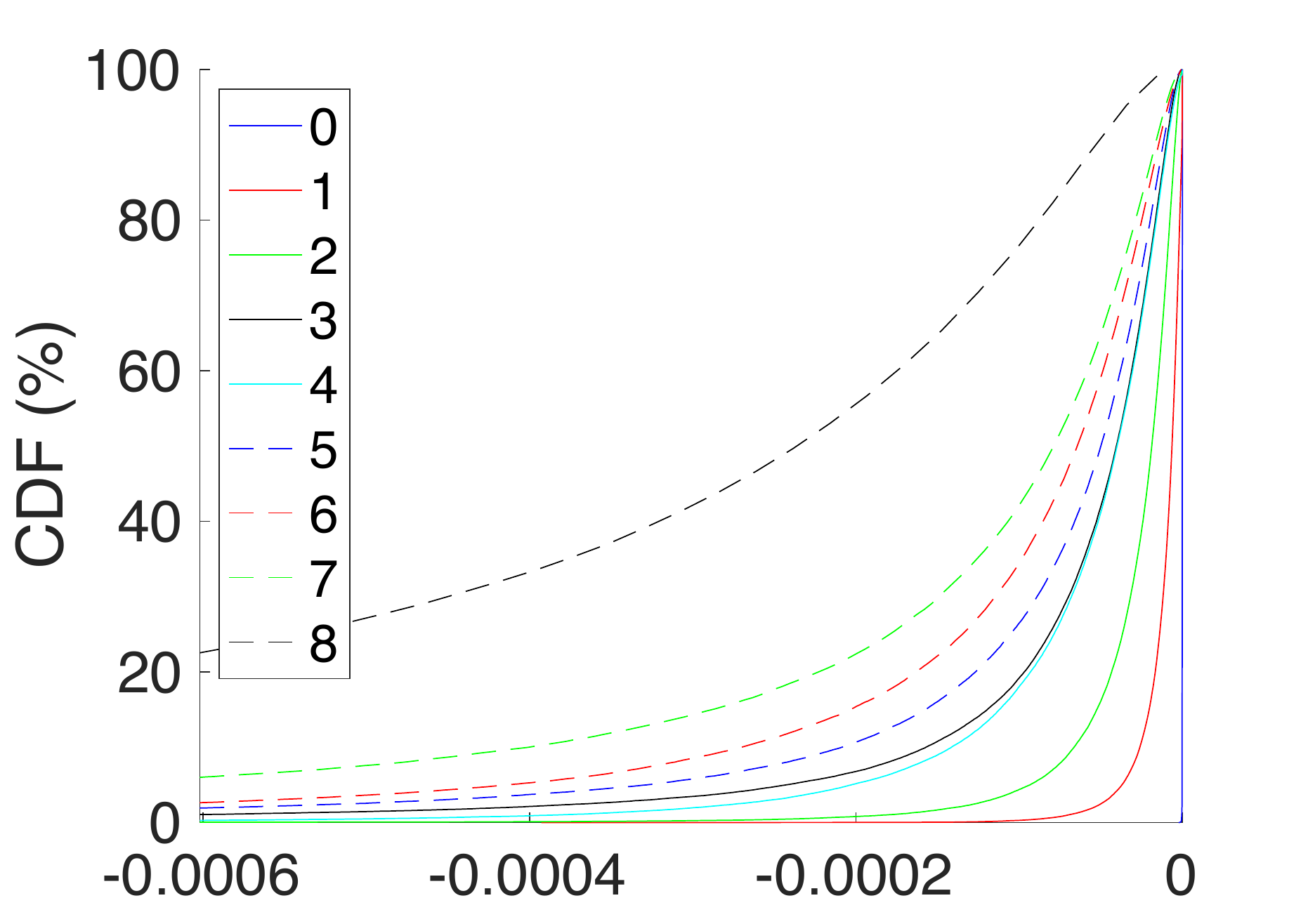}}
        \subfloat[$||\hat{u}^\prime_{\textrm{Rx},1}||$ (dB)\label{fig:rxDiffUnitPrime}]{\includegraphics[width=0.245\textwidth]{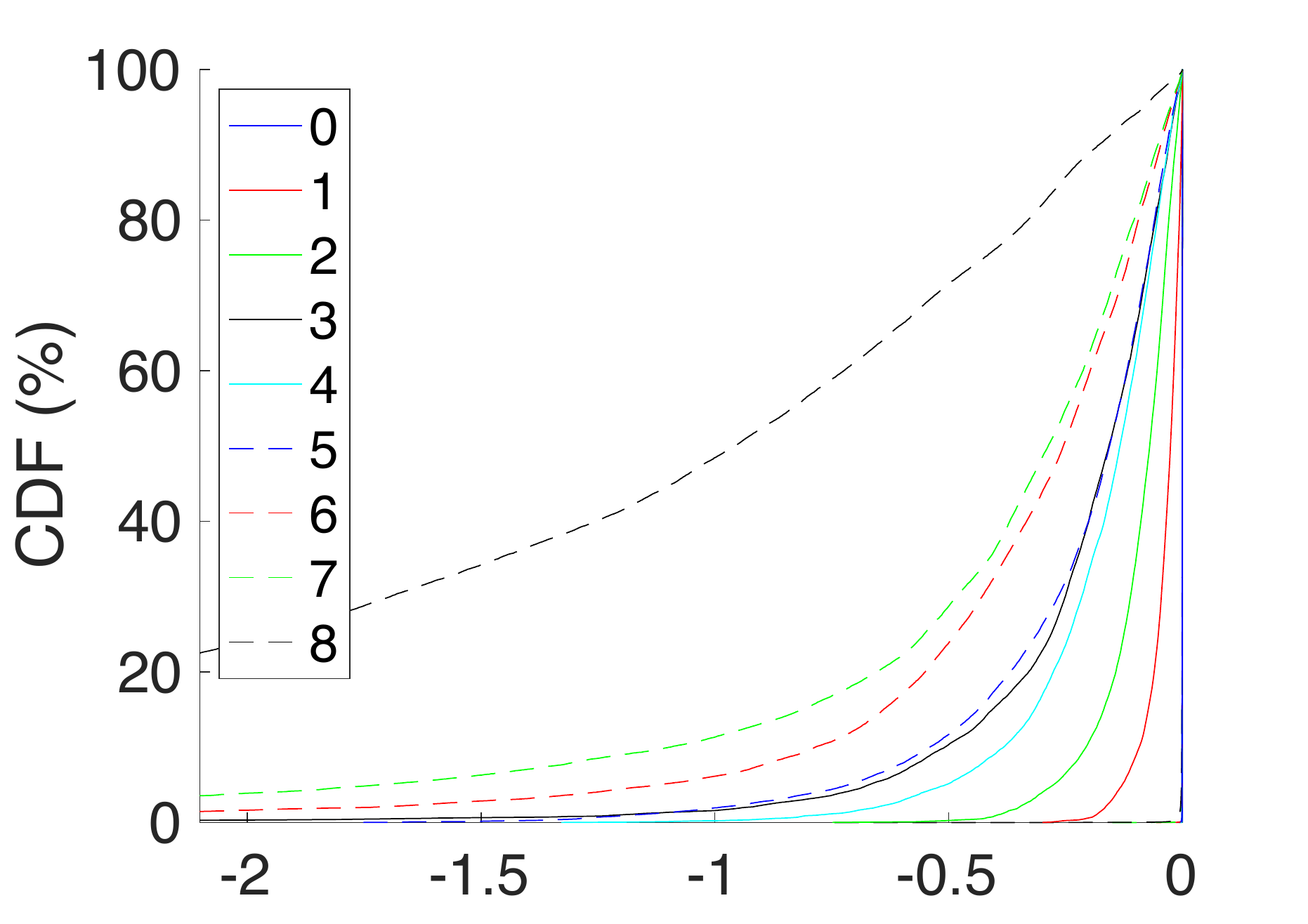}}
        \hspace{2mm}
        \subfloat[$||\hat{u}_{\textrm{Dy},1}||$ (dB)\label{fig:dyDiffUnit}]{\includegraphics[width=0.245\textwidth]{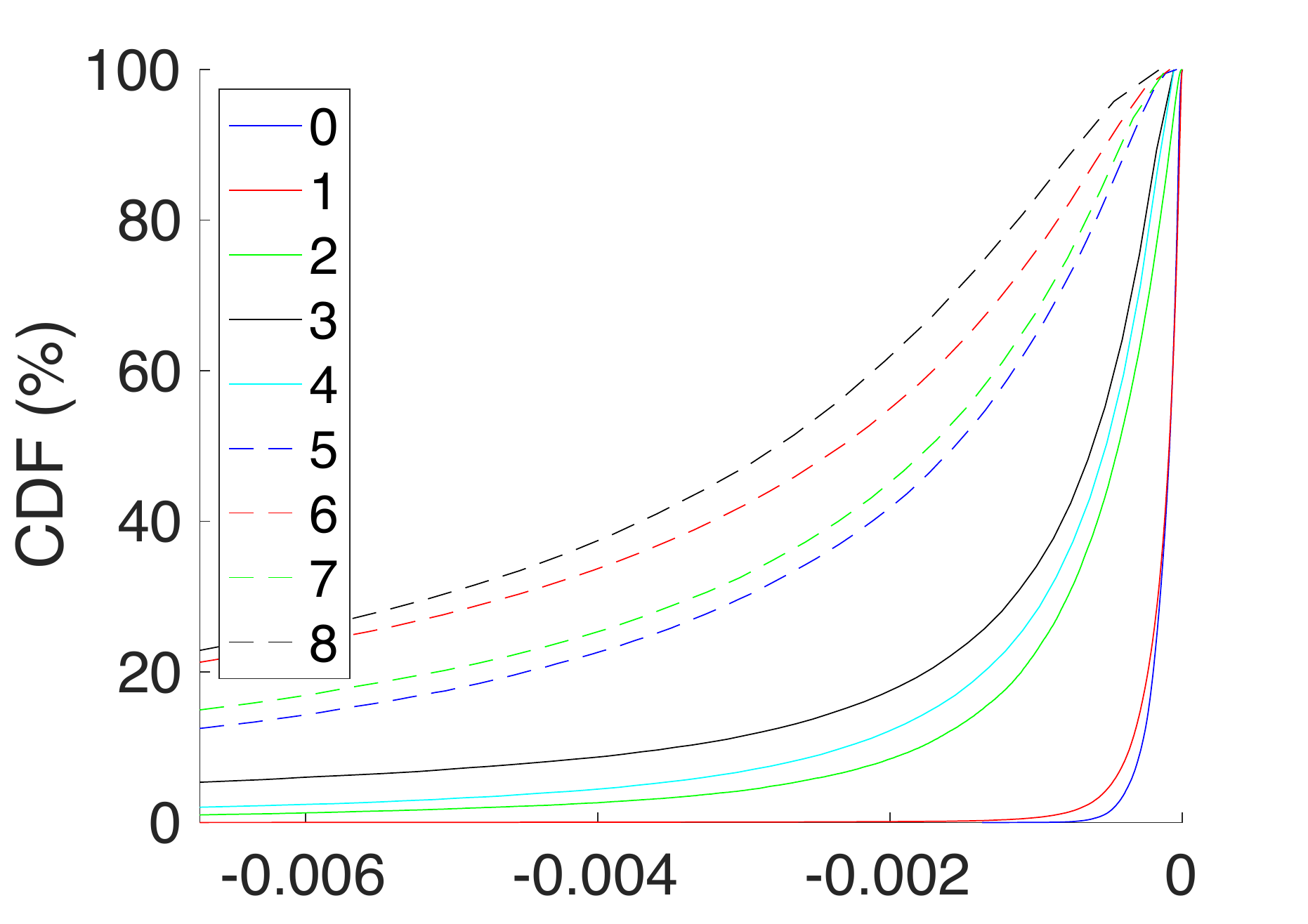}}
        \subfloat[$||\hat{u}^\prime_{\textrm{Dy},1}||$ (dB)\label{fig:dyDiffUnitPrime}]{\includegraphics[width=0.245\textwidth]{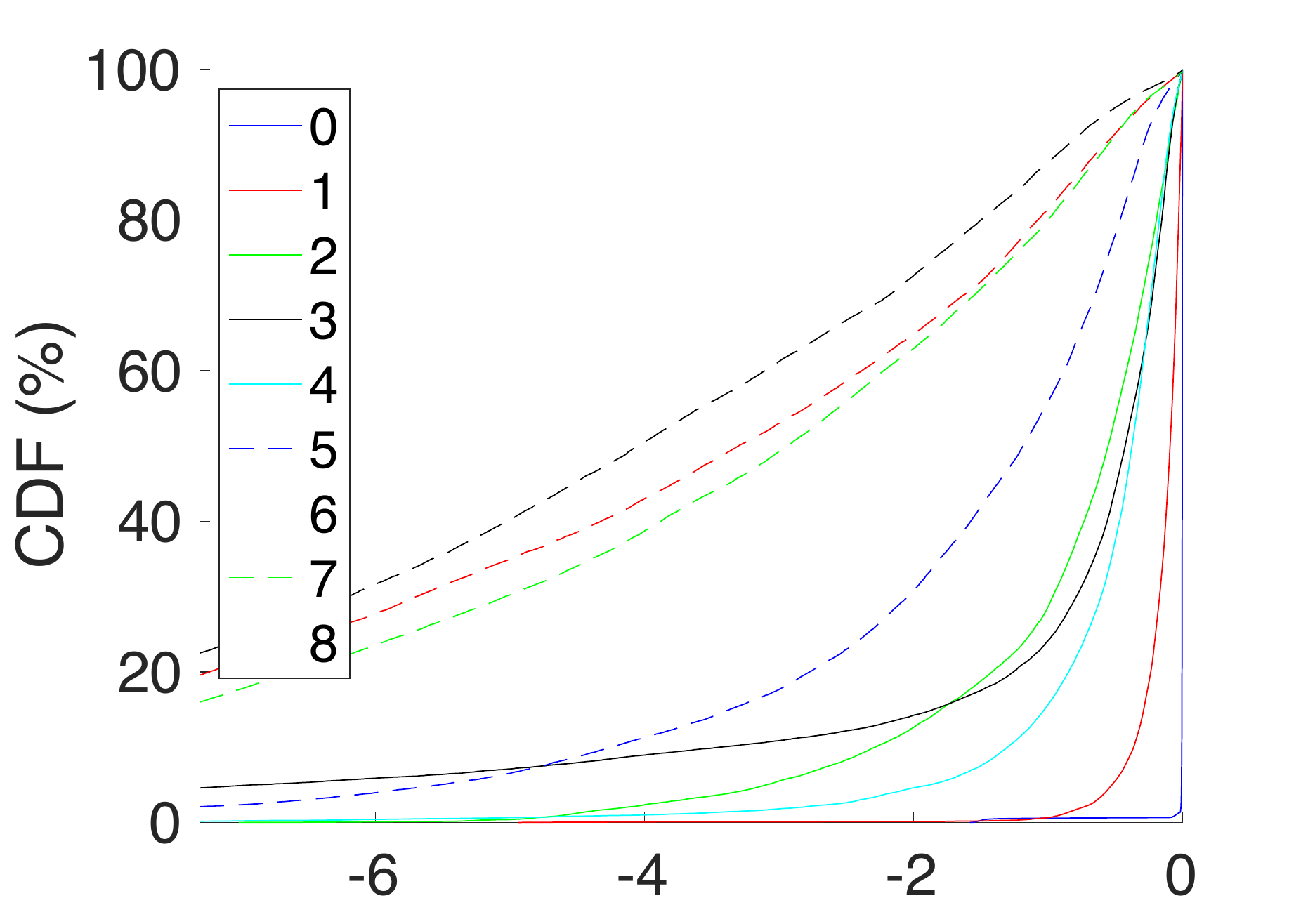}}
    \vspace{-0.10cm}
    \caption{\small Differential unitarity cumulative distribution functions across 9 occupancy scenarios for $\textrm{Rx}$ and $\textrm{Dy}$. Legend denotes how many moving people are present. (a) and (c) correspond to the successive pairwise correlations, while (b) and (d) measure the rate of range. The rate of change is referred to by the $^\prime$ operator.}
    \label{fig:differentialUnitarityChannelVolatility}
    \vspace{-0.10cm}
\end{figure*}

\begin{figure}[h]
  \centering
    \subfloat[pairwise\label{fig:diffUnitPairwise}]{\includegraphics[width=0.45\textwidth]{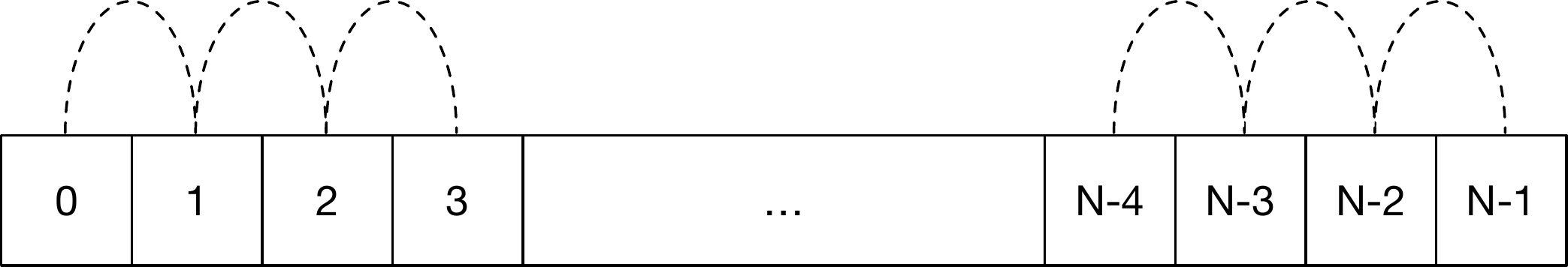}}
    \vspace{0.25cm} \\
    \vspace{-0.15in}
    \subfloat[slope\label{fig:diffUnitSlope}]{\includegraphics[width=0.45\textwidth]{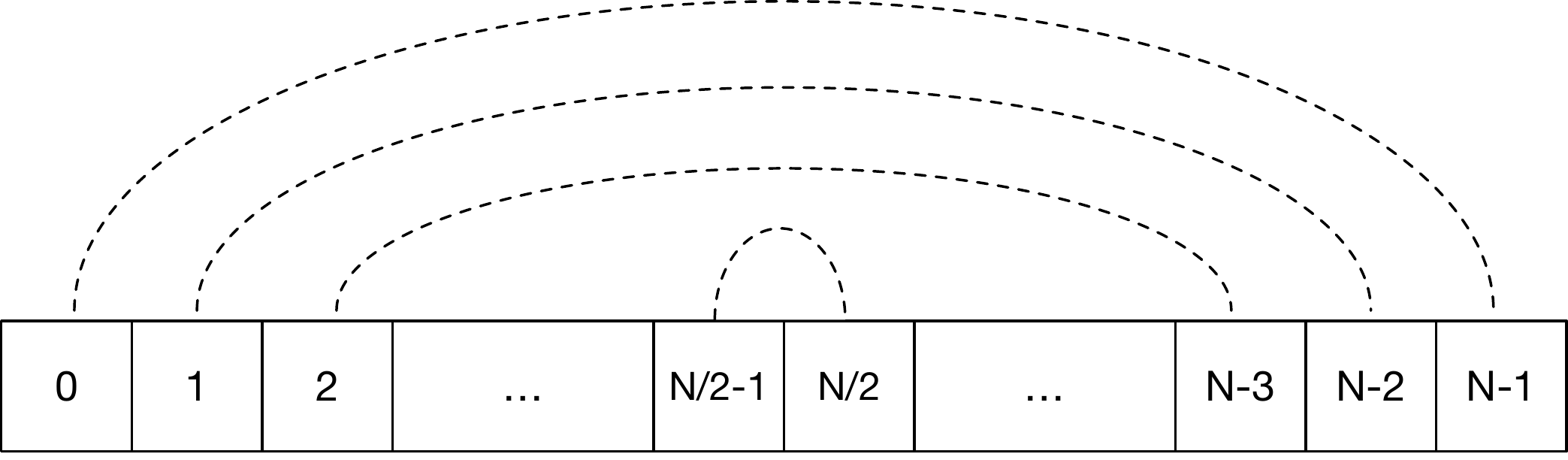}}
    \vspace{-0.10in} 
    \caption{Interrogating rate of change of differential unitarity by correlating farther apart eigenvectors with successively decreasing distance.}
    \label{fig:diffUnitPairwiseVsSlope} 
\end{figure}
 
Equations~\eqref{eq:dirrentialUnitarity_rx}~\&~\eqref{eq:dirrentialUnitarity_dy} represent two degrees of freedom through which we can measure the volatility in the wireless channel as a result of human stressors: (1) spatial from multiple antennae and (2) temporal across the delayspread (or equivalently bandwidth). 
We next build intuition for these complementary differential unitarity metrics by presenting a series of concrete numerical examples.

We return to the uncontrolled movement dataset reported in Section~\ref{sec:subspace_characterisation}. 
Further, let us examine the behaviour of the differential unitarity for the $1$st subspace component of both the receive-side and delayspread subspaces i.e. $\hat{u}_{\textrm{Rx},1}$ and $\hat{u}_{\textrm{Dy},1}$, respectively. 
Figure~\ref{fig:differentialUnitarityChannelVolatility} plots the cumulative distribution functions (CDFs) for $\hat{u}_{\textrm{Rx},1}$ and $\hat{u}_{\textrm{Dy},1}$ for all 9 occupancy cases. 
Specifically, note the dispersive nature of the metric in figures~\ref{fig:rxDiffUnit}~\&~\ref{fig:dyDiffUnit} as a function of increased human-induced channel perturbations. 
It is clear that the dispersion in the statistics of the magnitude of differential unitarity---corresponding to the $1$st subspace components---monotonically increases, generally, with increased human movement.

The diagram in figure~\ref{fig:diffUnitSlope} shows that the subspace bases relating to $N$ time periods are buffered so that comparison can be made across a wider time window.  
Thus, for example, the \textcolor{highlight_clr}{eigenbases} at $t=0$ can be compared with those at $t=N-1$, the eigenbases at $t=1$ can be compared with the eigenbases at $t=N-2$ and the eigenbases at $t=2$ can be compared with the eigenbases at $t=N-3$. 
Such an arrangement may enable the changes in channel statistics to be viewed across a wider time period and may enable the rate of change of eigenvector unitarity to be determined.

Deriving a measure of the rate of change of differential unitarity has the advantage of increasing the separation of the CDFs of figures~\ref{fig:rxDiffUnit}~\&~\ref{fig:dyDiffUnit}. 
To this end, we apply the scheme depicted in the lower diagram of figure~\ref{fig:diffUnitPairwiseVsSlope} (termed ``slope'') to the same experiment and obtain the CDFs shown in figures~\ref{fig:rxDiffUnitPrime}~\&~\ref{fig:dyDiffUnitPrime}. 
It is readily evident that the CDFs corresponding to the rate of change in the differential unitarity extracted over a window of time experience increased dispersion as a result of human occupancy. 
This may allow for learning and/or calibrating better discrimination boundaries in the inference logic.

\begin{figure*}[t]
    \centering
        \subfloat[falling -- batch]{\includegraphics[width=0.245\textwidth]{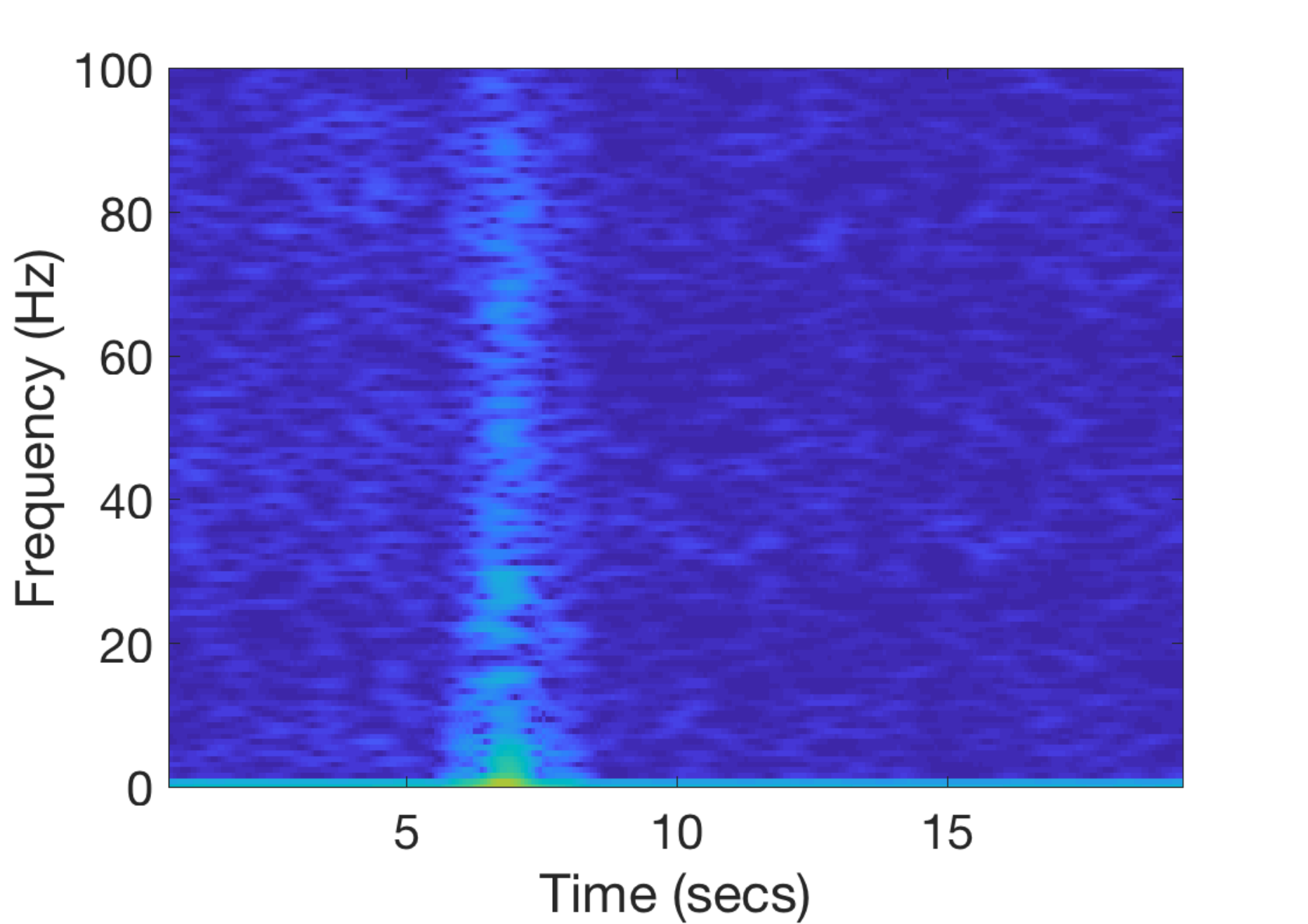}}
        \subfloat[lying down -- batch]{\includegraphics[width=0.245\textwidth]{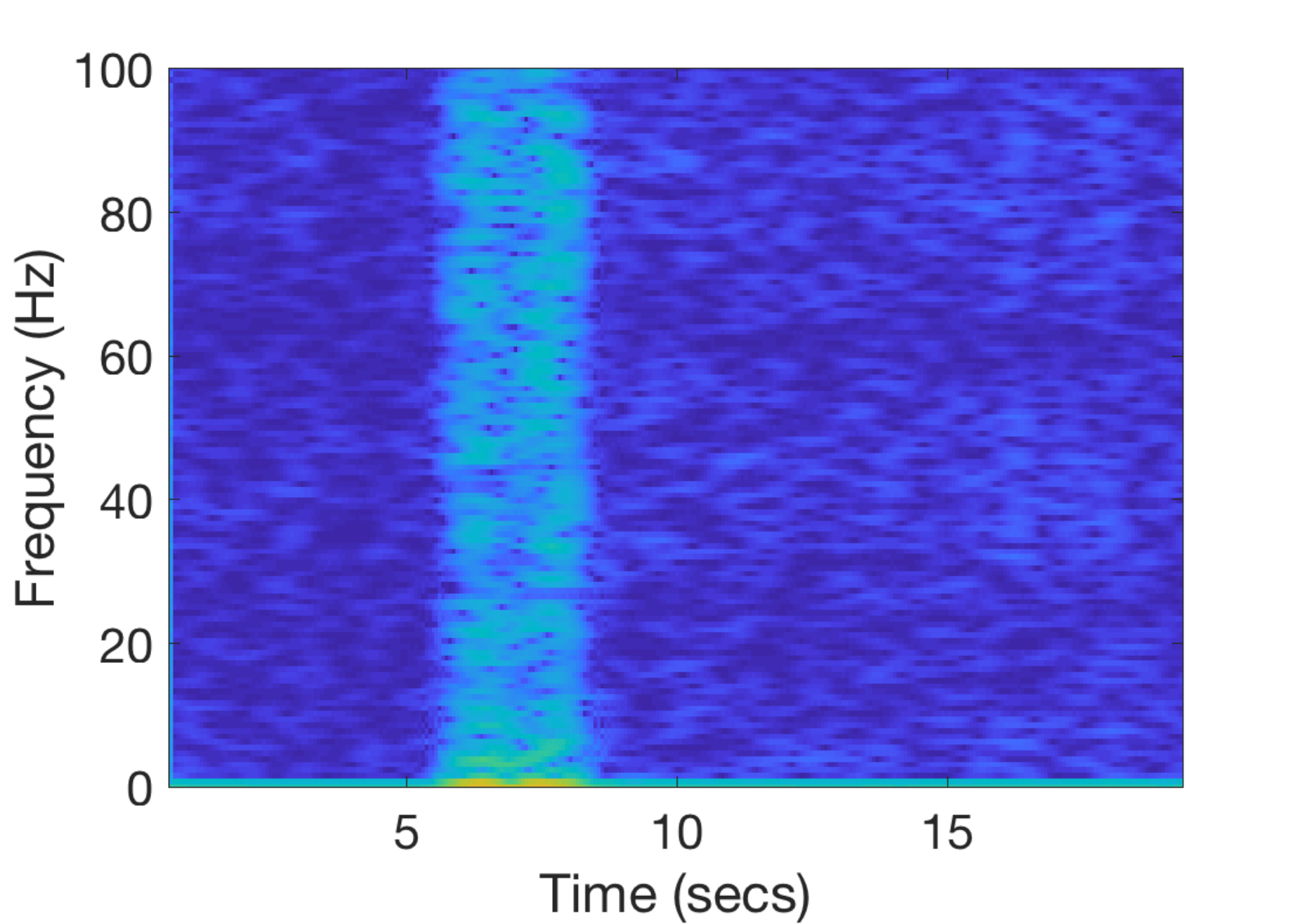}}
        \subfloat[walking -- batch]{\includegraphics[width=0.245\textwidth]{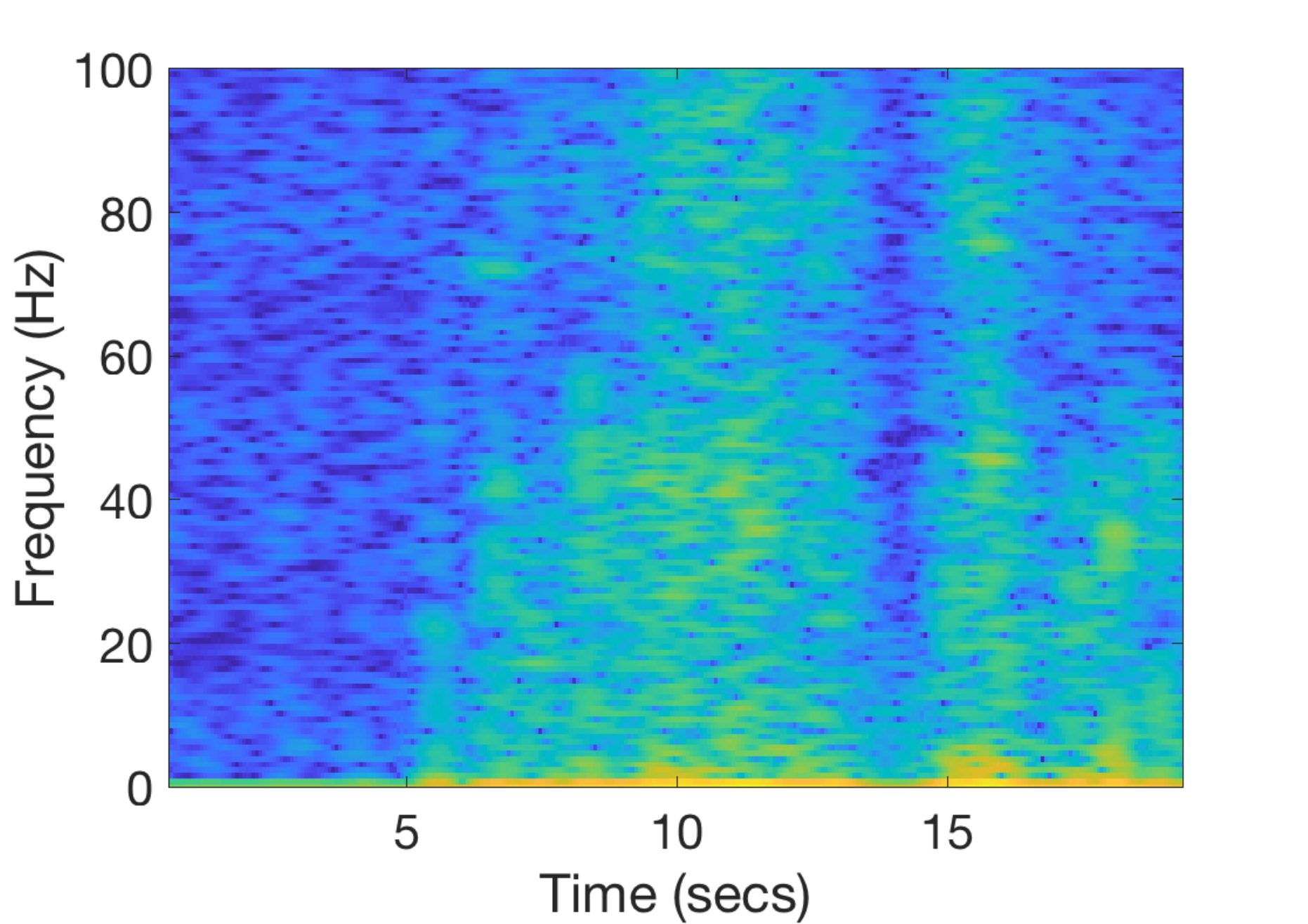}}
        \subfloat[running -- batch]{\includegraphics[width=0.245\textwidth]{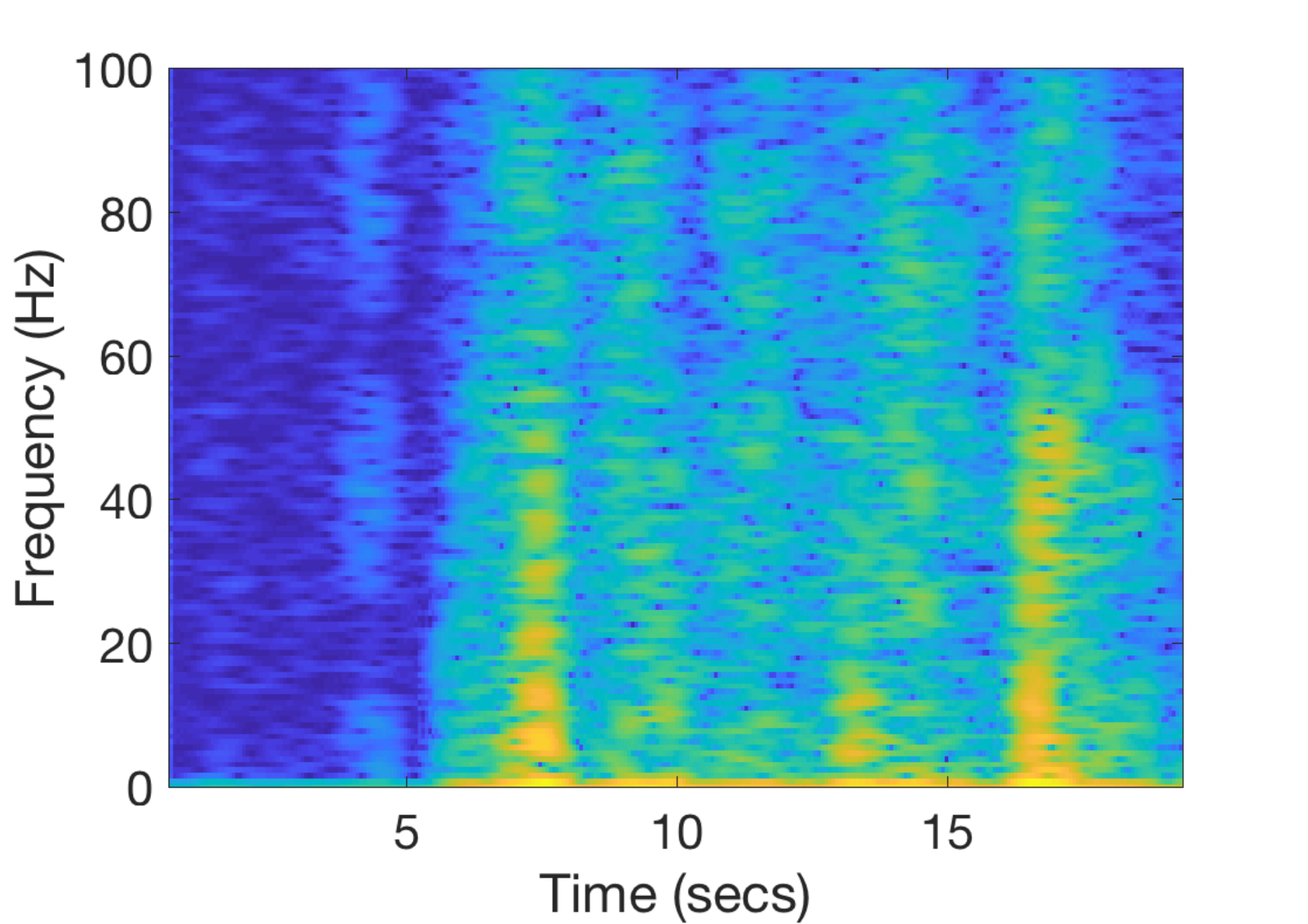}} \\
        \vspace{-0.15in}
        \subfloat[falling -- stochastic]{\includegraphics[width=0.245\textwidth]{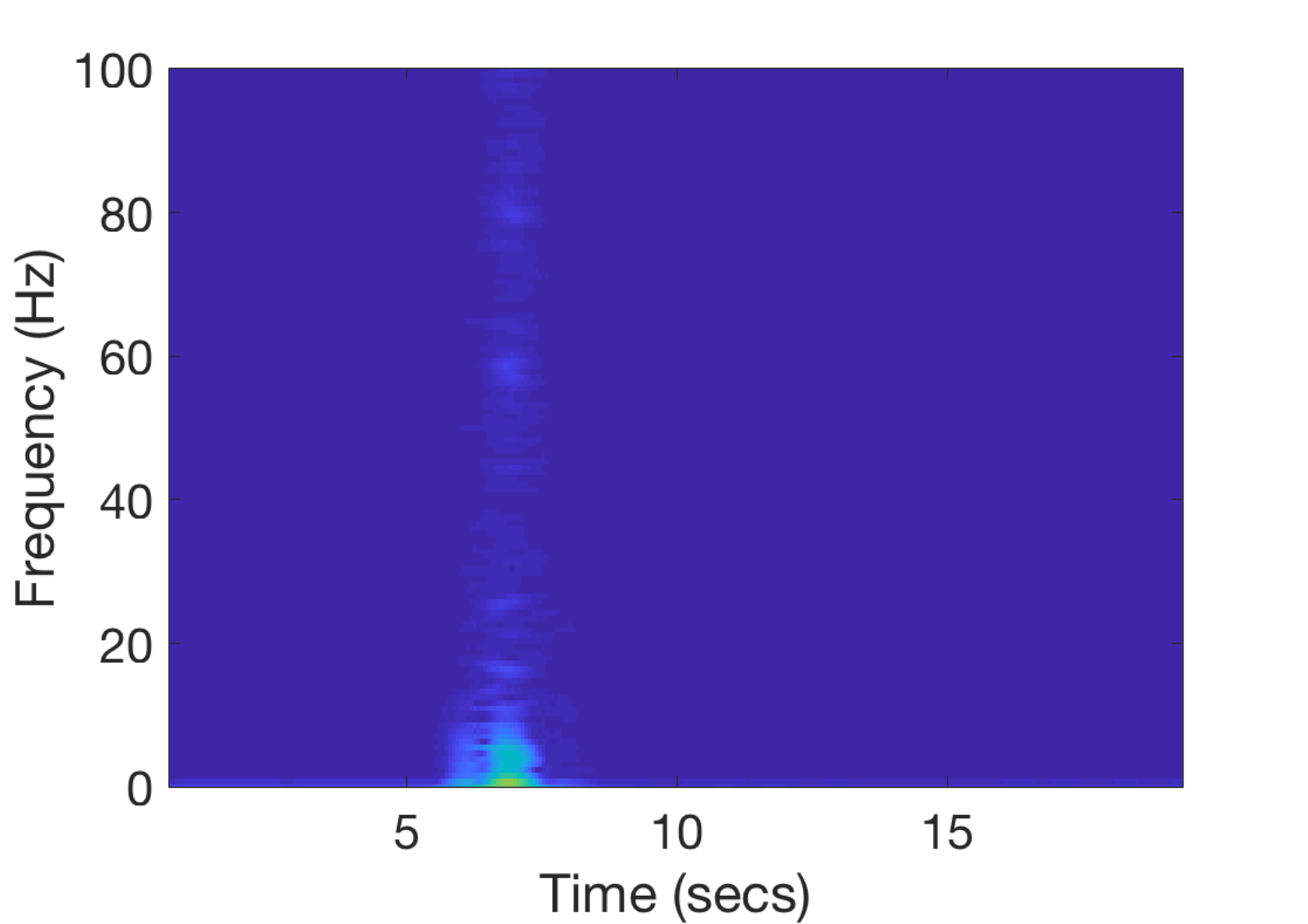}}
        \subfloat[lying down -- stochastic]{\includegraphics[width=0.245\textwidth]{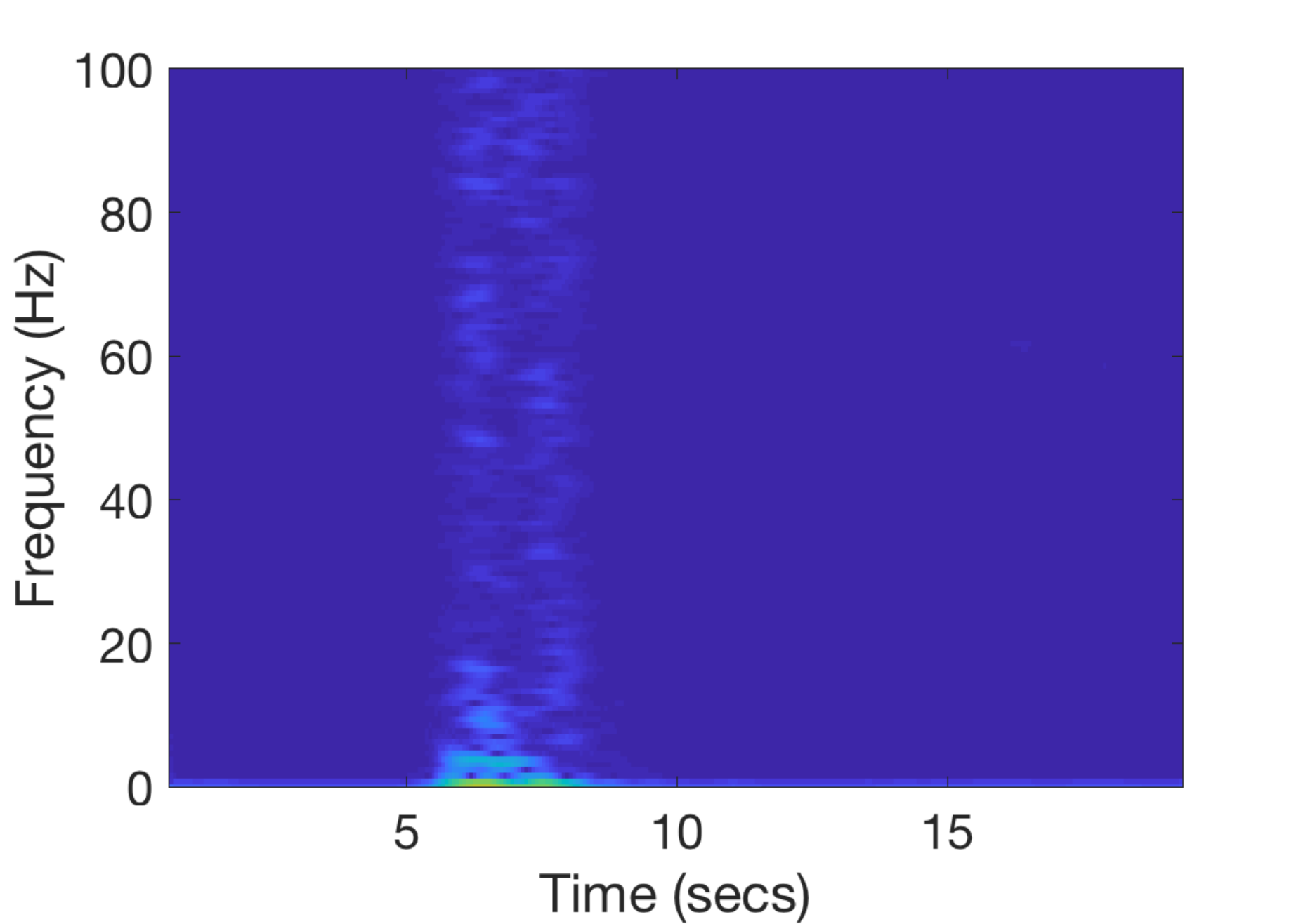}}
        \subfloat[walking -- stochastic]{\includegraphics[width=0.245\textwidth]{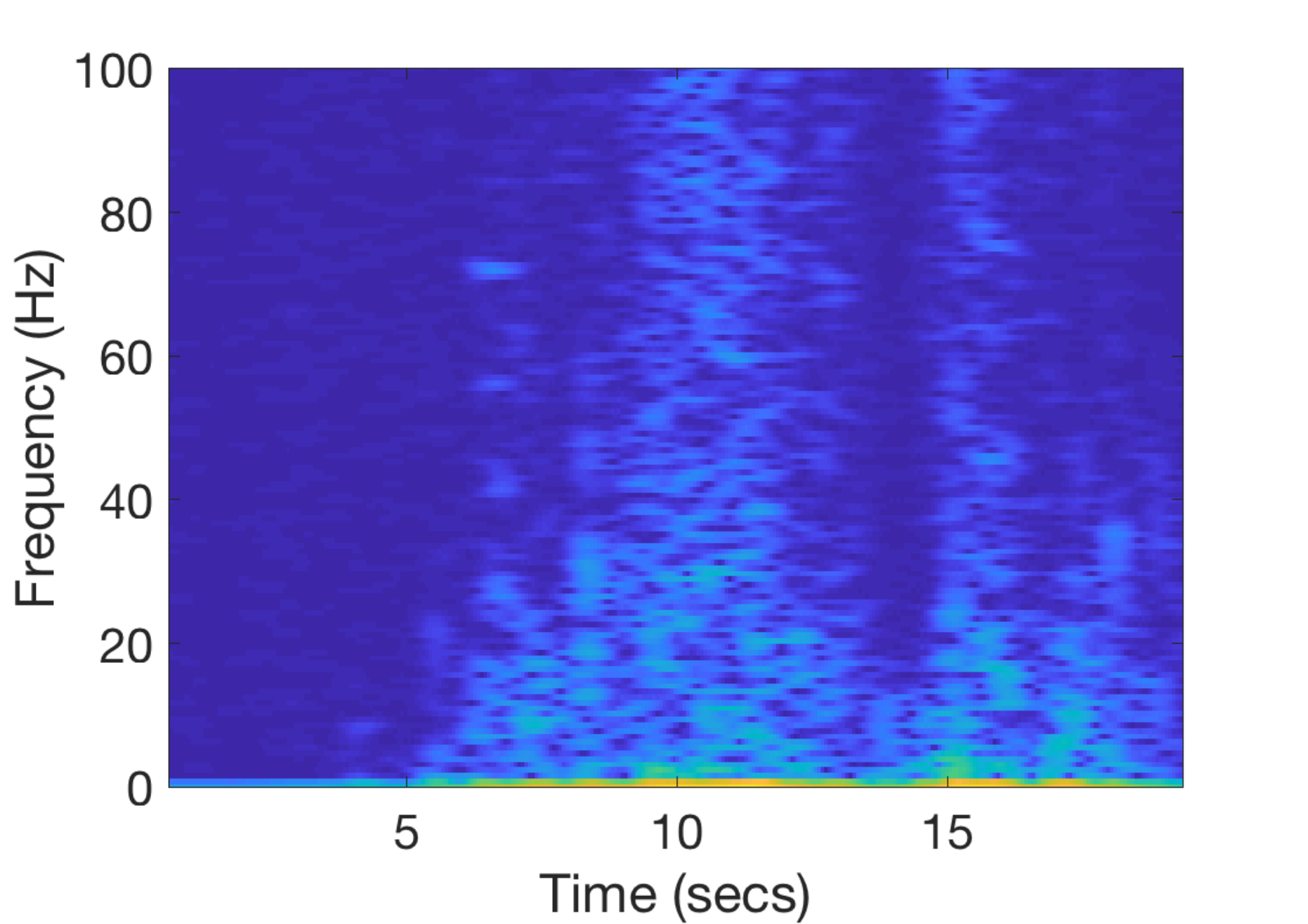}}
        \subfloat[running -- stochastic]{\includegraphics[width=0.245\textwidth]{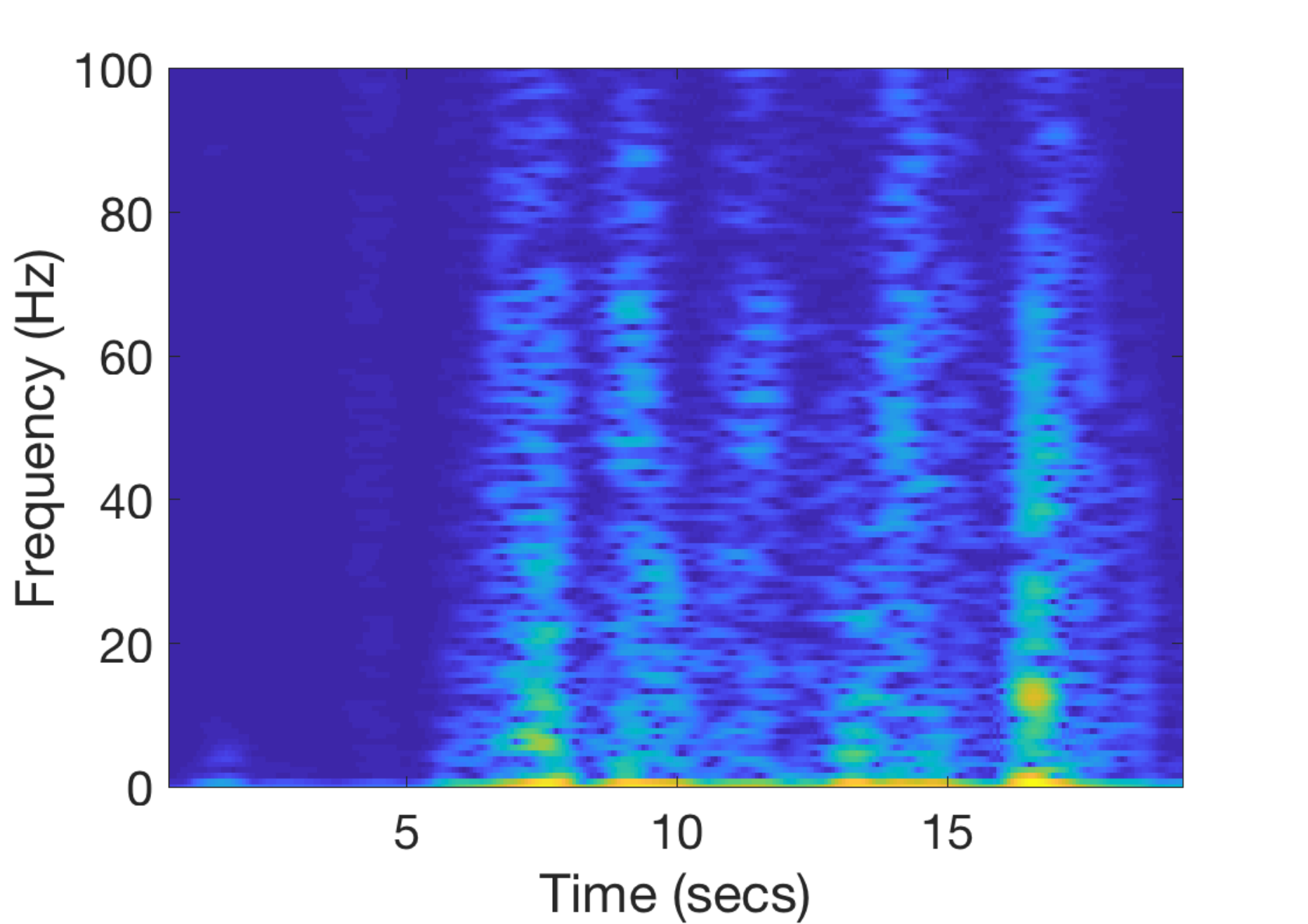}}
    \vspace{-0.10cm}
    \caption{\small Spectrograms for 4 activities (falling, lying down, walking, and running) using two subspace update variants: batch and stochastic. It is readily that stochastic has a filtering effect on the time-frequency localisation of pairwise differential unitarity subspace tracking metric.}
    \label{fig:subspaceTrackingSpectrogram}
\end{figure*}

\begin{figure*}[t]
    \centering
        \subfloat[falling\label{fig:subspace_slope_falling}]{\includegraphics[width=0.33\textwidth]{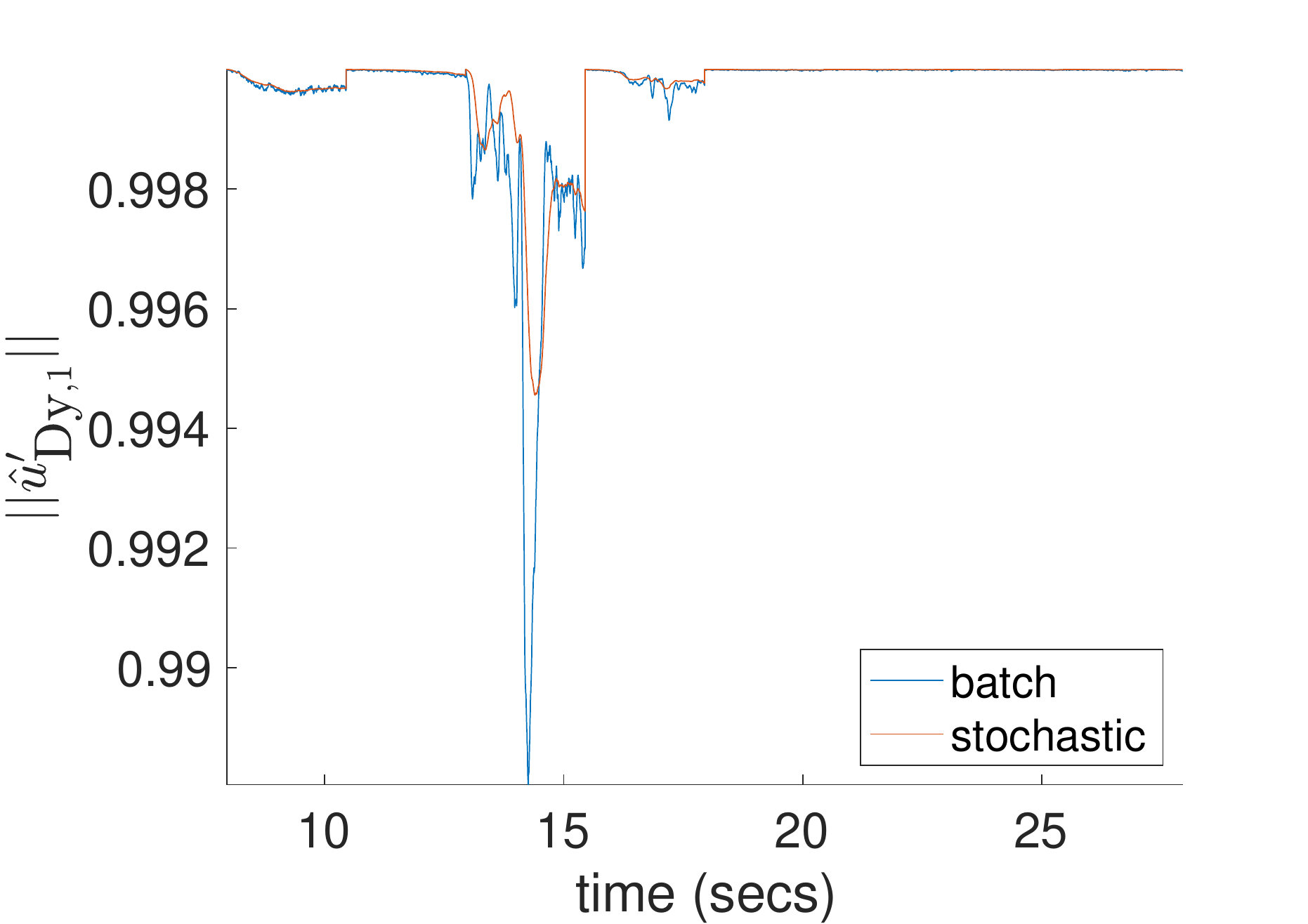}}
        \subfloat[lying down]{\includegraphics[width=0.33\textwidth]{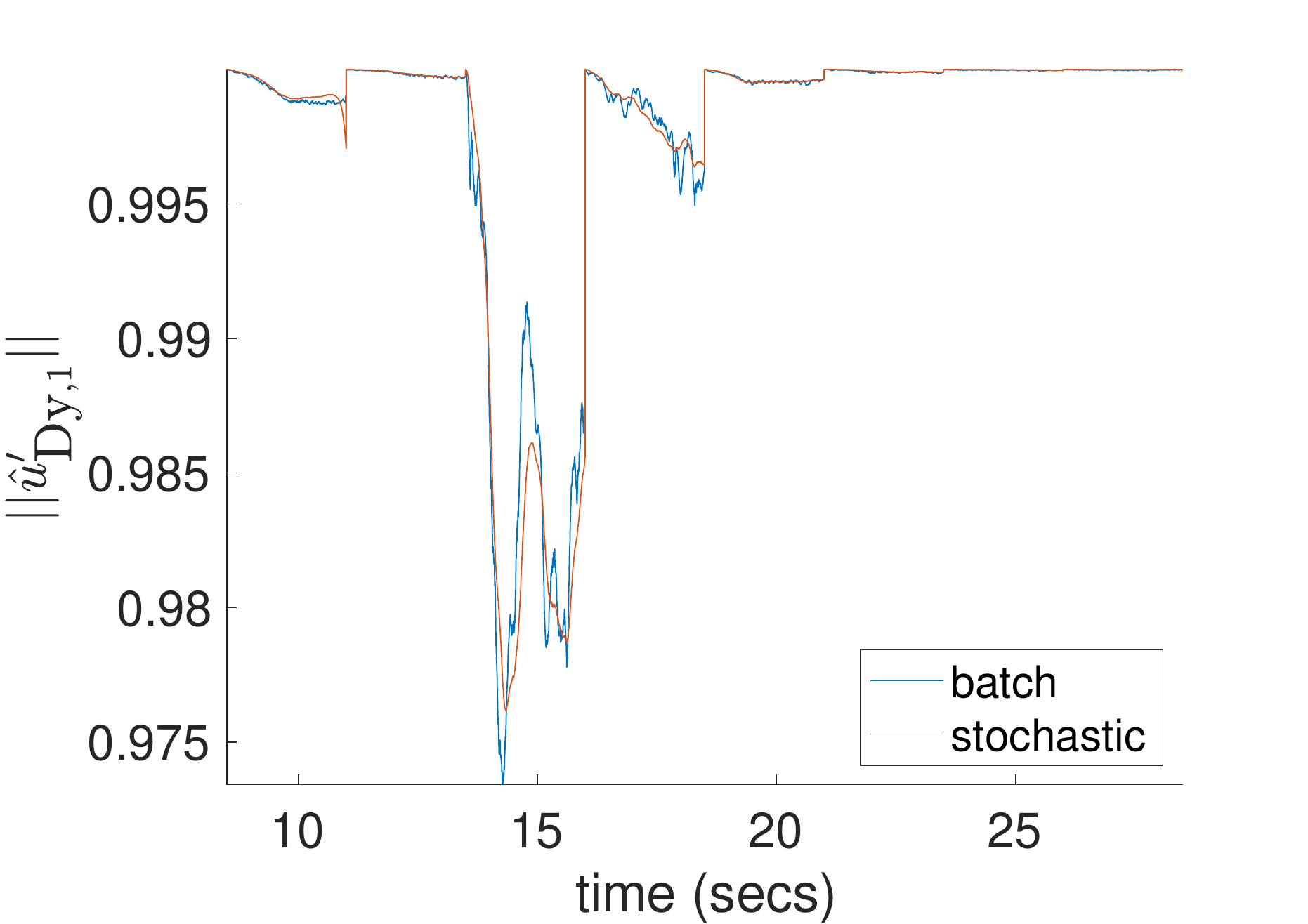}}
        \subfloat[sitting down]{\includegraphics[width=0.33\textwidth]{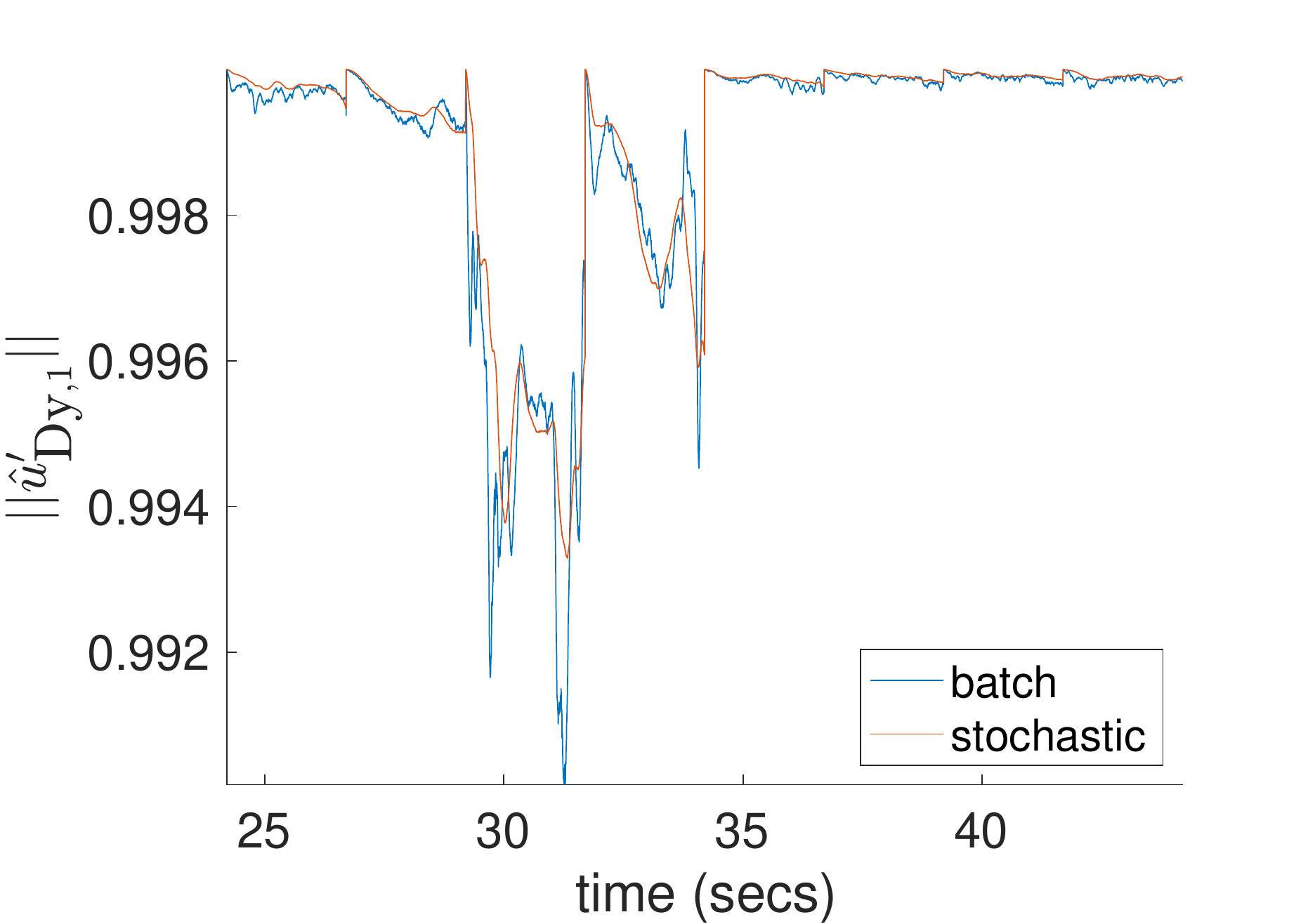}} \\
        \vspace{-0.15in}
        \subfloat[standing up]{\includegraphics[width=0.33\textwidth]{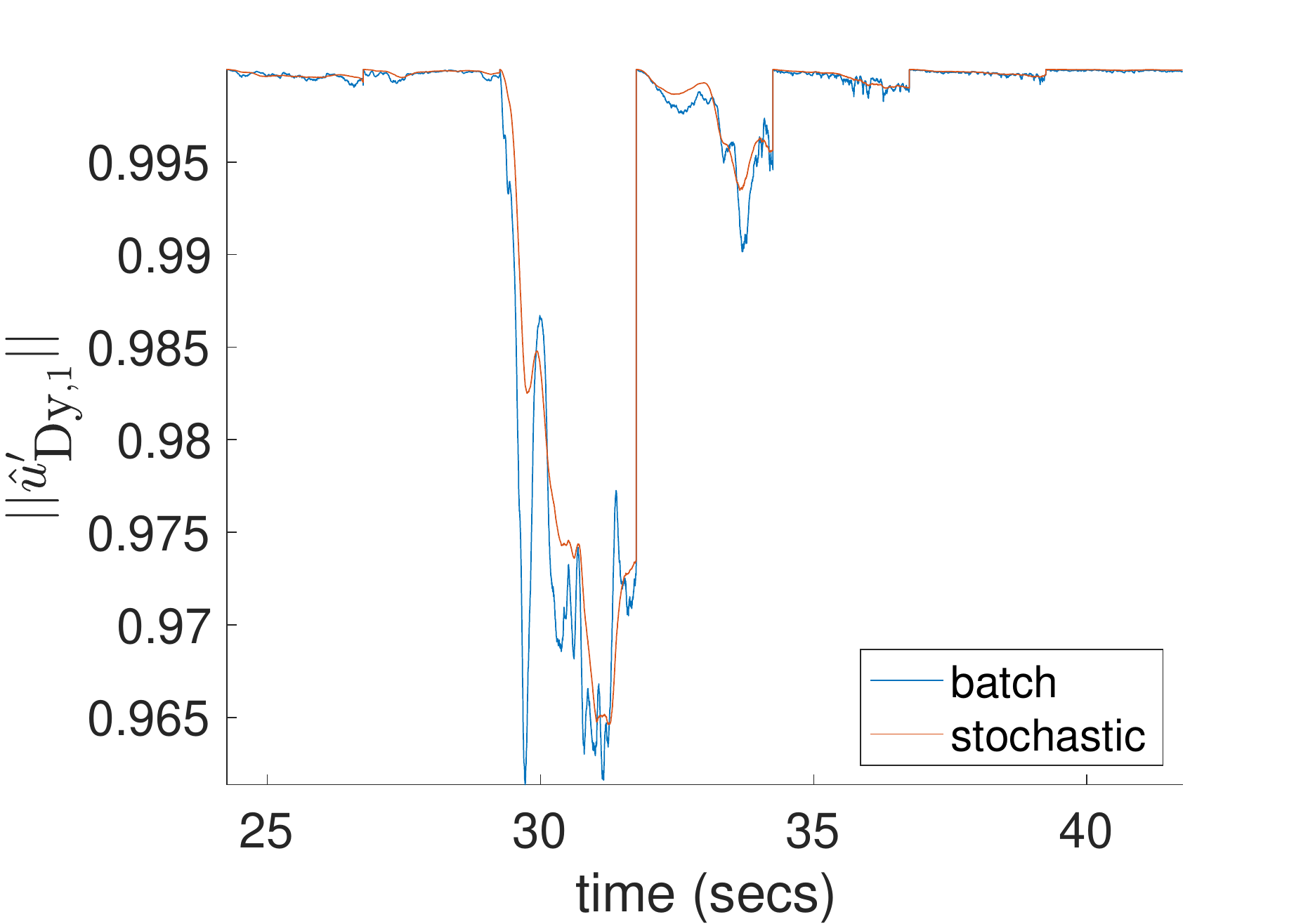}}
        \subfloat[walking]{\includegraphics[width=0.33\textwidth]{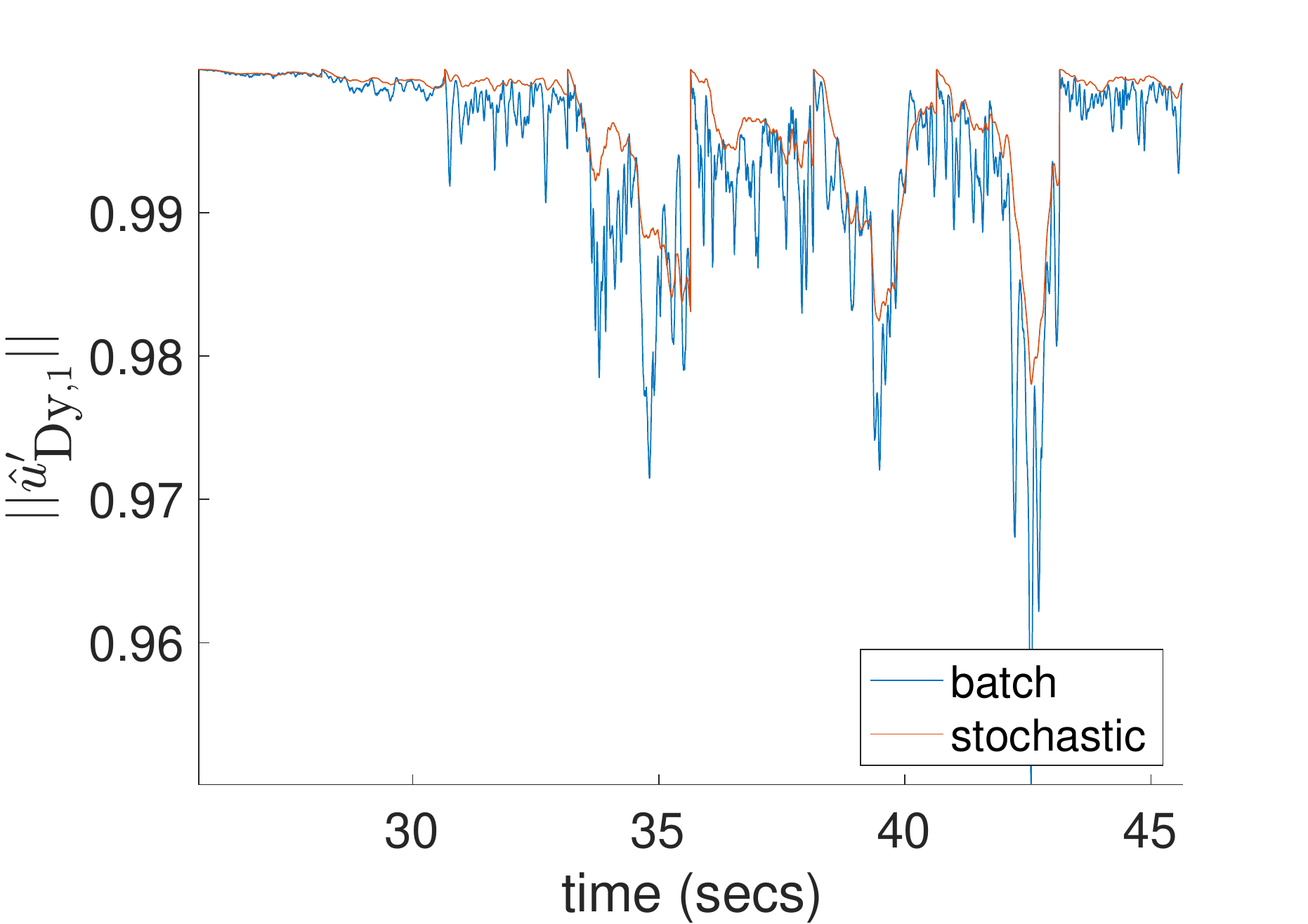}}
        \subfloat[running\label{fig:subspace_slope_running}]{\includegraphics[width=0.33\textwidth]{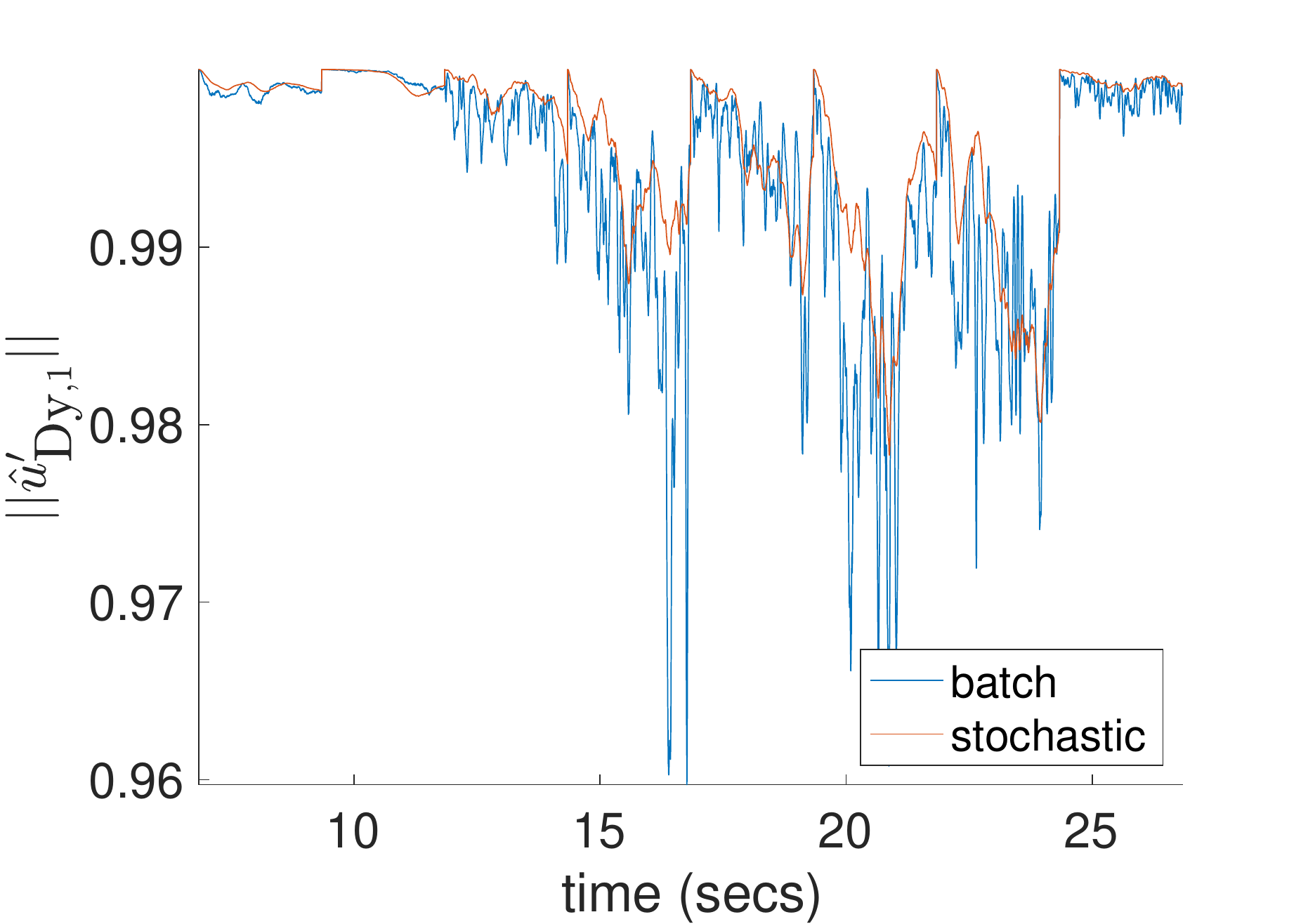}}
    \vspace{-0.10cm}
    \caption{\small Waveforms for 6 activities (standing up, sitting down, lying down, falling, walking, and running) using two subspace update variants: batch and stochastic. It is readily noticeable that stochastic has a filtering effect on the slope differential unitarity subspace tracking metric.}
    \label{fig:subspaceTracking_batchVsStochastic}
\end{figure*}

\noindent \textbf{Subspace sampling.} Recalling equation~\eqref{eq:channel-covariance}, we note that the expectation operator implies an averaging effect. 
Earlier we have elaborated on the notion of stationarity period and its connection to CSI sampling and application granularity requirements. 
Yet another pertinent aspect for consideration lies in how to realise the expectation. 
Broadly, there are two methods often employed in classic signal processing literature for updating the signal subspace: (i) stochastic approximation, and (ii) batch averaging. These two variants have implications on signal subspace tracking, which we discuss next. 

An unbiased stochastic expectation estimator is given by
\begin{align}
    \mtrx{R}_{\textrm{x}}[k] &= (1-\lambda) \sum_{n=0}^{k} \lambda^{k-n} \tnsr{H}_{(m)}[n] \tnsr{H}^H_{(m)}[n]
    \label{eq:stochastic_expectation_0}
\end{align}
where $\textrm{x} \in [\textrm{Rx}, \textrm{Tx}, \textrm{Dy}]$ and $m \in [1, 2, 3]$, respectively.
This estimator reduces to the recursive expression
\begin{align}
    \mtrx{R}_{\textrm{x}}[k] &= \lambda \mtrx{R}_{\textrm{x}}[k-1] + (1-\lambda) \tnsr{H}_{(m)}[k] \tnsr{H}^H_{(m)}[k]
    \label{eq:stochastic_expectation}
\end{align}
where $\lambda \in \textcolor{highlight_clr}{[0, 1)}$ is a forgetting factor often chosen close to 1.
This stochastic estimator accounts for a long channel history, albeit while de-emphasising far away events. Such subspace update tends to ``dampen'' the effect of abrupt channel changes on the signal subspace. Alternatively, these abrupt changes can also be preserved and admitted into the subspace using the sliding window (a.k.a batch) approach given by
\begin{align}
    \mtrx{R}_{\textrm{x}}[k] &= \frac{1}{L} \sum_{n=k-L+1}^{k} \tnsr{H}_{(m)}[n] \tnsr{H}^H_{(m)}[n]
    \label{eq:batch_expectation}
\end{align}
where $L$ is the window size determined by the assumed stationarity period.

We now compare and contrast between these two subspace update variants. 
An activity recognition dataset available publicly is used~\cite{Yousefi17_SurveyOnWifiBehaviourRecognition}. 
The dataset is comprised of 6 single-user activities; namely, standing up, sitting down, lying down, falling, walking, and running. 
SIMO CSI data from three receivers is sampled at 1 ksps rate. 
For added tracking responsiveness and resolution, we choose a stationarity period of $25$ms and proceed to update the covariance matrix with $95$\% CSI overlap from previous stationarity period. This results in around $800$ Hz subspace update rate.

In Figure~\ref{fig:subspaceTrackingSpectrogram}, we perform time-frequency localisation on the pairwise differential unitarity subspace tracking metric. 
The localisation uses a window of $1.28$ seconds with $95$\% content overlap between two windows for finer time-frequency resolution. 
In the interest of space, only four single-user activities are shown corresponding to falling, lying down, walking, and running. 
The spectrograms of the upper row of Figure~\ref{fig:subspaceTrackingSpectrogram} were generated using the batch subspace update variant; while those in the lower row utilised the stochastic variant with a forgetting factor $\lambda = 0.99$. 
The colour coding of the spectrograms in each row was group-harmonised in order to \textcolor{highlight_clr}{convey} correct information about the differential intensity of the time-frequency bins across activities. 
We therefore safely omit the colour maps from the spectrograms. 
As touched upon previously, the batch subspace update is more responsive to background disturbances in the channel and would admit these into signal subspace. 
We can readily observe more background variations across all activities in the upper series of spectrograms. 
Despite this, we can still see distinctly individual behaviour across these activities---falling being the most concentrated in time-frequency and running being the most dispersed.
However, it is interesting to see how the stochastic update was able to \emph{filter} out much of the background channel disturbances while preserving the discriminative features of the four activities; namely, the increased time-frequency dispersion from falling, lying down, through to walking and running---again the latter being the most dispersed. 

We have opted to conduct time-frequency localisation on the pairwise subspace tracker owing to its more intuitive association with speed i.e. $1$st-order derivative of subspace evolution. 
A justification for the correspondence between the rate of change in CSI and speed can be found in~\cite{Wang15_UnderstandingAndModelingWiFiHumanActivity}. 
Our $1$st-order differentiation of the subspace can be viewed as a generalised \emph{fusion} method for extracting information embedded in all subcarriers simultaneously. 
This fusion is a data-level fusion, rather than feature-level approaches involving ad hoc subcarrier selection strategies~\cite{Wang18_LowEffortDfl}. 
Some reported Wi-Fi sensing systems resort to selecting subcarriers of better SNR since frequency selectivity of wideband Wi-Fi channels causes some subcarriers to fall within the channel nulls---with obvious consequences for their reliability. 
Our subspace approach systematically fuses information contained in all subcarriers without the need to perform preconditioning. 
However, unlike the PCA-based approach~\cite{Wang15_UnderstandingAndModelingWiFiHumanActivity}, this fusion is principled, interpretable, and has its roots in formal wireless channel concepts~\cite{Weichselberger06_StochasticMimoChannelModelWithJointEndCorrelations,Costa08_NovelWidebandMimoChannelModel, Tulino06_Capacity-achievingMimoCovariance, Lopez-Martinez15_EigenvalueDynamicsOfCentralWishartForMimo}.

As illustrated in Figure~\ref{fig:diffUnitSlope}, we use a robust sampling technique to obtain clean statistics from the differential unitarity measurements. 
In Figure~\ref{fig:subspaceTracking_batchVsStochastic}, we illustrate the effect of this choice. 
Six single-user activities---standing up, sitting down, lying down, falling, walking, and running---for the slope tracker are depicted. 
Quick inspection of these plots corroborate the earlier findings of the spectrograms analysis; namely, that stochastic filters high-frequency channel perturbations compared to batch. 
That is, stochastic tracks the \emph{envelope} of the activity rather than its and/or the channel's background high-frequency fluctuations.
We have alluded to this \emph{tunable} channel detail in the signal subspace, be it channel background- or activity-related, by the hat accent in Equation~\eqref{eq:sig-noise-subspaces}. 
The abrupt activity of falling has an impulse-like acceleration content, while running is the richest in such $2$nd-order rate of change moments.

\begin{figure*}[t]
    \centering
        \subfloat[magnitude\label{fig:dualityMagnitude}]{\includegraphics[width=0.25\textwidth]{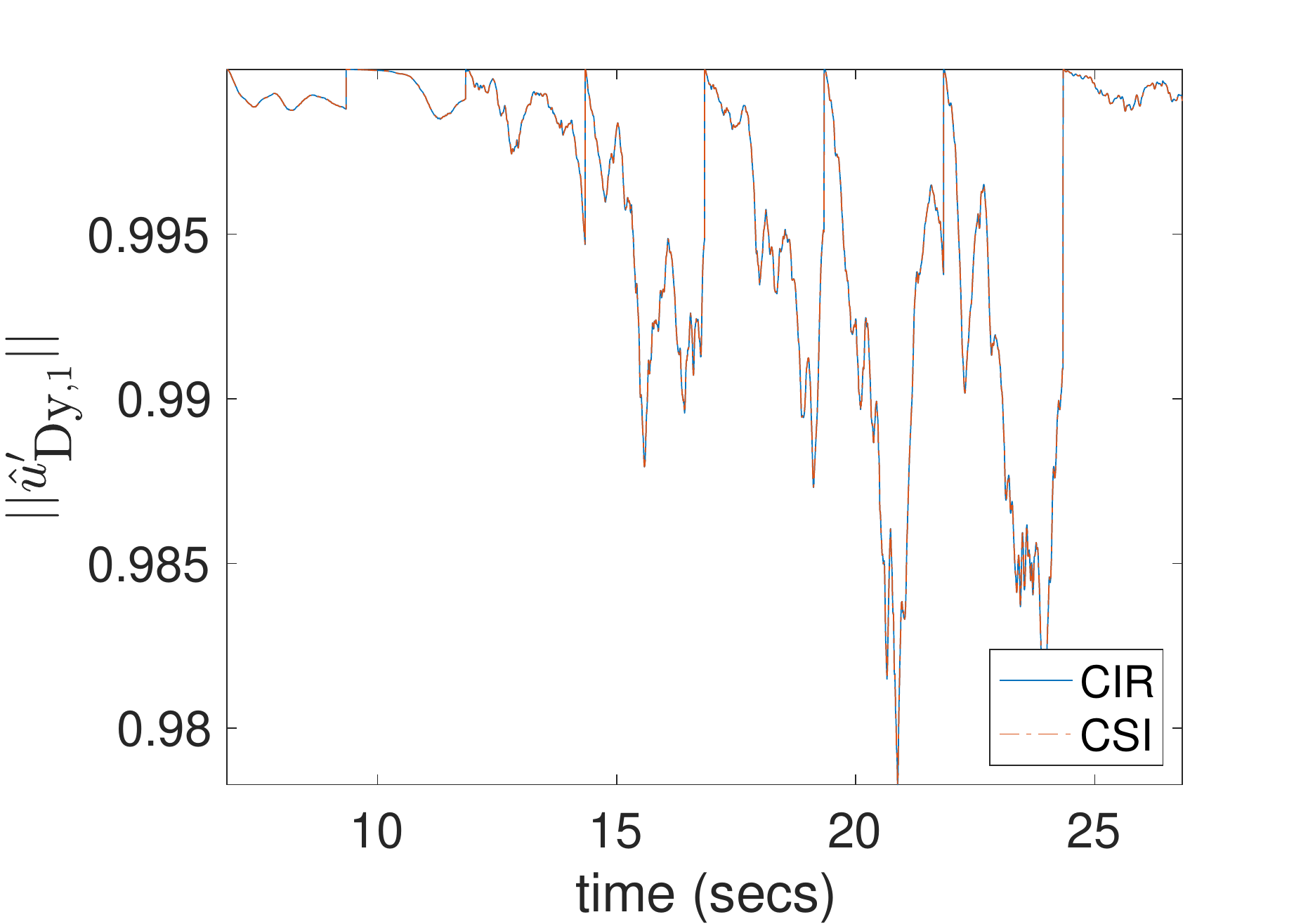}}
        \subfloat[error power\label{fig:dualityError}]{\includegraphics[width=0.25\textwidth]{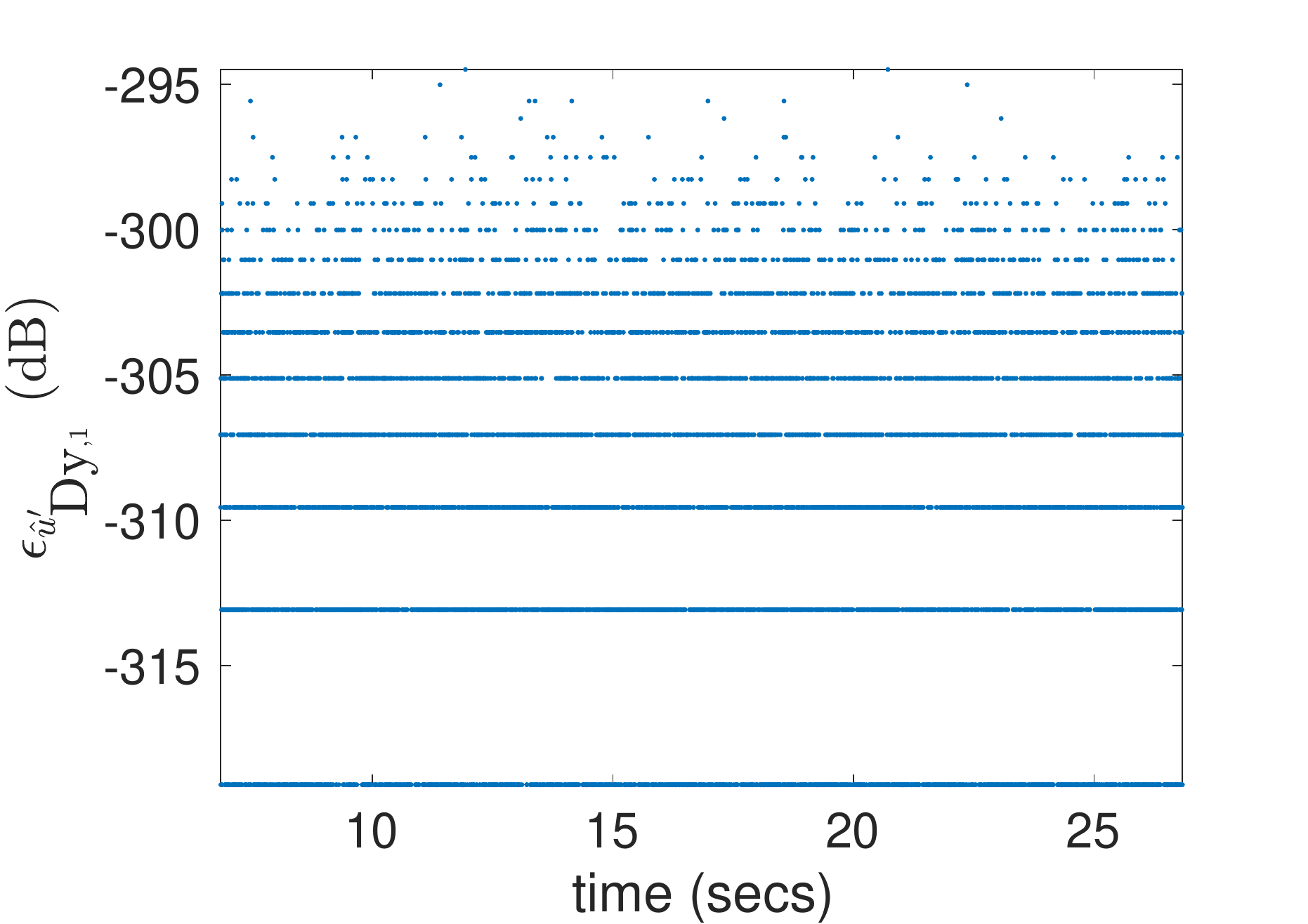}}
        \subfloat[phase\label{fig:dualityPhase}]{\includegraphics[width=0.25\textwidth]{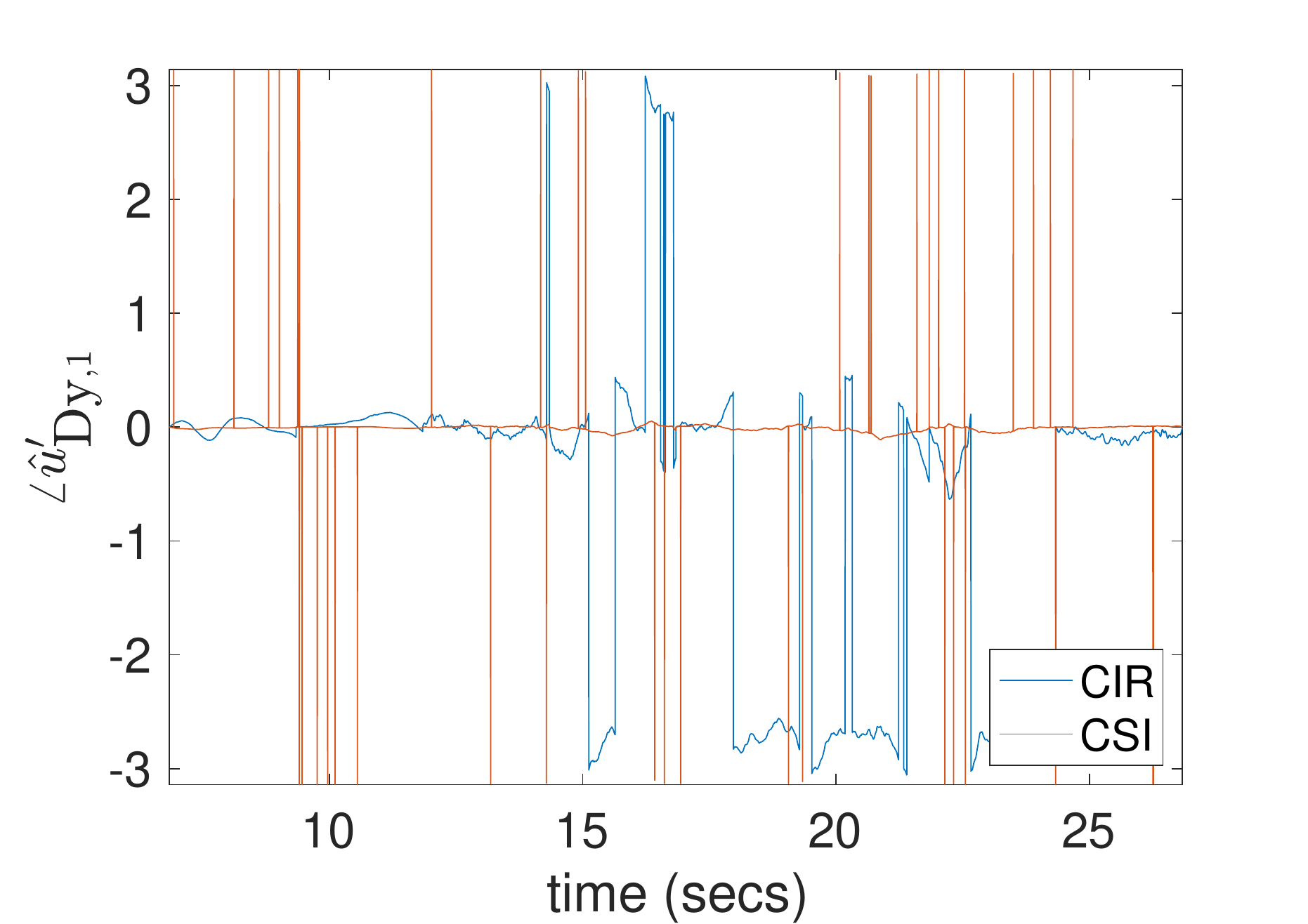}}
        \subfloat[scatter\label{fig:dualityScatter}]{\includegraphics[width=0.25\textwidth]{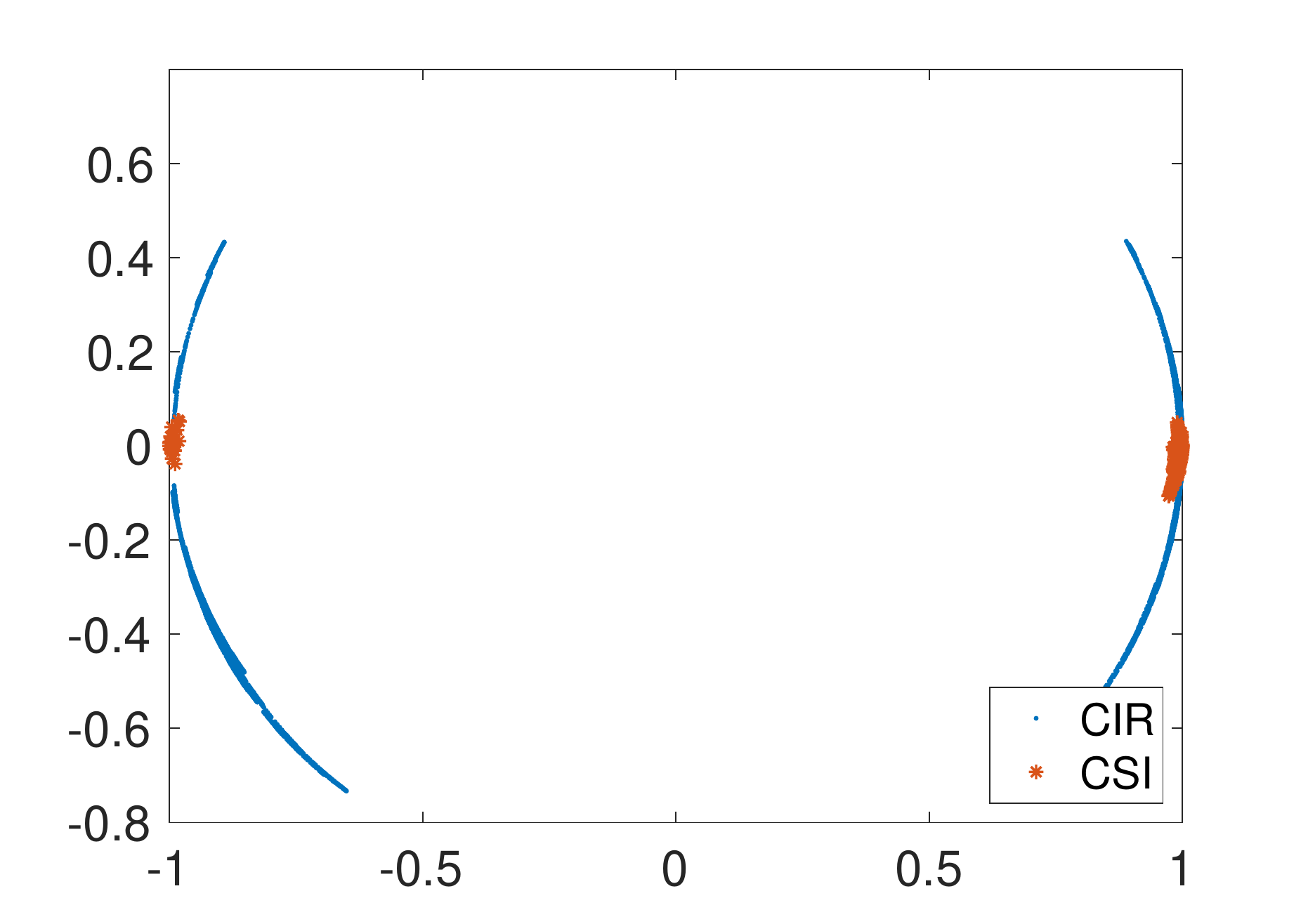}}
    \vspace{-0.10cm}
    \caption{\small Waveforms for slope tracker corresponding to the running activity. 
    Differential metric is identical in magnitude, but subspace phase stabilities exhibit interesting variations that are different depending on whether subspace decomposition is performed in the time-domain or frequency-domain.}
    \label{fig:subspaceTrackingDuality_timeVsFreq}
\end{figure*}

\noindent \textbf{Duality.} For completeness, we provide commentary on the pertinent issue of choosing a channel representation: time- versus frequency-domain. 
The structured model we introduced in Section~\ref{sec:mimo_model} has been validated with empirical channel impulse response (CIR) measurements i.e. in the time-domain. 
Identical eigenspace formulation has been applied in the frequency-domain for CSI instead~\cite{Zhang12_GeneralCoupling-basedModelForWidebandMimoChannels}, and also validated with empirical capacity measurements. 
Since our subspace trackers are \emph{differential} in nature, tracking is insensitive to the representation of the channel be it time- or frequency-domain. 
That said, a salient point in relation to the \emph{phase} behaviour of the trackers is worth making for completeness of treatment. 
The numerical perturbations experienced in the time-domain---as a function of human motion---differ to those experienced in the frequency-domain. 
Classic work on the stability of subspaces provides bounds on their trigonometric (i.e. angular) behaviour as a function of technical mathematical issues ranging from eigenvalue spectral gap to numerical residuals~\cite{Davis70_RotationOfEigenvectorsByPerturbation}.

To highlight this point, we revisit the waveform of the slope subspace tracker for the running activity depicted in Figure~\ref{fig:subspace_slope_running}. 
We perform channel decomposition through to differential unitarity calculations both for the CIR and the CSI versions of measurements (i.e. time \& frequency domains). The results are shown in Figure~\ref{fig:subspaceTrackingDuality_timeVsFreq}. As illustrated in Figures~\ref{fig:dualityMagnitude}~\&~\ref{fig:dualityError}, it is intuitive to note that the differential tracker performs identically in time and frequency domains. After all, a linear operator (i.e. [I]DFT) translates between one domain to another. The occasional polarity switch in the phase of the differential tracker (Figures~\ref{fig:dualityPhase}~\&~\ref{fig:dualityScatter}) can be explained by the effects studied in~\cite{Davis70_RotationOfEigenvectorsByPerturbation}. 
However, it is interesting to note the increased phase instabilities when running the differential metric on top of CIR measurements over those obtained from CSI measurements. 
This phenomenon can be readily seen in Figures~\ref{fig:dualityPhase}~\&~\ref{fig:dualityScatter}. 
We conjecture that the sparsity in the CIR measurements (i.e. impulse-like nature) compared to the smoother CSI measurements causes numerical instabilities which give rise to added phase instabilities in the subspace. 
The scatter plot of Figure~\ref{fig:dualityScatter} supports this hypothesis as can be seen by the tighter clustering in the CSI case. 
However, further investigations are needed to fully illuminate this issue before solid conclusions can be drawn.

%%%%%%%%%%%%%%%%%%%%%%%%%%%%%%%%%%%%%%%%%%%%%%%%%%%%%%%%%%%
\section{Evaluation} \label{sec:evaluation}
%%%%%%%%%%%%%%%%%%%%%%%%%%%%%%%%%%%%%%%%%%%%%%%%%%%%%%%%%%%

In what follows, we showcase how specialised occupancy and activity sensing can be built atop our featurisation.

\vspace{-0.25cm}
%++++++++++++++++++++++++++++++++++++++++++++++++++++++++++
\subsection{Occupancy Detection}
%++++++++++++++++++++++++++++++++++++++++++++++++++++++++++

\begin{figure}
    \begin{center}
        \includegraphics[width=0.4\textwidth]{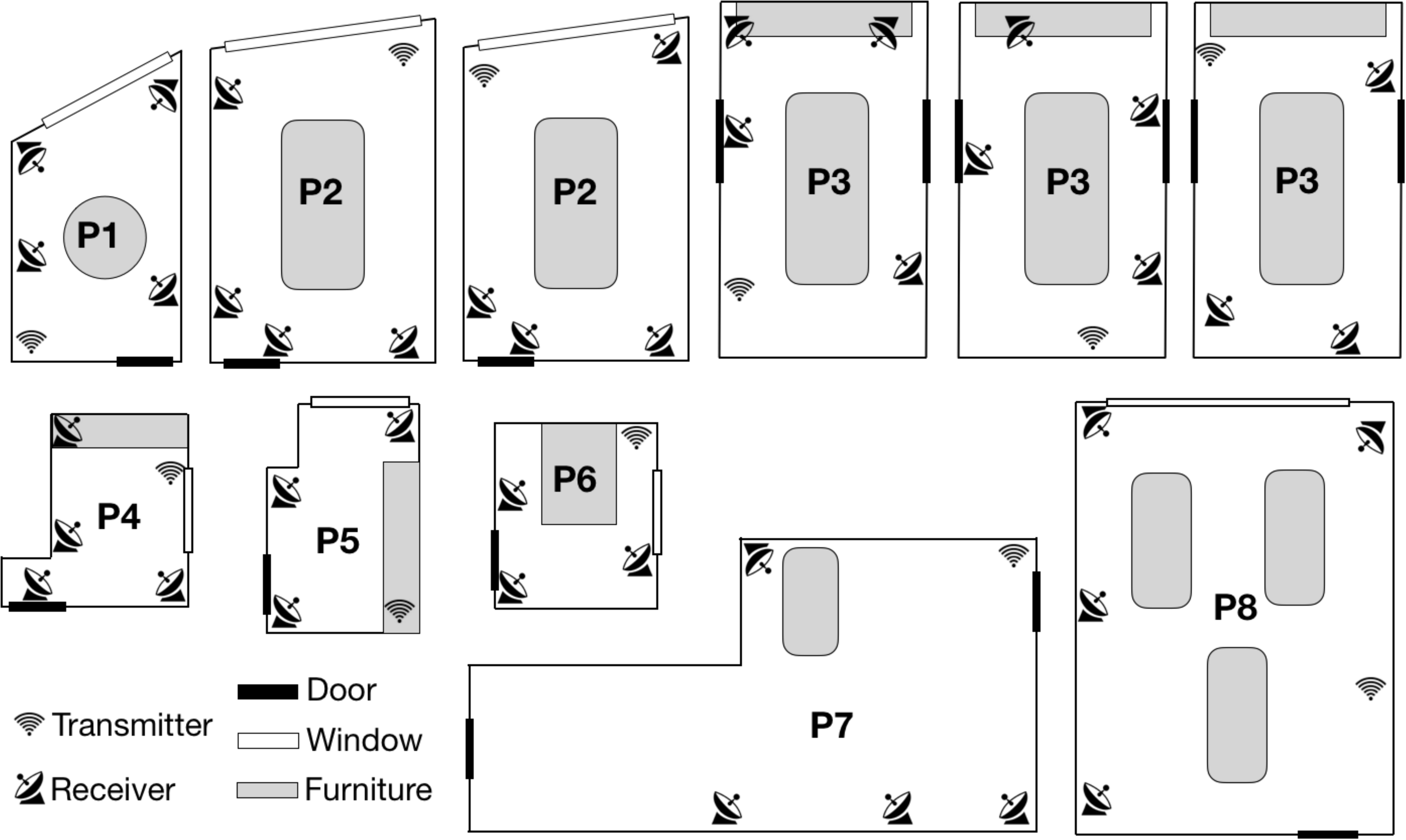}
        \caption{Device placements.\label{fig:device_placement}}
    \end{center}
\end{figure}

\subsubsection{Experimental setup}

We evaluate the performance of subspace tracking in terms of the robustness of occupancy detection. To evaluate the robustness, we investigate the accuracy of the classification model in new environments. More specifically, we trained the classification model using CSI data obtained from a certain placement and tested its accuracy on different placements.

\noindent \textbf{Data.} We collected the CSI data in 8 places and on 41 placements in total. As depicted in figure~\ref{fig:device_placement}, the places include six rooms, one lobby, and one lounge and have different characteristics such as room layout and furniture position. We collected the CSI data while varying the number of moving people from zero to 2 (P4, P5, P6) and to 3 (the rest). Each session lasted five minutes and participants were asked to freely move during the session. Figure~\ref{fig:device_placement} shows room layouts and device placements. 
\textcolor{highlight_clr}{The purpose of multiple placements are to investigate what is a realistic \emph{upper-bound} on the classification performance of a \emph{single} device under different training and testing conditions.}
MIMO CSI data were sampled at a nominal 500Hz rate. A stationarity period of $50$ms was used and the subspace update was performed in a sliding window fashion with no overlap as in Equation~\eqref{eq:batch_expectation}.

\noindent \textbf{Pipeline.} For the occupancy detection, we developed an inference pipeline using a long short-term memory (LSTM) classifier. We chose LSTM as a classifier to leverage spatio-temporal variation of our differential unitarity features from subspace tracking. In the current implementation, we adopted two hidden LSTM layers, each of which has 50 nodes. Some prior presence detection work dwells on the signal much longer with distribution-based approach while using a diversity of frequency channels~\cite{Guo17_WifiSmartHumanDynamicsMonitoring}. In contrast, we define a short $5$ seconds inference window and with no channel frequency diversity. In this paper, our objective is to showcase how to specialise various subspace tracking-based applications rather than demonstrate best-in-class performance.

\noindent \textbf{Comparison.} For comparison, we implemented the baseline pipeline from~\cite{domenico2016wpa}. It takes temporal variations of CSI data as feature values and uses linear discriminant analysis as a classifier.

\noindent \textbf{Training and test.} For training, we selected a receiver located at a diagonal position of the transmitter, thereby maximising the RF coverage. Accordingly, we have 11 different models. For the evaluation, we considered three environment variations, \textit{same}, \textit{minor}, and \textit{major}. \textit{Same} refers where the data from the same receiver, i.e., same placement, is used both for training and test. \textit{Minor} and \textit{major} use the CSI data from different receivers placed in the same room and different room, respectively. \textit{Same} represents the upper bound of the performance that the inference logic can achieve in a specific environment. \textit{Minor} and \textit{major} show how robust the inference pipeline is in unseen environments.

\subsubsection{Experimental results}

\begin{figure}[h]
  \vspace{-0.10in}
  \centering
    \subfloat[Environment variations\label{fig:eval_overall}]{\includegraphics[width=0.24\textwidth]{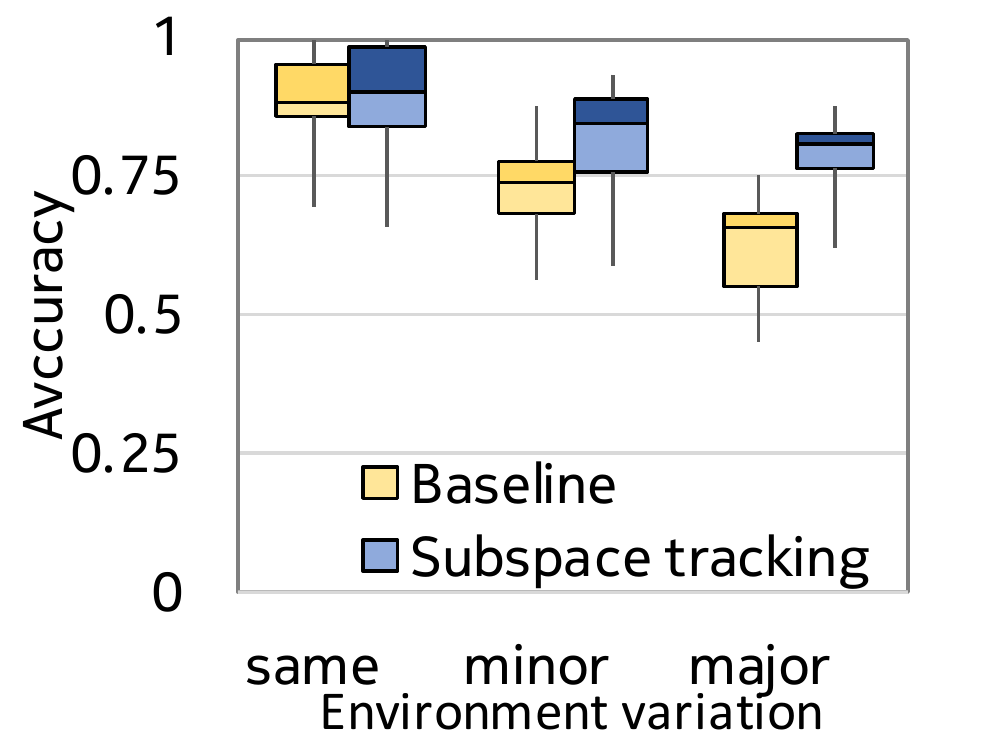}}
    \subfloat[Number of classes\label{fig:eval_classes}]{\includegraphics[width=0.24\textwidth]{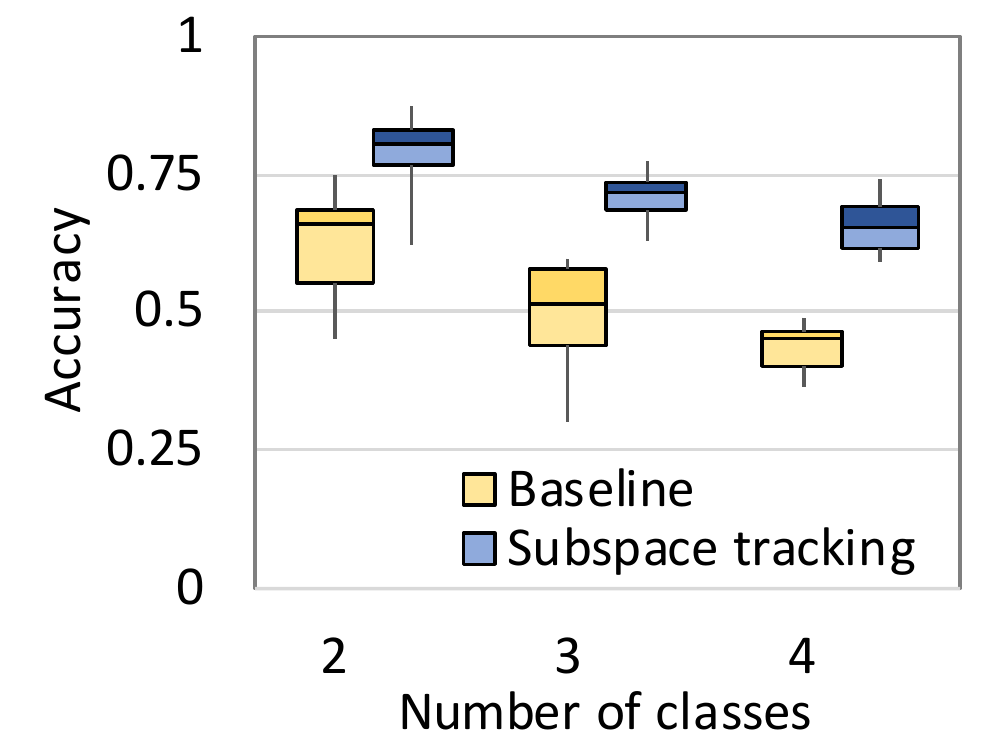}}
  \vspace{-0.075in}
  \caption{Occupancy performance.}
  \label{fig:occupancy_perf}
\end{figure}

We investigate how the subspace tracking effectively mitigates the environmental effect of CSI on the occupancy detection. Figure \ref{fig:eval_overall} shows the box plots of the accuracy of 11 models for different variations. Although the accuracy of both pipelines is similar in \textit{same} variation, the subspace tracking retains more competitive accuracy as we introduce minor and major environmental changes compared to the baseline. The accuracy in \textit{same} variation is 89\% and 88\% for the subspace tracking and baseline, respectively. However, in minor and major variations, the subspace tracking decreases to 82\% and 78\%, whereas the baseline does to 73\% and 62\%. 

We further investigate the effect of the number of classes on the occupancy detection on major variation. Figure \ref{fig:eval_classes} shows the box plots of the accuracy while varying the number of classes. \textit{2} classes represent presence detection, i.e., empty or occupied. \textit{3} and \textit{4} classes are for the number of people as [0, 1, 2+] and [0, 1, 2, 3], respectively.\footnote{By 2+ we mean 2 or greater than.} The results show that the subspace tracking achieves reasonable performance even with higher number of classes. Our pipeline shows 85\%, 70\% and 65\% for 2, 3, and 4 classes, respectively, whereas the baseline does 62\%, 49\%, and 43\%.

\begin{figure*}
  \centering
    \subfloat[ours: $\hat{u}_{\textrm{Dy},1}$ + DTW + K-nearest neighbours\label{fig:activityOurs}]{\includegraphics[width=0.40\textwidth]{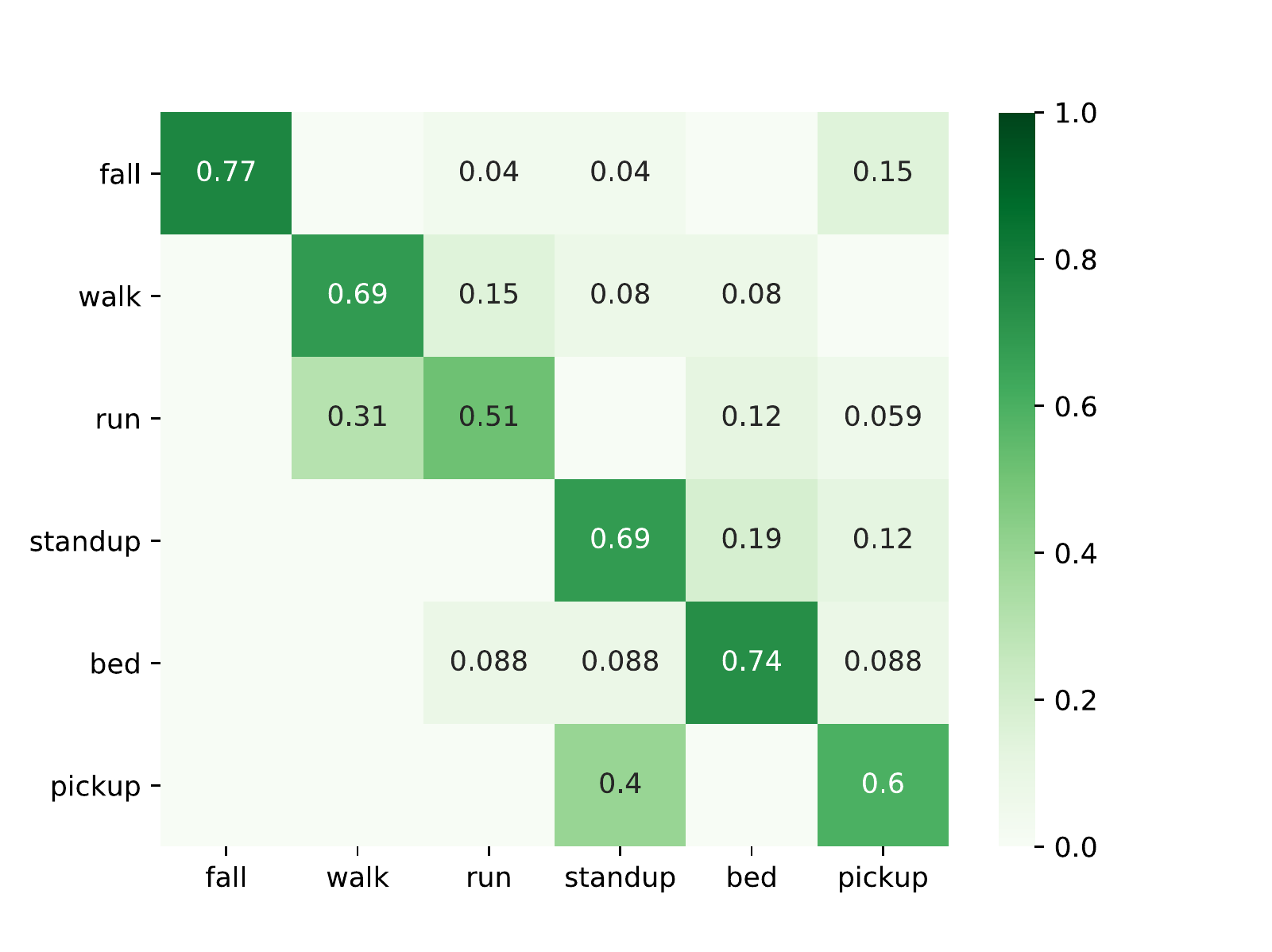}} \ 
    \subfloat[Yousef et al: PCA + STFT + HMM\label{fig:activityYousef}]{\includegraphics[width=0.40\textwidth]{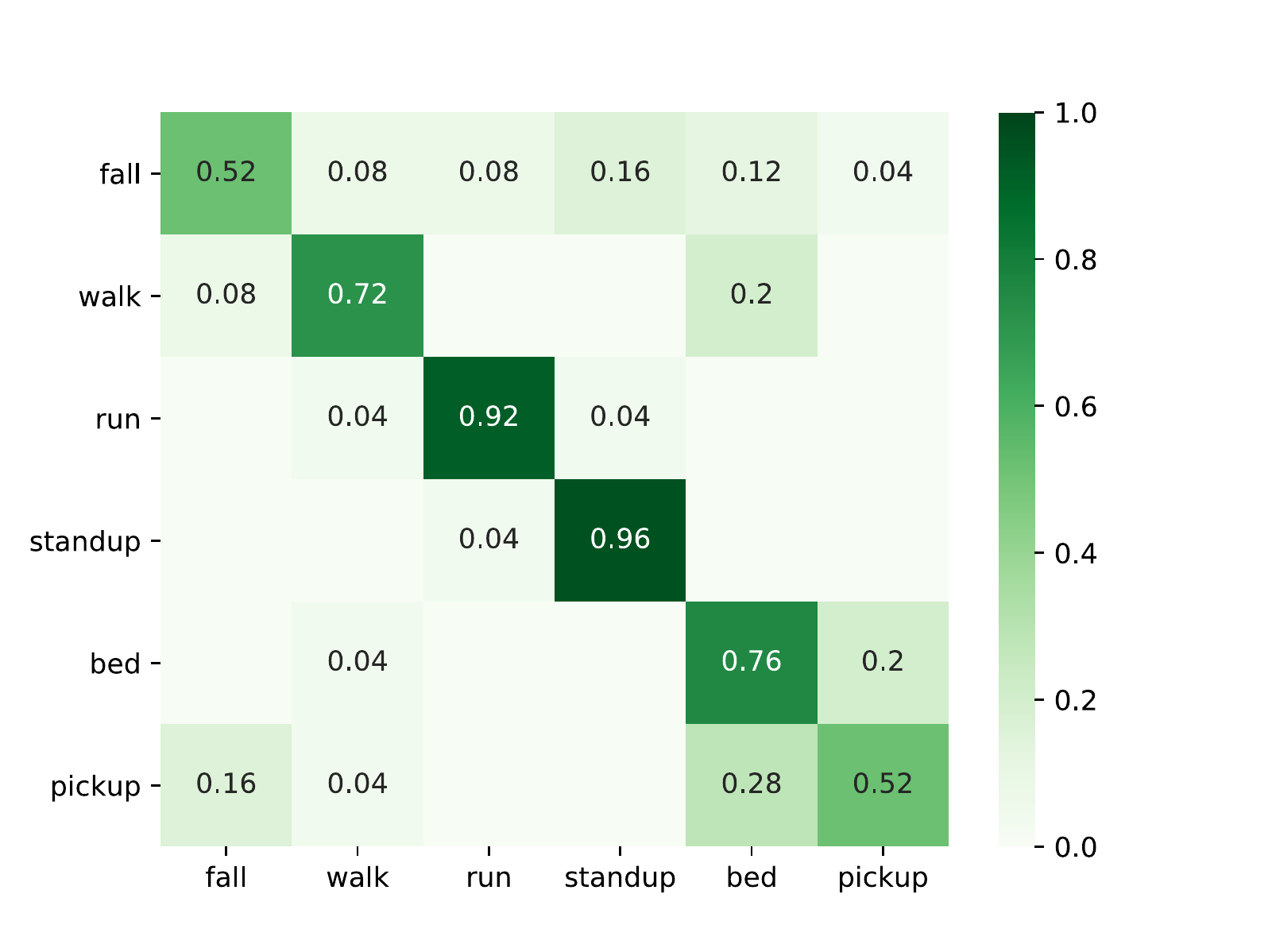}}
  \vspace{-0.1cm}
  \caption{Activity recognition performance.}
  \label{fig:activity_eval}
\end{figure*}

\vspace{-0.25cm}
%++++++++++++++++++++++++++++++++++++++++++++++++++++++++++
\subsection{Physical Activity}
%++++++++++++++++++++++++++++++++++++++++++++++++++++++++++

We use the activity recognition dataset available publicly by Yousef et al.~\cite{Yousefi17_SurveyOnWifiBehaviourRecognition} to demonstrate the applicability of our subspace tracking technique on the problem domain of activity classification.
The dataset is comprised of 6 single-user activities; namely, standing up, sitting down, lying down, falling, walking, and running. 
SIMO CSI data from three receiving multiple antennae is sampled at 1 ksps rate. 
We choose a stationarity period of $25$ms and proceed to update the covariance matrix with $95$\% CSI overlap from previous stationarity period with $\lambda = 0.99$ for recursive subspace update as in Equation~\eqref{eq:stochastic_expectation}. This gives around $800$ Hz subspace update rate. As illustrated previously in figure~\ref{fig:subspaceTracking_batchVsStochastic}, recursive subspace tracking filters background channel activity and/or subspace noise. This unwanted channel activity has been alluded to in Equations~\eqref{eq:channel-covariance}~\&~\eqref{eq:sig-noise-subspaces}.

In a preliminary evaluation, we build a simple classifier based around dynamic time warping (DTW) and K-nearest neighbours. This is applied to a single-dimensional ${\text{Dy}}$ slope differential unitarity (see figure~\ref{fig:diffUnitSlope}). We evaluate our classifier against the author's mid-range hidden Markov model (HMM) which uses a combination of PCA and the short-time Fourier transform (STFT) time-frequency localisation pre-processing. The results are shown in figure~\ref{fig:activity_eval}. Capability-wise, there is an asymmetry in that featurisation based around 2D STFT + HMM is in principle far stronger than our 1D DTW + K-nearest. Nonetheless, on the whole, the performance of our simple classifier is not far from that reported by Yousef et al, albeit with different characteristics. For instance, while 2D STFT + HMM outperforms our 1D DTW + K-nearest in nearly all activities, our fall activity performance is substantially better. We attribute this to the high acceleration content of fall which our slope metric is able to capture easily as shown in figure~\ref{fig:subspace_slope_falling} due to native acceleration sensing. Perhaps our pairwise metric with 2D time-frequency localisation would perform much better. Since our focus in this paper is to only showcase a generic formal featurisation suited for many applications, we leave improved classification for future work.

\vspace{-0.25cm}
%%%%%%%%%%%%%%%%%%%%%%%%%%%%%%%%%%%%%%%%%%%%%%%%%%%%%%%%%%%
\textcolor{highlight_clr}{\section{Discussion} \label{sec:discussion}}
%%%%%%%%%%%%%%%%%%%%%%%%%%%%%%%%%%%%%%%%%%%%%%%%%%%%%%%%%%%

\textcolor{highlight_clr}{
In this section, we provide commentary on the limitations of our work and discuss relevance to other wireless systems, thereby exposing items of future research.
}

\noindent \textbf{\textcolor{highlight_clr}{Applicability to other 802.11 standards.}} \textcolor{highlight_clr}{Physical propagation behaviour will differ depending on the frequency band. Such behaviour will be mirrored when viewed through the lens of the signal and noise subspaces. Our proposed featurisation provides sensing primitives to track the variations in propagation dynamics that are induced by human motion. However, it is the role of the machine learning (ML) component to capture such behaviour in a robust sensing model. Thus, when operating within different frequency bands, it is important to ensure that the back-end ML component is trained for the respective human-modulated propagation behaviour corresponding to that specific band. Our experimental results in this paper are for the 5GHz Wi-Fi band with 40MHz bandwidth. Nonetheless, other wireless standards---such as 802.11ah operating in the sub-1GHz band and 802.11ad/ay operating in the 60GHz band---could benefit from identical featurisation, albeit after specialising the back-end ML component to capture their individual propagation characteristics as a function of human motion. Moreover, we have shown in Section~\ref{sec:subspace_tracking} that the magnitude of our differential subspace tracking behaves identically irrespective of the representation of the channel response, be it in time or frequency. This means that both the single-carrier and OFDM variants of WiGig would benefit from our subspace-based featurisation. It is also worth pointing out that in relation to WiGig, 60GHz frequencies are quasi-optical and are less able to diffract around objects. The subspace will mirror this behaviour; however, increased coverage of the environment may be possible by considering the beam training procedure that 802.11ad/ay implements. Specifically, recent work has shown that such beam training procedure from infrastructure access points can be used to localise a mobile user~\cite{Palacios17_JadeMmwaveLocalisation}. It would be interesting in this particular example to see if tracking the subspace would allow for inferring finer-grained details on the nature of the mobile node's movement. OFDMA systems such as 802.11ax can also benefit from the proposed subspace tracking; however, care should be taken to handle instances of transition in user-assigned subcarriers and their implications on the subspace. 
}

\noindent \textbf{\textcolor{highlight_clr}{ML model coverage and vectors of variation.}} \textcolor{highlight_clr}{There are many variables that impact the robustness of the back-end ML model. We call these the vectors of variation of the ML model. Exhaustive training across these vectors of variation is needed for sufficient coverage of the sampling space in order to ensure the ML model generalises in the real-world. One such vector of variation is that arising from the individualised way in which different users perform activities. Broadly, there are two methods in prior art for dealing with such variations: design-based and learning-based. In design-based methods, hand-crafted features by an expert designer---such as careful frequency binning in~\cite{Wang18_SecurityByCreepingWave} and coarser wavelet spectral bins in~\cite{Wang15_UnderstandingAndModelingWiFiHumanActivity}---are engineered to absorb the expected variations in the real-world. In contrast, learning-based approaches rely on automatic coverage of these natural variations by the inference component through the sheer amount of empirical data used for training. In this paper, we focused on a formal and interpretable low-dimensional featurisation of the wireless channel, with our evaluation (cf. figure~\ref{fig:occupancy_perf}) falling under the latter learning-based approach.
}

\noindent \textbf{\textcolor{highlight_clr}{Axes of resolution.}} \textcolor{highlight_clr}{The performance of sensing applications built atop channel tracking is fundamentally limited by the spatio-temporal resolutions of channel measurements. Specifically, the utilised bandwidth and number of antennae have a large bearing on what can be perceived unambiguously in the environment i.e. without \emph{over-fitting} inference. To see this, consider the environmental imaging capability of the covariance $\mtrx{R}_{\textrm{x}}$ through its \emph{beamspace} representation $\mtrx{F}\mtrx{R}_{\textrm{x}}\mtrx{F}^H$, where $\mtrx{F}$ is the Fourier transform matrix~\cite{Weichselberger06_StochasticMimoChannelModelWithJointEndCorrelations,Costa08_NovelWidebandMimoChannelModel,Brady13_BeamspaceMIMO}. Clearly, for meaningful imaging, the number of antennae needs to be high in order to resolve environmental spatial scatterers. Similarly, bandwidth delivers the temporal resolution necessary for measuring the channel's delayspread (or frequency selectivity) more accurately. It is customary to see in related literature prolonged signal dwell times in order to compensate for the lack of spatio-temporal resolution as supplied by current research testbeds e.g. of the order of minutes dwell time to estimate occupancy in~\cite{Guo17_WifiSmartHumanDynamicsMonitoring,Depatla15_OccupancyEstimationUsingWifi}. To put it in wireless terms, clearly the ``coherence'' time of crowd movement indoors is much shorter than 5 minutes. We, therefore, would argue that practical indoor channel sensing systems are likely to appear once we begin to see the roll-out of wireless infrastructure of enhanced spatio-temporal resolutions such as indoor massive MIMO in the millimetre-wave band. 
}

\vspace{+0.25cm}
%%%%%%%%%%%%%%%%%%%%%%%%%%%%%%%%%%%%%%%%%%%%%%%%%%%%%%%%%%%
\section{Conclusion} \label{sec:conclusion}
%%%%%%%%%%%%%%%%%%%%%%%%%%%%%%%%%%%%%%%%%%%%%%%%%%%%%%%%%%%

In this paper, we formalise the problem of Wi-Fi-based human sensing and cast it as a channel signal subspace tracking task. We demonstrate the equivalence of the two problems. We posit the optimality of such formulation citing prior established work from wireless literature. We conclude by providing evidence for the applicability of our subspace tracking across two usage scenarios: presence detection and activity recognition with promising early results. Future work will focus on machine learning classification using our subspace-based featurisation.


\vspace{-0.25cm}
{\footnotesize \bibliographystyle{IEEEtran}
\bibliography{./rsense_abrv}}

\begin{thebibliography}{10}
\providecommand{\url}[1]{#1}
\csname url@samestyle\endcsname
\providecommand{\newblock}{\relax}
\providecommand{\bibinfo}[2]{#2}
\providecommand{\BIBentrySTDinterwordspacing}{\spaceskip=0pt\relax}
\providecommand{\BIBentryALTinterwordstretchfactor}{4}
\providecommand{\BIBentryALTinterwordspacing}{\spaceskip=\fontdimen2\font plus
\BIBentryALTinterwordstretchfactor\fontdimen3\font minus
  \fontdimen4\font\relax}
\providecommand{\BIBforeignlanguage}[2]{{%
\expandafter\ifx\csname l@#1\endcsname\relax
\typeout{** WARNING: IEEEtran.bst: No hyphenation pattern has been}%
\typeout{** loaded for the language `#1'. Using the pattern for}%
\typeout{** the default language instead.}%
\else
\language=\csname l@#1\endcsname
\fi
#2}}
\providecommand{\BIBdecl}{\relax}
\BIBdecl

\bibitem{Yousefi17_SurveyOnWifiBehaviourRecognition}
S.~Yousefi, H.~Narui, S.~Dayal, S.~Ermon, and S.~Valaee, ``{A Survey on
  Behavior Recognition Using WiFi Channel State Information},''
  \url{https://github.com/ermongroup/Wifi_Activity_Recognition}, pp. 98--104,
  OCTOBER 2017.

\bibitem{Xi14_ElectronicFrogEyeCrowdCounting}
W.~Xi, J.~Zhao, X.~Y. Li, K.~Zhao, S.~Tang, X.~Liu, and Z.~Jiang, ``{Electronic
  frog eye: Counting crowd using WiFi},'' in \emph{IEEE INFOCOM}, April 2014,
  pp. 361--369.

\bibitem{Depatla15_OccupancyEstimationUsingWifi}
S.~Depatla, A.~Muralidharan, and Y.~Mostofi, ``{Occupancy Estimation Using Only
  WiFi Power Measurements},'' \emph{IEEE J. Sel. Areas Commun.}, vol.~33,
  no.~7, pp. 1381--1393, July 2015.

\bibitem{Wang18_LowEffortDfl}
J.~Wang, J.~Xiong, H.~Jiang, K.~Jamieson, X.~Chen, D.~Fang, and C.~Wang, ``{Low
  Human-Effort, Device-Free Localization with Fine-Grained Subcarrier
  Information},'' \emph{IEEE Trans. Mobile Comput.}, 2018.

\bibitem{Wang15_UnderstandingAndModelingWiFiHumanActivity}
W.~Wang, A.~X. Liu, M.~Shahzad, K.~Ling, and S.~Lu, ``{Understanding and
  Modeling of WiFi Signal Based Human Activity Recognition},'' in \emph{Proc.
  of MobiCom '15}.\hskip 1em plus 0.5em minus 0.4em\relax New York, NY, USA:
  ACM, 2015, pp. 65--76.

\bibitem{Costa08_NovelWidebandMimoChannelModel}
N.~Costa and S.~Haykin, ``{A novel wideband MIMO channel model and experimental
  validation},'' \emph{IEEE Trans. Antennas Propag.}, vol.~56, no.~2, pp.
  550--562, 2008.

\bibitem{Weichselberger06_StochasticMimoChannelModelWithJointEndCorrelations}
W.~Weichselberger, M.~Herdin, H.~Ozcelik, and E.~Bonek, ``{A stochastic MIMO
  channel model with joint correlation of both link ends},'' \emph{IEEE Trans.
  Wireless Commun.}, vol.~5, no.~1, pp. 90--100, 2006.

\bibitem{Tulino06_Capacity-achievingMimoCovariance}
A.~M. Tulino, A.~Lozano, and S.~Verd{\'u}, ``{Capacity-Achieving Input
  Covariance for Single-User Multi-Antenna Channels},'' \emph{IEEE Trans.
  Wireless Commun.}, vol.~5, no.~3, 2006.

\bibitem{Lopez-Martinez15_EigenvalueDynamicsOfCentralWishartForMimo}
F.~J. Lopez-Martinez, E.~Martos-Naya, J.~F. Paris, and A.~Goldsmith,
  ``{Eigenvalue dynamics of a central Wishart matrix with application to MIMO
  systems},'' \emph{{IEEE Trans. Inf. Theory}}, vol.~61, no.~5, pp. 2693--2707,
  2015.

\bibitem{Delmas10_SubspaceTrackingForSigProc}
J.~P. Delmas, ``{Subspace Tracking for Signal Processing},'' \emph{{Adaptive
  Signal Processing: Next Generation Solutions}}, pp. 211--270, 2010.

\bibitem{3GPP_TU6}
\emph{3rd Generation Partnership Project; Technical Specification Group
  GSM/EDGE Radio Access Network; Radio transmission and reception (Release
  1999), Annex C.3 Propagation models}, 3GPP, 1999, v8.20.0.

\bibitem{Smith15_ChannelModelingForWBAN}
D.~B. Smith and L.~W. Hanlen, ``Channel modeling for wireless body area
  networks,'' in \emph{Ultra-Low-Power Short-Range Radios}.\hskip 1em plus
  0.5em minus 0.4em\relax Springer, 2015, pp. 25--55.

\bibitem{Smith13_PropagationModelsForBAN}
D.~B. Smith, D.~Miniutti, T.~A. Lamahewa, and L.~W. Hanlen, ``Propagation
  models for body-area networks: A survey and new outlook,'' \emph{IEEE
  Antennas Propag. Mag.}, vol.~55, no.~5, pp. 97--117, 2013.

\bibitem{Fort07_IndoorBanModelForNarrowbandComms}
A.~Fort, C.~Desset, P.~Wambacq, and L.~V. Biesen, ``Indoor body-area channel
  model for narrowband communications,'' \emph{IET Microwaves, Antennas
  Propagation}, vol.~1, no.~6, pp. 1197--1203, Dec 2007.

\bibitem{Scharf94_MatchedSubspaceDetectors}
L.~Scharf and B.~Friedlander, ``{Matched Subspace Detectors},'' \emph{IEEE
  Trans. Signal Process.}, vol.~42, no.~8, pp. 2146--2157, Aug. 1994.

\bibitem{Kraut01_AdaptiveSubspaceDetectors}
S.~Kraut, L.~Scharf, and L.~McWhorter, ``{Adaptive Subspace Detectors},''
  \emph{IEEE Trans. Signal Process.}, vol.~49, no.~1, pp. 1--16, Jan. 2001.

\bibitem{Studer18_ChannelCharting}
C.~Studer, S.~Medjkouh, E.~G{\"o}n{\"u}lta{\c{s}}, T.~Goldstein, and
  O.~Tirkkonen, ``Channel charting: Locating users within the radio environment
  using channel state information,'' \emph{IEEE Access}, vol.~6, pp.
  47\,682--47\,698, 2018.

\bibitem{Zhang12_GeneralCoupling-basedModelForWidebandMimoChannels}
Y.~Zhang, O.~Edfors, P.~Hammarberg, T.~Hult, X.~Chen, S.~Zhou, L.~Xiao, and
  J.~Wang, ``A general coupling-based model framework for wideband {MIMO}
  channels,'' \emph{IEEE Trans. Antennas Propag.}, vol.~60, no.~2, pp.
  574--586, 2012.

\bibitem{Davis70_RotationOfEigenvectorsByPerturbation}
C.~Davis and W.~M. Kahan, ``{The Rotation of Eigenvectors by a Perturbation.
  III},'' \emph{SIAM Journal on Numerical Analysis}, vol.~7, no.~1, pp. 1--46,
  1970.

\bibitem{Guo17_WifiSmartHumanDynamicsMonitoring}
X.~Guo, B.~Liu, C.~Shi, H.~Liu, Y.~Chen, and M.~C. Chuah, ``{WiFi-Enabled Smart
  Human Dynamics Monitoring},'' in \emph{Proc. of SenSys '17}.\hskip 1em plus
  0.5em minus 0.4em\relax New York, NY, USA: ACM, 2017, pp. 16:1--16:13.

\bibitem{domenico2016wpa}
S.~Di~Domenico, M.~De~Sanctis, E.~Cianca, and G.~Bianchi, ``{A Trained-once
  Crowd Counting Method Using Differential WiFi Channel State Information},''
  in \emph{Proc. of WPA '16}.\hskip 1em plus 0.5em minus 0.4em\relax New York,
  NY, USA: ACM, 2016, pp. 37--42.

\bibitem{Palacios17_JadeMmwaveLocalisation}
J.~Palacios, P.~Casari, and J.~Widmer, ``{JADE: Zero-knowledge device
  localization and environment mapping for millimeter wave systems},'' in
  \emph{IEEE INFOCOM 2017}, pp. 1--9.

\bibitem{Wang18_SecurityByCreepingWave}
W.~Wang, L.~Yang, Q.~Zhang, and T.~Jiang, ``{Securing On-Body IoT Devices By
  Exploiting Creeping Wave Propagation},'' \emph{IEEE J. Sel. Areas Commun.},
  2018.

\bibitem{Brady13_BeamspaceMIMO}
J.~Brady, N.~Behdad, and A.~M. Sayeed, ``{Beamspace MIMO for millimeter-wave
  communications: System architecture, modeling, analysis, and measurements},''
  \emph{IEEE Trans. Antennas Propag.}, vol.~61, no.~7, pp. 3814--3827, 2013.

\end{thebibliography}

\balance

\end{document}